\documentclass[goettingen, submitted, gauss, englishtitle, print]{thesis}					

\usepackage{graphicx}
\usepackage[percent]{overpic}
\usepackage{xcolor}
\usepackage{color}
\usepackage{tikz}
\usetikzlibrary{shapes,arrows,shapes.multipart,positioning,fit,calc}
\usepackage{pdfpages}

\usepackage{txfonts}
\usepackage{wasysym}
\usepackage{fixltx2e}
\usepackage[
	colorlinks=true,
	linkcolor=black,
	citecolor=black,
	filecolor=black,
	urlcolor=black,
	bookmarks=true,
	bookmarksopen=true,
	bookmarksopenlevel=3,
	plainpages=false,
	pdfpagelabels=true,
]{hyperref}

\usepackage{siunitx}
\DeclareSIUnit\au{au}
\DeclareSIUnit\mag{mag}
\newcommand*{\ra}[2][]{%
	\ang[
	math-degree=\textsuperscript{h},
	text-degree=\textsuperscript{h},
	math-arcminute=\textsuperscript{m},
	text-arcminute=\textsuperscript{m},
	math-arcsecond=\textsuperscript{s},
	text-arcsecond=\textsuperscript{s},
	#1]{#2}%
}

\usepackage[ugly]{units}
\usepackage{chemformula}
\usepackage[english=british]{csquotes}
\usepackage{longtable}
\usepackage[para,online]{threeparttable}
\usepackage{dpfloat}
\usepackage{booktabs}
\usepackage{subcaption}
\usepackage{blindtext}
\usepackage{makecell}

\newcommand{\diff}[2]{\frac{\mathrm{d}#1}{\mathrm{d}#2}}
\newcommand{\dif}[1]{\mathrm{d}#1}

\newcommand*\asymmnumerr[3]{%
	\ensuremath{\num{#1}%
		_{\num[retain-explicit-plus,explicit-sign=-]{#2}}%
		^{\num[retain-explicit-plus,explicit-sign=+]{#3}}%
}}

\usepackage{natbib}
\bibpunct{(}{)}{;}{a}{}{,}	

\usepackage{textalpha}
\newcommand{\textgreek}[1]{\begingroup\fontencoding{LGR}\selectfont#1\endgroup}

\usepackage[hang,splitrule]{footmisc}
\usepackage{rotating}
\usepackage{microtype}


\title{Timing of stellar pulsations to search for sub-stellar companions beyond the main sequence}
\author{Felix Mackebrandt}
\town{Brandenburg/Havel, Deutschland}  

\thesisadvisorycommitteea{Dr. Sonja Schuh}
\instthesisadvisorycommitteea{Max-Planck-Institut f\"ur Sonnensystemforschung, G\"{o}ttingen, Germany}

\thesisadvisorycommitteeb{Prof. Dr. Laurent Gizon}
\instthesisadvisorycommitteeb{Max-Planck-Institut f\"ur Sonnensystemforschung, G\"{o}ttingen, Germany;\\
Institut f\"ur Astrophysik, Georg-August-Universit\"{a}t G\"{o}ttingen, Germany}

\thesisadvisorycommitteec{Prof. Dr. Stefan Dreizler}
\instthesisadvisorycommitteec{Institut f\"{u}r Astrophysik, Georg-August-Universit\"{a}t G\"{o}ttingen, Germany}

\refereea{Dr. Sonja Schuh}
\instrefereea{Max-Planck-Institut f\"ur Sonnensystemforschung, G\"{o}ttingen, Germany}

\refereeb{Prof. Dr. Stefan Dreizler}
\instrefereeb{Institut f\"{u}r Astrophysik, Georg-August-Universit\"{a}t G\"{o}ttingen, Germany}

\commissiona{Prof. Dr. Laurent Gizon}
\instcommissiona{Max-Planck-Institut f\"ur Sonnensystemforschung, G\"{o}ttingen, Germany;\\
Institut f\"ur Astrophysik, Georg-August-Universit\"{a}t G\"{o}ttingen, Germany}

\commissionb{Dr. Roberto Silvotti}
\instcommissionb{INAF-Osservatorio Astrofisico di Torino, Pino Torinese, Italy}

\commissionc{Prof. Dr. Ariane Frey}
\instcommissionc{II. Physikalisches Institut, Georg-August-Universit\"{a}t G\"{o}ttingen, Germany}

\commissiond{Prof. Dr. Laura Covi}
\instcommissiond{Institut f\"ur Theoretische Physik, Georg-August-Universit\"{a}t G\"{o}ttingen, Germany}

\submitteddate{17.03.2020}
\submittedyear{2020}
\examinationdate{22.06.2020}
\publicationyear{2020}

\definecolor{cb-red}{HTML}{E41A1C}
\definecolor{cb-blue}{HTML}{377EB8}
\definecolor{cb-green}{HTML}{4DAF4A}
\definecolor{cb-purple}{HTML}{984EA3}
\definecolor{cb-orange}{HTML}{FF7F00}

\begin{document}
	
\pagenumbering{Roman}

\includepdf[pages={1}]{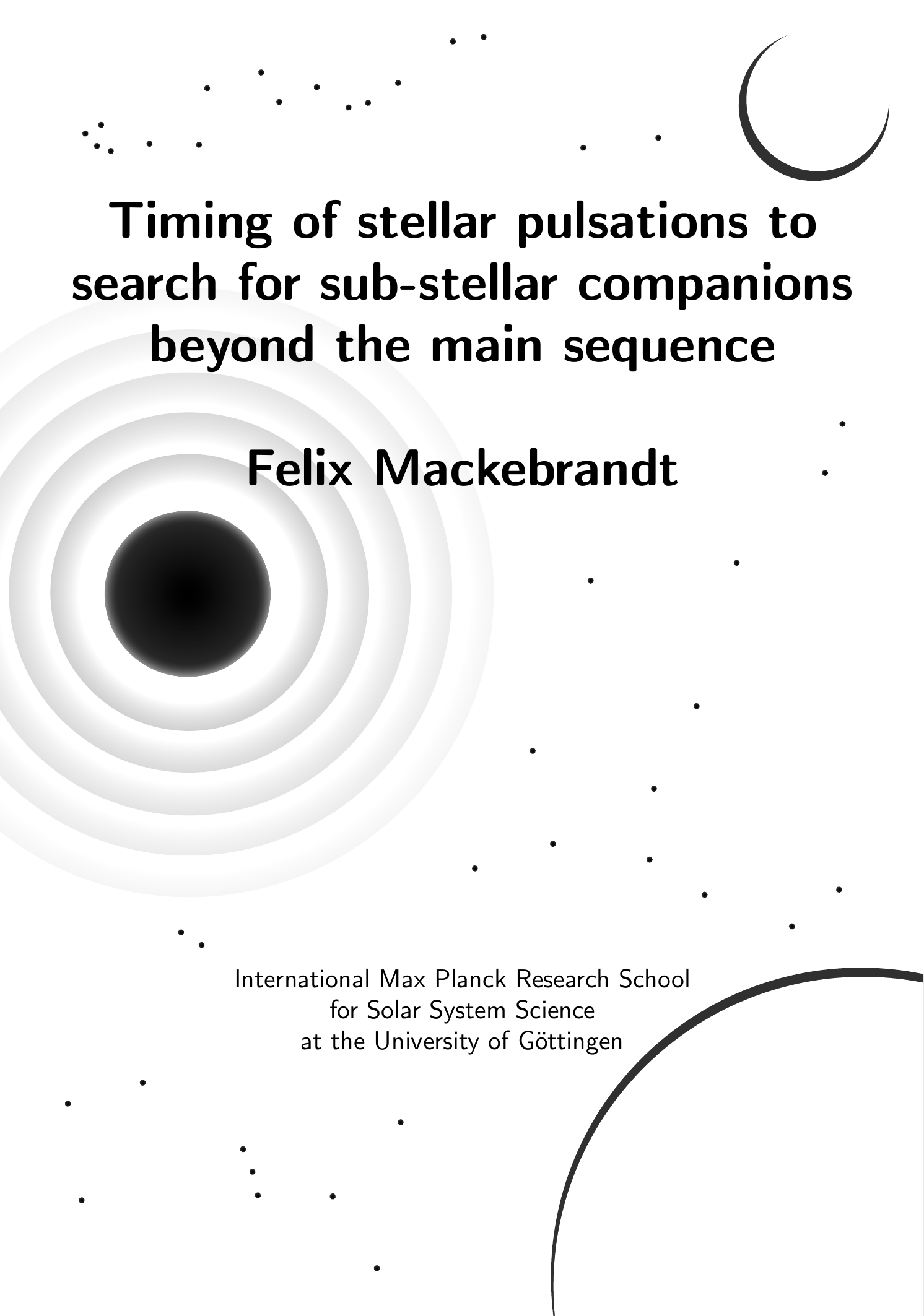}

\maketitle

\tableofcontents \newpage
\listoffigures \newpage
\listoftables \newpage

\cleardoublepage
\pagenumbering{arabic}

\chapter*{Abstract\markboth{Abstract}{Abstract}}
\addcontentsline{toc}{chapter}{Abstract}
Stars spend most of their life on the main sequence (MS). But their most substantial changes occur off the MS stage, either before on the pre-MS or beyond at the post-MS phase. 
Due to very complex and varied dynamical processes, the evolution of planetary systems orbiting non-MS stars significantly differs from those of MS planetary systems.

This work focusses on the search for sub-stellar companions in post-MS systems and determination of the evolutionary state of their host stars, especially subdwarf B stars (sdB stars). These are stripped Helium-burning cores of red giants with a thin hydrogen atmosphere. The canonical model involves binary evolution to explain the existence of sdB stars. Formation scenarios for single sdBs are more controversially discussed and can be hard to reconcile with observational properties. Besides the merger of two helium white dwarfs or other merger processes for apparently single sdB stars, an alternative formation channel involves planetary systems. During the red giant phase, the star would develop a common envelope with a giant planet that leads to the loss of the envelope. Thus, sdB stars are laboratories to test how planets survive and influence the late phases of stellar evolution.

The rapid pulsations of sdB stars can be used to detect sub-stellar companions from periodic variations in the expected arrival times of the pulsation maxima. This timing method is particularly sensitive to companions at large distances and complementary to other exoplanet detection methods because they are not efficient for stars with small radii and high gravities. Thus, the timing method opens up a new parameter range in terms of the host stars and helps to understand the formation process of single sdBs. 

In this work I implemented, tested and applied the pulsation timing analysis to search for sub-stellar companions in late evolutionary stage stellar systems. The method is already established in the literature but not to an extent which is capable of automatically processing long-time series of high-cadence data, i.e., from space born observations. 

Part~\ref{sec:intro} provides an introduction to extrasolar planets, and to the formation and properties of sdB stars.

Part~\ref{sec:exotime-project} and \ref{sec:exotime} describe the long-term ground-based observations of four rapidly pulsating sdB stars DW~Lyn, V1636~Ori, QQ~Vir and V541~Hya. The data are used to measure the secular drifts in pulsation periods. The results constrain the evolutionary state of these stars and are compared to theoretical predictions of stellar evolutionary models. Furthermore, the measurements set limits to masses and orbital periods of sub-stellar companions. In contrast to previous studies, tentative companion detections are not confirmed.  
  
Part~\ref{sec:other-targets} describes the application of the implemented timing analysis to other pulsating stars and data sets. Compared to ground-based observatories, satellite-based telescopes offer the advantage of uninterrupted observations. Observatories like \textit{Kepler}, TESS or the upcoming PLATO mission provide a large sample of targets. Besides sdB stars, \textdelta~Scuti (\textdelta~Sct) pulsators are excellent stars to apply the timing method on. \textdelta~Sct stars are evolved beyond the MS and of spectral type A. From \textit{Kepler} observations, previous studies revealed a planetary companion orbiting the \textdelta~Sct star KIC~791748. The implemented timing analysis of this work is applied to these data and can recover the planetary signature, validating the implementation at hand and independently confirming the planetary companion discovery.

Part~\ref{sec:summary} discusses the results of this thesis and provides an outlook to further applications.

\chapter*{Zusammenfassung\markboth{Zusammenfassung}{Zusammenfassung}}
\addcontentsline{toc}{chapter}{Zusammenfassung}
Sterne verbringen den Gro\ss teil ihres Lebens auf der Hauptreihe (HR). Die gravierendsten Ver\"anderungen geschehen jedoch abseits der HR, entweder zuvor auf der pre-HR oder folgend auf der post-HR. Durch sehr komplexe und variierende dynamische Prozesse, unterscheidet sich die Entwicklung von Planeten in nicht-HR Sternsystemen signifikant von der Entwicklung von Planeten in HR Sternsystemen.

Diese Arbeit legt ihren Fokus auf die Suche nach sub-stellaren Begleitern in post-HR Systemen und auf die Bestimmung des evolution\"aren Stadiums derer Zentralsterne, insbesondere bei hei\ss en Unterzwergen des Spektraltyps B (engl.: subdwarf B, kurz: sdB), auch Blaue-Unterzwerge genannt. Solche sdB-Sterne sind freigelegte, heliumbrennende Kerne, von vormals Roten-Riesen, mit einer d\"unnen Wasserstoffatmosph\"are. Deren kanonische Entstehung beschreibt die Entwicklung in einem Doppelsternsystem. Entstehungsszenarien f\"ur einzelne sdB-Sterne werden jedoch kontrovers diskutiert und sind nur schwer mit Beobachtungen zu verifizieren. Neben der Verschmelzung von zwei heliumdominierten Wei\ss en Zwergen, k\"onnten auch Planetensysteme zur Bildung von sdB-Sternen beitragen. W\"ahrend der Roten-Riesen-Phase, w\"urde der Stern einen seiner Riesenplaneten in eine gemeinsame H\"ulle einschlie\ss en, was letztendlich zum Verlust der H\"ulle f\"uhrt. Somit erweisen sich sdB-Sterne als Testobjekte, das \"Uberleben von Planeten zu pr\"ufen, sowie den Einfluss der Planeten auf die Sp\"atphasen stellare Entwicklung zu beobachten.

Die schnellen Pulsationen von sdB Sternen k\"onnen, mit Hilfe von periodischen Ver\"anderungen in den zu erwarteten Ankunftszeiten der Pulsationsmaxima, zur Detektion von sub-stellaren Begleitern genutzt werden. Diese Methode der Lichtlaufzeitmessungen ist besonders sensitiv f\"ur Begleiter mit gro\ss er Distanz zu ihrem Zentralstern und damit komplement\"ar zu anderen Detektionsmethoden. Denn andere Detektionsmethoden sind bei Sternen mit hoher Oberfl\"achengravitation und kleinen Radien weniger aussagekr\"aftig. Damit er\"offnet die Methode der Lichtlaufzeitmessungen eine M\"oglichkeit f\"ur Entdeckungen von bisher nicht erforschbaren Planetensystemen und hilft somit den Entstehungsprozess von sdB-Sternen besser zu verstehen.

In dieser Arbeit habe ich die Methode der Pulsations-Lichtlaufzeitmessungen implementiert, getestet und zur Suche nach sub-stellaren Begleitern bei Systemen in den Sp\"atphasen stellarer Entwicklung angewandt. Die Methode ist bereits etabliert, bisher jedoch nicht f\"ur eine automatische Anwendung auf lange Beobachtungsreihen in hoher zeitlicher Aufl\"osung, wie z.B. von Weltraumteleskopen, optimiert.

Teil~\ref{sec:intro} dieser Arbeit gibt eine Einleitung zu extrasolaren Planeten, sowie auch zur Entstehung und Eigenschaften von sdB-Sternen.

Teil~\ref{sec:exotime-project} und \ref{sec:exotime} beschreiben die bodengebundenen Langzeitbeobachtungen der vier schnell pulsierenden sdB-Sterne DW~Lyn, V1636~Ori, QQ~Vir und V541~Hya. Mit Hilfe dieser Daten werden die Ver\"anderungen der Pulsationsperioden gemessen. Diese Ergebnisse lassen auf das evolution\"are Stadium der Sterne schlie\ss en und werden mit theoretischen Sternentwicklungsmodellen verglichen. Weiterhin setzen die Ergebnisse der Beobachtungen Grenzwerte f\"ur die Masse und Umlaufdauer m\"oglicher sub-stellarer Begleiter. Im Widerspruch zu vorhergehenden Studien, k\"onnen potentielle Detektionen von Begleitern nicht best\"atigt werden.

Teil~\ref{sec:other-targets} beschreibt die Anwendung der implementierten Methode der Pulsations-Lichtlaufzeitmessungen auf andere Sterne und Datens\"atze. Im Vergleich zu bodengebundenen Observatorien bieten weltraumgest\"utzte Teleskope den Vorteil der ununterbrochenen Beobachtung. Observatorien wie das \textit{Kepler} Weltraumobservatorium, TESS oder die geplante PLATO Mission, liefern eine gro\ss e Anzahl von Beobachtungszielen. Neben sdB-Sternen ist die Klasse der pulsierenden  \textdelta~Scuti (\textdelta~Sct) Sterne ein hervorragendes Ziel f\"ur Lichtlaufzeitmessungen. \textdelta~Sct Sterne sind in einem Entwicklungsstadium jenseits der HR und in der Spektralklasse A zu finden. Vorhergehende Studien haben aus Beobachtungen des \textit{Kepler} Weltraumteleskops einen planetaren Begleiter um den Stern KIC~791748 offenbart. Mit der in dieser Arbeit implementierten Methode der Pulsations-Lichtlaufzeitmessungen kann das Signal dieses Begleiters aus den Daten gewonnen und somit die Implementierung validiert, sowie eine unabh\"angige Best\"atigung dieser Planetenentdeckung erbracht werden.

Teil~\ref{sec:summary} diskutiert die vorhergehenden Ergebnisse und liefert einen Ausblick auf weitere Anwendungen.

\part{Introduction} \label{sec:intro}
\chapter{Extrasolar planets}

\section{Historical overview}

Speculations about the existence of other worlds or solar systems other than our own go as far back as the ancient times. Greek philosophers debated about the possibility of countless worlds: 
\begin{quote}
	And he maintained worlds to be infinite, and varying in bulk; and that in some there is neither sun nor moon, while in others that they are larger than with us, and with others more numerous. [...] And that some worlds are destitute of animals and plants, and every species of moisture.
	
	\hfill Democritus \citep[\numrange{460}{370} BCE;][]{litwa_refutation_2016}
\end{quote}
He was supported by Epicurus (\numrange{314}{270} BCE) who debated about an \enquote{infinite number of worlds both like and unlike this world of ours} \citep{bailey_epicurus_1926} in a letter to Democritus. But the atomists of this time believed in the geocentric universe, and thus the work of Aristotle (\numrange{341}{270} BCE) outshone their ideas. This started to change with the introduction of the Heliocentric universe theory of Nicolaus Copernicus in 1543. However, the Italian philosopher Giordano Bruno was burned during the Inquisition because of his belief in an infinite number of stars which have an infinite number of terrestrial worlds orbiting them. Finally, Galileo Galilei discovered with his telescope, invented in 1606, the true nature of the Moon and planets and paved the way for the discovery of the remaining planets in our solar system. 

Our solar system shaped the understanding of planetary evolution and system architecture, until \citet{wolszczan_planetary_1992} discovered the first exoplanetary system later confirmed by \citet{wolszczan_confirmation_1994}. The two Earth-mass like planets orbit the millisecond pulsar PSR~1257+12, a rather unusual host star. With the discovery of the first extrasolar planet, 51~Pegasi\,b orbiting a sun-like star \citep{mayor_jupiter-mass_1995}, the architecture of exoplanetary systems remained very different than expected from the solar system. The planet has a mass comparable to the mass of Jupiter but a very close-in orbit of about \SI{0.05}{\au}, leading to a new category of planets called \enquote{hot Jupiters}. 

\begin{figure}[t]
	\centering
	\input{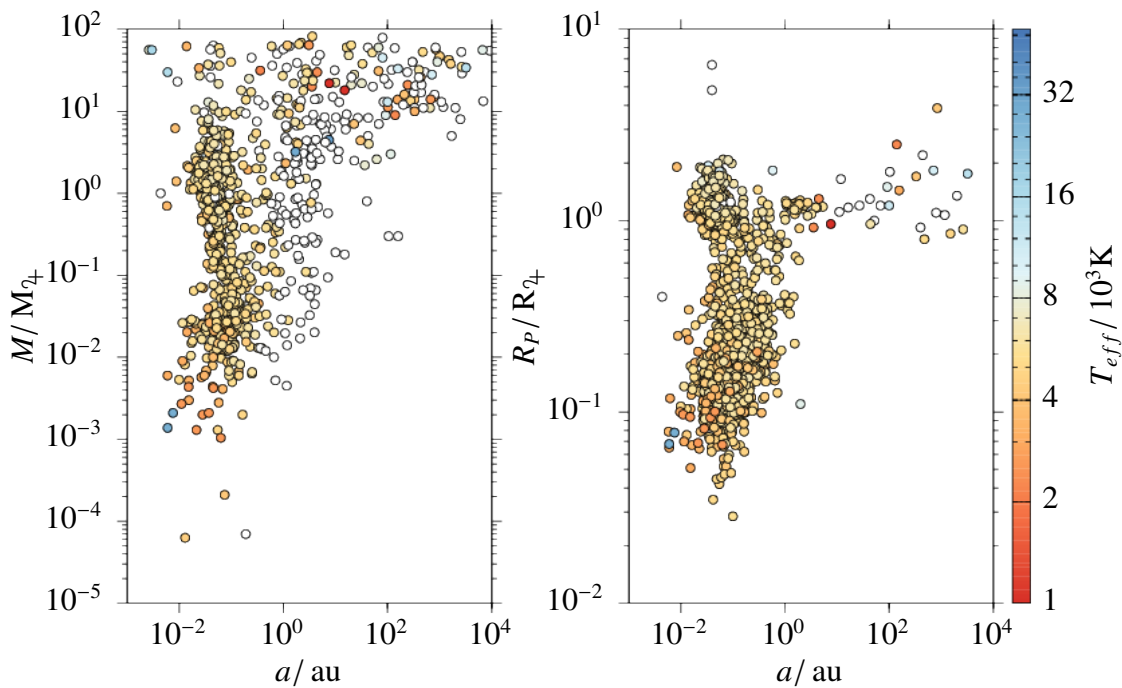}
	\caption[Exoplanet mass and radius distribution.]{Exoplanet mass and radius distribution as function of their orbital period and effective temperature of their host star (color coded). Out of the \num{4130} discovered exoplanets, there are \num{994} mass and \num{1631} radius measurements \citep[as of November 2019;][]{schneider_defining_2011}.}
	\label{fig:intro:exoplanets}
\end{figure}

So far, more than \num{4000} exoplanets have been discovered by various detection methods. They exhibit a wide range of planetary masses and radii as shown in Figure~\ref{fig:intro:exoplanets}. Besides hot Jupiters, Neptune- and Saturn-sized planets as well as super-Earths were discovered. 

The first transiting exoplanet HD~209458\,b was discovered in 1999 \citep{charbonneau_detection_2000} which later became the first planet with a spectroscopically analysed atmosphere \citep{charbonneau_detection_2002}. HD~28185\,b was discovered as the first exoplanet in the habitable zone, a region around a star in which a planet may retain liquid water on its surface. Using the phase-mapping technique \citet{knutson_map_2007} reconstructed a rough map of HD~189733\,b showing the temperature of the cloud deck. An image of Formalhaut\,b in 2008 was the first direct image of an exoplanet \citep{kalas_optical_2008}. With Kepler-186\,f, an Earth-sized planet orbiting within its star's habitable zone was discovered \citep{quintana_earth-sized_2014}. With the discovery of Kepler-452\,b and its \num{1.6} Earth-radii, another category not yet present in our solar system called \enquote{super Earths} was created. Another super Earth orbiting our closest neighbour star, Proxima Centauri, even in the habitable zone was discovered by \citet{anglada-escude_terrestrial_2016}. Future missions aim to characterise exoplanetary systems and to understand the formation, evolution and chemical composition in more detail.

\section{Definition of an exoplanet}

The word \enquote{planet} was introduced by the ancient Greeks as the term 
\textgreek{\`{a}st\'{e}res plan\~{h}ta} 
(\emph{asteres planetai}) which can be translated as \enquote{wandering star}. This was inspired by the apparent movement of the planets in the solar system as opposed to the fixed stars. In a more scientific way the International Astronomical Union (IAU) defined a planet in the solar system in its resolution 5A as follows:
\begin{quote}
	A planet is a celestial body that
	\begin{enumerate}
		\item is in orbit around the Sun,
		\item has sufficient mass for its self-gravity to overcome rigid body forces so that it assumes a hydrostatic equilibrium (nearly round) shape, and
		\item has cleared the neighbourhood around its orbit.
	\end{enumerate}
	\hfill \citet{iau_resolution_2006}
\end{quote}
Additionally the working group on extrasolar planets of the IAU published the following statement, not yet adopted as a resolution: 
\begin{quote}
	\begin{enumerate}
		\item Objects with true masses below the limiting mass for thermonuclear fusion of deuterium (currently calculated to be 13 Jupiter masses for objects of solar metallicity) that orbit stars or stellar remnants are \enquote{planets} (no matter how they formed). The minimum mass/size required for an extrasolar object to be considered a planet should be the same as that used in our Solar System. 
		\item Substellar objects with true masses above the limiting mass for thermonuclear fusion of deuterium are \enquote{brown dwarfs}, no matter how they formed nor where they are located. 
		\item Free-floating objects in young star clusters with masses below the limiting mass for thermonuclear fusion of deuterium are not \enquote{planets}, but are \enquote{sub-brown dwarfs} (or whatever name is most appropriate). 
	\end{enumerate}
	\hfill \citet{boss_working_2005}, reproduced with permission.
\end{quote}

The designation of exoplanets consists of the name of the host star followed by a lowercase letter. The first part can be an astronomical catalogue, the scientific instrument or project that discovered the exoplanet. The following letter indicates the order of the exoplanets discovery around its host star usually beginning with \enquote{b}. This is analogous to the designation of multiple star systems.

\section{Observational techniques}

\begin{sidewaysfigure}
	\centering
	\includegraphics[width=0.95\textwidth]{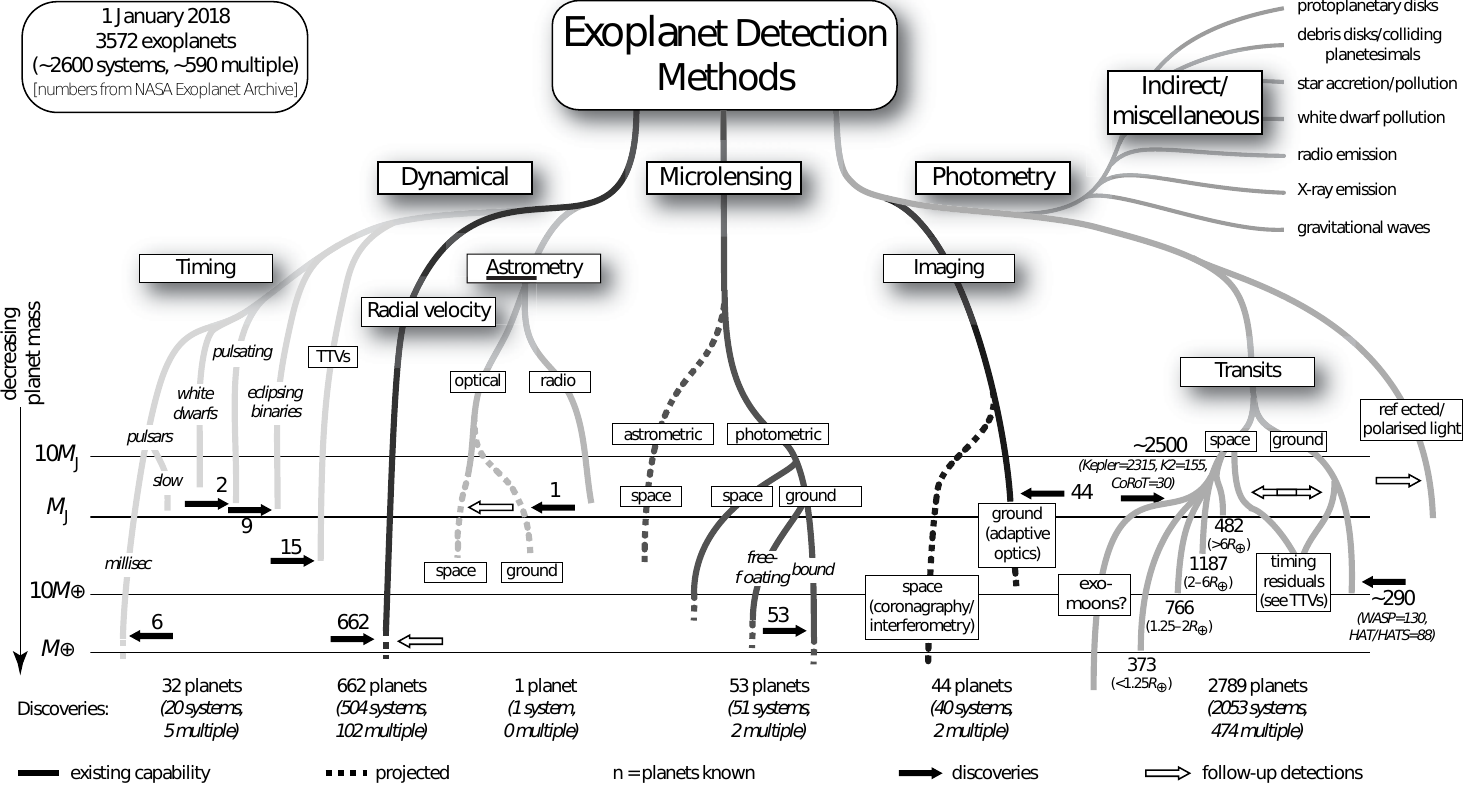}
	\caption[Exoplanet detection methods.]{Exoplanet detection methods. The lower limits of the lines indicate masses within reach of present measurements	(solid lines), and those that might be expected within the next few years (dashed). The (logarithmic) mass scale is shown on the left.	Miscellaneous signatures to the upper right are less well quantified in mass terms. Solid arrows show relevant discoveries. Open arrows indicate measurements of previously-detected systems. Numbers are from the NASA Exoplanet Archive, 2018 January 1 \citep[Reproduced with permission of The Licensor through PLSclear.]{perryman_exoplanet_2018}.}
	\label{fig:intro:perryman}
\end{sidewaysfigure}

The vast variety of exoplanets necessitates the need for different detection techniques. They can be separated into three broad categories as the \enquote{Perryman tree}, a visual overview on the detection methods, illustrates in Fig.~\ref{fig:intro:perryman}.

Widely separated exoplanets can be imaged directly. Also, protoplanetary disks can be detected this way. In case the orbital plane of the exoplanet aligns with the line of sight of the observer, the planet transits in front of the stellar disk which can be detected as flux variations in primary and secondary eclipse.

Gravitational lensing is known for very massive objects like galaxy clusters but can also be detected using a single star as a lens for a background star. This induces a characteristic signal in the light curve. If the foreground star hosts an exoplanet, it will act as an additional lens.

Dynamical effects are based on the gravitational interaction of host star and planet. This can be measured directly as an astrometric change of position on the sky, or indirectly via Doppler shifts in the host star's spectrum or timing variations of transits, eclipses or stellar pulsations. The latter method is used in particular in this work and explained in detail in section~\ref{sec:methods}.

\chapter{Subdwarf B stars}

The majority of known exoplanets orbit cool host stars, i.e., G and M type stars. So far, only a small fraction of planets orbiting evolved host stars have been discovered (compared to effective temperatures of host stars in Fig.~\ref{fig:intro:exoplanets}). This raises questions on the evolution and fate of planetary systems. The following chapter revises the stellar evolution with focus on the post-main sequence phase of the host star, arriving at subdwarf B stars (sdB stars). These stars are laboratories to test how planets survive and influence the late phases of stellar evolution.

\section{Classification}

\begin{figure}[t]
	\centering
	\includegraphics[width=0.95\textwidth]{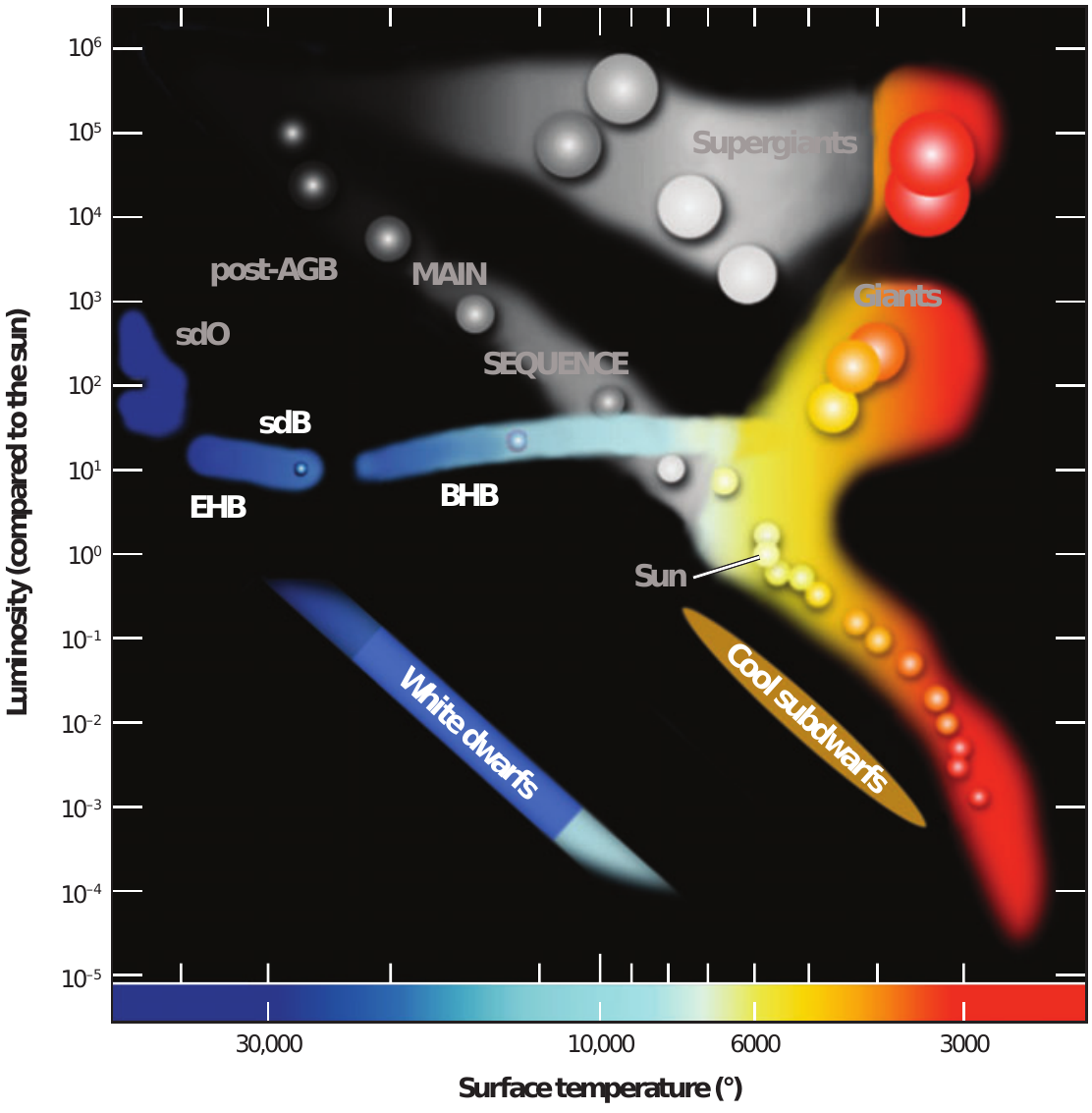}
	\caption[Sketch of a Hertzsprung-Russell diagram.]{Sketch of a Hertzsprung-Russell diagram highlighting the position of hot subdwarf (sdB and sdO) stars and the extreme horizontal branch (EHB) located to the left and below the hot end of the main sequence but above the white dwarf cooling sequence. The EHB is separated from the blue horizontal branch (BHB). The location of stars having evolved from the postasymptotic giant branch is shown for comparison. The hot subdwarf stars have nothing in common with traditional cool subdwarfs found below the lower main sequence \citep{heber_hot_2009-1}. Reprinted by permission from Copyright Clearance Center: Annual Reviews, Annual Review of Astronomy and Astrophysics \enquote{Hot Subdwarf Stars}, Heber, \textcopyright 2009.}
	\label{fig:intro:hrd}
\end{figure}

\citet{humason_search_1947} discovered sdB stars in a photometric survey of the North Galactic Pole region.
Their position in the Hertzsprung-Russel diagram (HRD) in Fig~\ref{fig:intro:hrd}, in between the main sequence (MS) and the white-dwarf (WD) sequence, was later determined by \citet{greenstein_nature_1974}. They show effective temperatures $ T_{\text{eff}} $ from \SIrange{\sim 20000}{50000}{K} with surface gravities $ \log (g/ \si{\centi\meter \per \second\squared}) $ of \numrange{5.0}{6.5}.
\citet{heber_atmosphere_1984,heber_atmosphere_1986} could connect the spectroscopic sdB class with the extreme horizontal branch (EHB) evolutionary stage.
Thus, sdB stars are stripped helium-burning cores ($ \SI{\sim 0.5}{M_{\odot}} $) of red giants with a thin hydrogen (\ch{H}) atmosphere. Their spectra are therefore dominated by the broad Balmer lines of H, while helium (\ch{He}) is usually depleted.
But their structure differs from horizontal-branch stars, as their H envelopes are too thin to sustain H burning.
They evolve directly on to the WD cooling sequence, as they lack the envelope mass to pass via the asymptotic giant branch (AGB).

\section{Canonical stellar and planetary evolution}

The progenitors of sdB stars are assumed to be solar-mass stars. They form in molecular clouds, consisting mostly of \ch{H}, about \numrange{20}{30} per cent of \ch{He} and a few percent of heavy elements. The gas remains in hydrostatic equilibrium as long as the gas pressure is in balance with gravitational force (virial theorem) but a gravitational collapse can be triggered by shock waves (e.g. nearby supernova). As the cloud contracts, it breaks into smaller fragments, in which the collapsing gas radiates the gravitational potential as heat. As temperature and pressure increase, the fragment condenses into a protostar which is in hydrostatic equilibrium. This way, stars with different masses form mostly in groups. Stellar winds of more massive stars ionize the remaining \ch{H} in between the stars, creating \ch{H} II regions, and ultimately disrupt the cloud which prevents further star formation.

\begin{figure}[t]
	\centering
	\begin{subfigure}[b]{0.49\textwidth}
		\centering
		\includegraphics[width=0.95\textwidth]{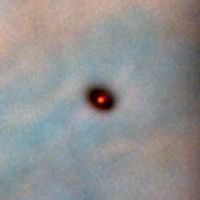}
	\end{subfigure}
	\begin{subfigure}[b]{0.49\textwidth}
		\centering
		\includegraphics[width=0.95\textwidth]{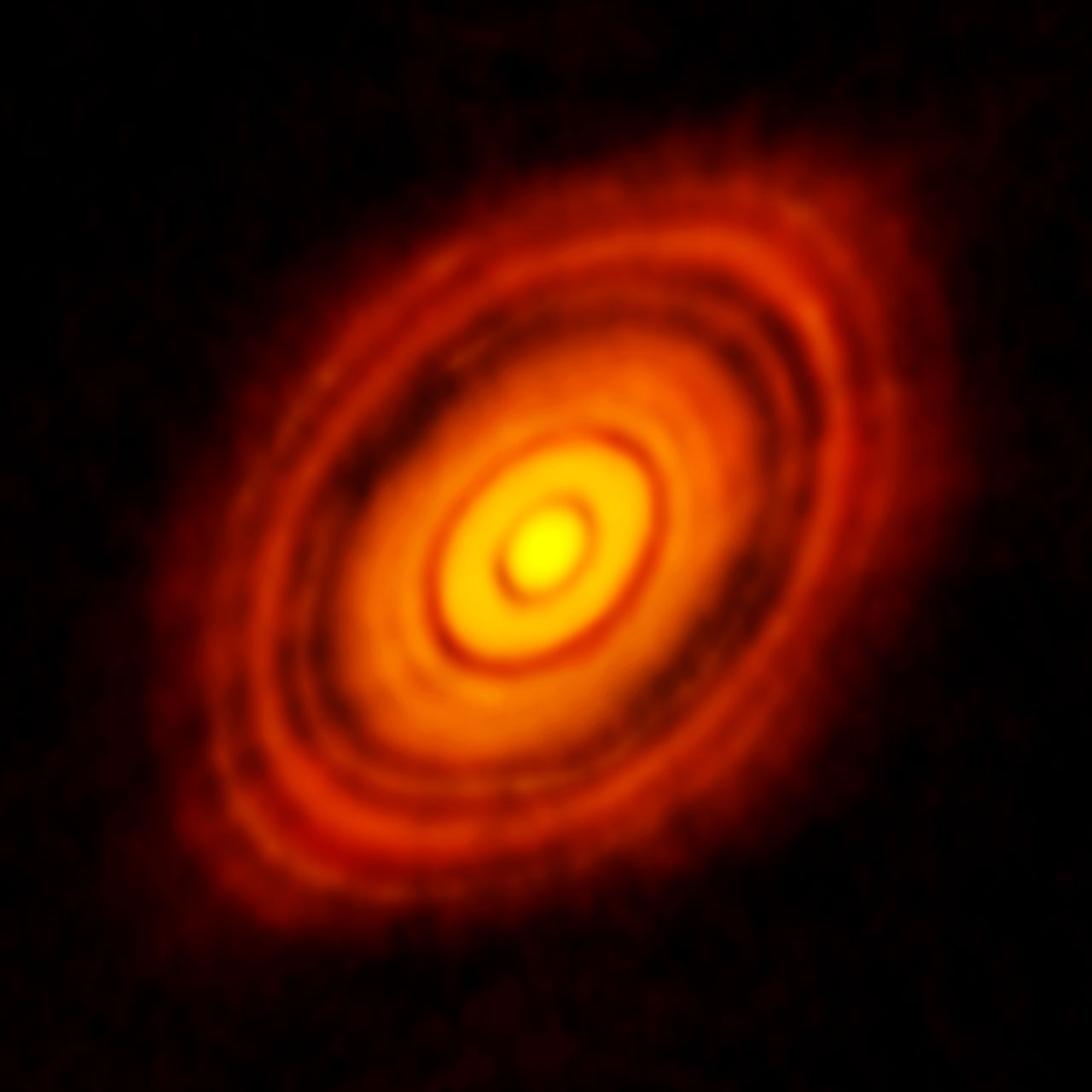}
	\end{subfigure}
	\caption[Protoplanetary discs.]{\textit{Left:} Protoplanetary disc imaged by the Hubble Space Telescope in the Orion nebula. The image is about \SI{1800}{\au} across. Credit: NASA, C.R. O'Dell and S.K. Wong. \textit{Right:} ALMA image of HL Tauri showing th protoplanetary disc at a wavelength of \SI{1.3}{mm}. The image is about \SI{250}{\au} across. Credit: ALMA (ESO/NAOJ/NRAO).}
	\label{fig:intro:alma}
\end{figure}

The protostar continues to accrete material from its protoplanetary disk (Fig.~\ref{fig:intro:alma}) until finally \ch{H} fusion starts in the core and the star begins its phase on the main-sequence of the HRD (Fig.~\ref{fig:intro:hrd}).
Inside the protoplanetary disk, planetesimals form from electrostatic and gravitational interactions as building blocks for planets. During their formation, planets migrate within the disk and accrete material until the disk is evaporated by the central stars radiation. On the main-sequence, the star will evolve only slowly and remain there for about \si{10^{10}} years. As the hydrogen in the core eventually depletes, the fusion rate cannot be maintained and the core contracts. This continues to fuse H in a shell outside the \ch{He} core. The core contracts and the outer layers begin to expand. The star becomes a red giant, moving up the red giant branch (RGB) in the HRD. The \ch{He} core compacts into degenerate matter and once core pressure and temperature are high enough, the \ch{He} starts to fuse with a \ch{He} flash at the tip of the RGB in the HRD. The core mass during the helium flash is about half a solar-mass. The star decreases in radius, increases its surface temperature, and moves to the horizontal branch (HB) of the HRD. After the He is fused in the core, outer shells fuse further He and the star moves to the asymptotic giant branch (AGB) and becomes red giant again. The produced carbon (\ch{C}) can fuse with He and oxygen (\ch{O}) to build a \ch{CO} core. The \ch{H} and \ch{He} shells are consecutively fusing in a cyclic process. The star ejects most of its envelope during these thermal pulses. The expelled material is ionized by the UV emission of the hot remnant core and can be observed as a planetary nebula. After this short phase, the remnant star cools and enters the white dwarf cooling sequence.

Planet occurrence rates, estimated from \textit{Kepler} observations, suggest that 30 per cent of Sun-like stars host \textit{Kepler}-like planetary systems\footnote{\textit{Kepler}-like planets are planets that have radii $ R_P \gtrsim R_{\earth} $ and orbital periods $ P > \SI{400}{\day} $. Our Solar system is not detectable by \textit{Kepler}. } \citep{zhu_about_2018}. Several studies investigated occurrence rates from radial velocity surveys. Jupiter-like planets\footnote{Planets with masses $ 0.15 R_{\jupiter} \lesssim M_P \lesssim 13 M_{\jupiter} $ and orbital periods of \SIrange{3}{7}{\au}.} may occur at six per cent of Solar-like stars \citep{wittenmyer_anglo-australian_2016}. Systems with giant planets from the \textit{Kepler} mission show outer gas giants within orbital periods of less than 400 days at a 5 per cent occurrence rate \citep{santerne_sophie_2016}. \citet{bryan_statistics_2016} suggest the total occurrence rate of companions\footnote{For companions in the range of \SIrange{1}{20}{M_{\jupiter}} and \SIrange{5}{20}{\au}} from radial velocity measurements and direct imaging with about 50 per cent. Recently \citet{grunblatt_giant_2019} estimated the occurrence rate of planets larger than Jupiter for low-luminosity red giant branch stars from the \textit{K2} mission to be 0.5 per cent.
Observations of polluted WDs, debris disks orbiting WDs like WD~1145+017 \citep{vanderburg_disintegrating_2015} or even in-situ accretion of a giant planet to WD J0914+1914 \citep{gansicke_accretion_2019} show that these disks must form during the WD phase since they are located within the preceding giant stellar radii. Spectroscopic analysis of polluted WD atmospheres find metals, as to be expected from planetary composition. Since the sinking times for such heavy metals are orders of magnitude shorter than the WD cooling age, the expected detection rate of polluted WDs is at about 0.1 per cent. But the actual observed rate of metals is at \numrange{25}{50} per cent. Thus, planetary systems appear to survive the RGB phase of their host star and may get disrupted and accreted during the WD phase. \citet{veras_post-main-sequence_2016,farihi_circumstellar_2016} give a broad overview on post-main-sequence planetary system evolution and polluted WDs.

\section{Subdwarf B stars formation scenarios}

\begin{figure}[t]
	\centering
	\includegraphics[width=0.95\textwidth]{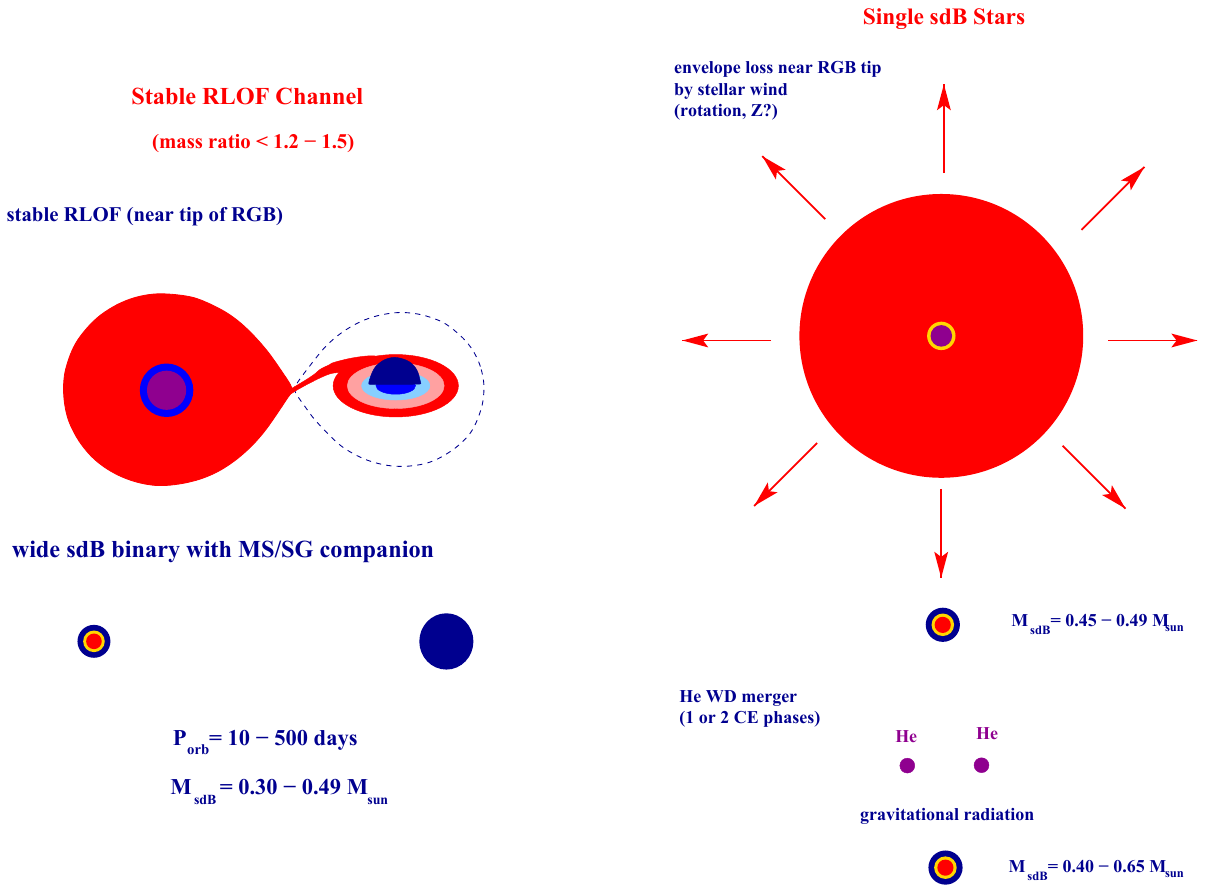}
	\caption[Formation of sdB stars I.]{Stable Roche-lobe channel (left) and single-star/merger channels (right) for the formation of sdB stars \citep{podsiadlowski_evolution_2008}. By the kind permission of ASP Conference Series, vol. 401, P. Podsiadlowski, p. 63.}
	\label{fig:intro:formationI}
\end{figure}

\begin{figure}[t]
	\centering
	\includegraphics[width=0.95\textwidth]{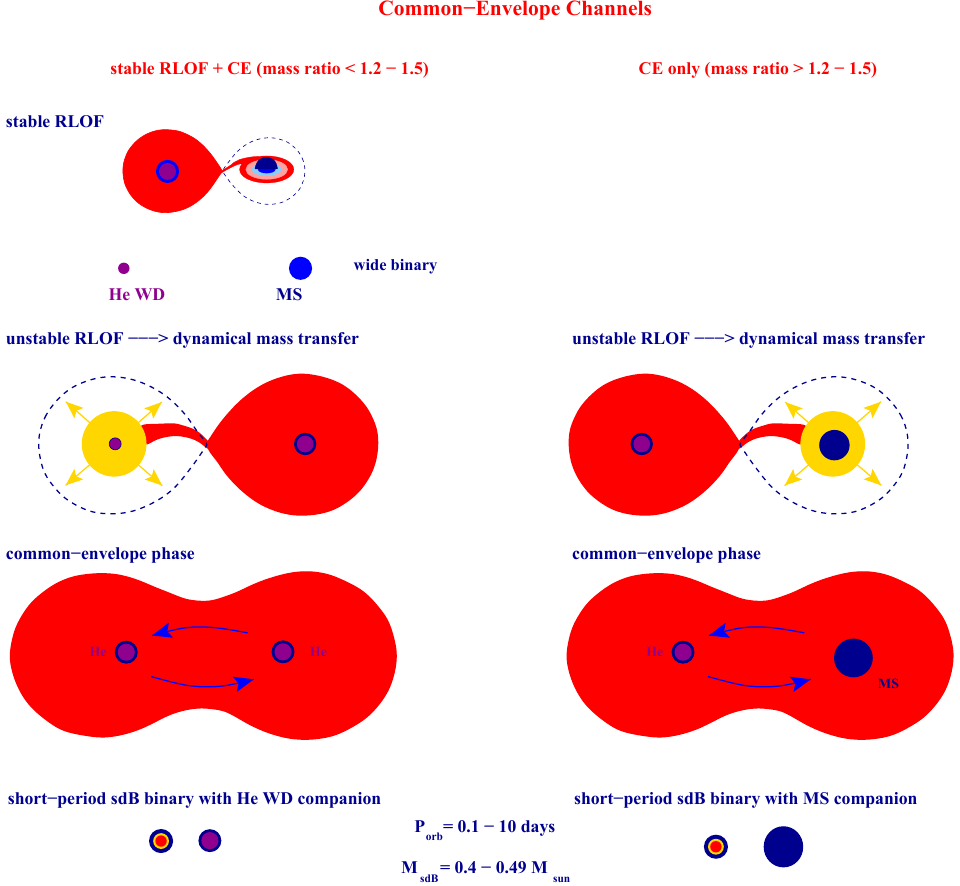}
	\caption[Formation of sdB stars II.]{Common-envelope (CE) channels for the production of sdB stars. The CE phase can be either the first (right panel) or the second (left panel) mass-transfer phase, producing a tight sdB binary with either a normal-star or white-dwarf companion, respectively \citep{podsiadlowski_evolution_2008}. By the kind permission of ASP Conference Series, vol. 401, P. Podsiadlowski, p. 63.}
	\label{fig:intro:formationII}
\end{figure}

The canonical formation of Subdwarf B stars is not explained by the typical evolution described in the section above since some mechanism during the progenitors giant phase must have removed enough of the envelope to not sustain a H burning after the He flash. 
The formation scenario in a binary system, first proposed by \citet{mengel_binary_1976}, gives rise to different formation channels \citep{han_origin_2002,han_origin_2003,podsiadlowski_evolution_2008} depending on the initial mass ratios. The scenarios are illustrated in Fig.~\ref{fig:intro:formationI} and \ref{fig:intro:formationII}. The most simple scenario is a stable Roche-lobe over flow (RLOF), as in the left panel of Fig.~\ref{fig:intro:formationI}. The sdB progenitor fills its Roche lobe near the tip of the RGB. The mass transfer is dynamically stable and the companion accretes the matter. This will form a sdB in a long-period binary system with a main-sequence companion. Observations find periods from \SIrange{700}{1300}{\day}, leading to an improvement of the RLOF model \citep{chen_orbital_2013}. If the mass transfer during  the RLOF is too high, the companion cannot accrete all the matter and a common envelope (CE) is formed (see Fig.~\ref{fig:intro:formationII}). Friction in the CE let the stars spiral inwards and until the CE gets ejected. This will form a close binary system.

While about 50 per cent of sdB stars are found in close binary systems with periods of less than ten days, observations show about 30 per cent appear as apparent single sdB stars. There are several possible formation scenarios including atmosphere loss due to stellar winds, the merger of two \ch{He} WD (right panel of Fig.~\ref{fig:intro:formationI}), or undetected low-mass companions.

\citet{hall_hydrogen_2016} estimated the \ch{H} mass in the remnants of He WD mergers. In the effective temperature-surface gravity plane, they found a region occupied by these stars during the core-\ch{He}-burning phase and located the majority of apparently single sdB stars inside this region. Nevertheless, this conclusion depends on assumptions made to the model, regarding the initial \ch{H} mass and loss of \ch{H} during the merger. Subdwarf B stars located outside of this parameter region are found to be rapidly rotating low-gravity sdB stars. 

Alternatively, the sdB progenitor could lose mass near the RGB tip due to strong stellar winds or internal rotation. 
While the \ch{He} flash occurs typically at the tip of the RGB, a sufficient mass loss can lead a star to depart from the RGB so that the \ch{He} flash is delayed \citep{castellani_mass_1993}.
The remnants of these \enquote{hot flashers} are located close to the He main sequence, which coincides with the sdB location. Although an \enquote{early hot flasher} scenario may explain \ch{He} and \ch{C} abundances in sdB stars, observations show some inconsistencies \citep{moehler_hot_2011,latour_helium-carbon_2014}.


\begin{figure}[t]
	\centering
	\includegraphics[width=0.95\textwidth]{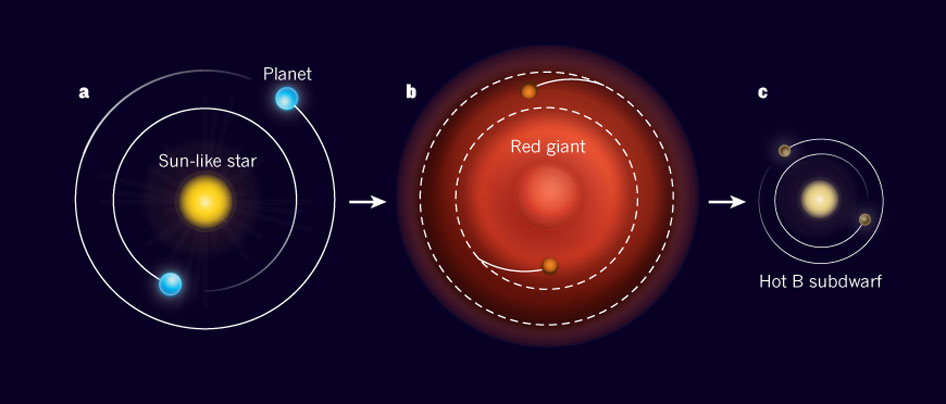}
	\caption[Possible formation mechanism of close-in planets.]{Possible formation mechanism of close-in planets. a, The two planets identified by \citet{charpinet_compact_2011} were probably massive gas-giant planets during the early part of their host star's life, when the star burned hydrogen into helium as our Sun does today. At this time, the planets resided farther away from their host star than they do today. b, When the star exhausted the supply of hydrogen in its core, it expanded to become a red-giant star. The outer layers of the star then engulfed the orbits of the two planets. The planets lost their outer gaseous layers as their orbits spiralled inwards. c, For unknown reasons, the red giant expelled its cool outer layers, leaving behind a small hot B subdwarf star. The small rocky cores of the initial planets were left behind in close-in orbits with periods of 5.8 and 8.2 hours. (Figure not drawn to scale.) \citep{kempton_planetary_2011-2}. Reprinted by permission from  Copyright Clearance Center: Springer Nature, Nature, Planetary science: \enquote{The ultimate fate of planets}, Kempton, \textcopyright 2011.}
	\label{fig:intro:sdb-planet}
\end{figure}

Finally, substellar objects, e.g., brown dwarfs or even planets, can lead to enough mass loss in order to form sdB stars. As described in the previous section, the abundance of planets is high enough to be considered to survive and influence the late stellar evolution.	For a single red giant, planets with orbital separations smaller than about \SI{5}{\au}, will enhance the mass loss when the planet is engulfed by the red giant's atmosphere. Since planets have a comparable high angular momentum, parts of it can be transferred to the envelope. In consequence of this momentum transfer, the planet spirals inwards and stellar mass loss increases, which leads to the formation of a sdB star. In this scenario, the planet would either evaporate, merge with the core, or survive while most of the planet's envelope is lost \citep{soker_can_1998}. With the merger scenario, an enhanced stellar rotation could mix more \ch{He} to the envelope and thus increase the RGB tip luminosity and further enhance the total mass loss \citep{sweigart_effects_1997}. The hypothesis of surviving planetary systems is supported by planetary-mass companions like the candidates V391 Peg b \citep{silvotti_giant_2007}, KIC 05807616 b,c \citep{charpinet_compact_2011}, KIC 10001893 b,c,d \citep{silvotti_kepler_2014} or brown dwarf companions like V2008-1753 B \citep{schaffenroth_eclipsing_2015} or CS 1246 \citep{barlow_fortnightly_2011}. Figure~\ref{fig:intro:sdb-planet} illustrates the process of forming a sdB planetary system.

From observational data it cannot be ruled out that these planets may have formed as second generation planets. After the merger of two \ch{He} WDs, the remaining circumstellar disk could provide enough material to form planets, in analogy to the pulsar planets \citep{wolszczan_planetary_1992,thorsett_psr_1993}.

Finally, there is evidence that some sdB stars are not core \ch{He}-burning objects. \citet{heber_discovery_2003} discovered a sdB with insufficient mass to ignite \ch{He}-burning in the core.

\section{Asteroseismology}

\begin{figure}[t]
	\centering
	\includegraphics[width=0.85\textwidth]{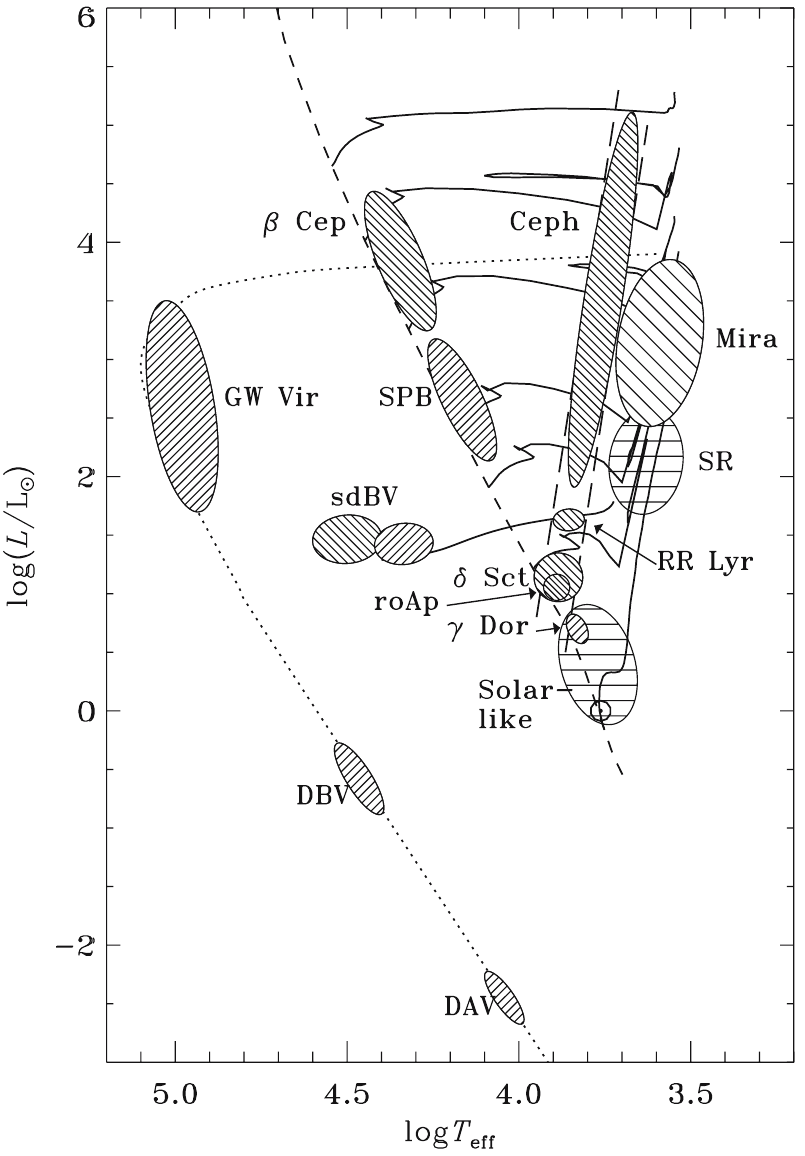}
	\caption[Hertzsprung-Russell diagram showing different classes of pulsating stars.]{Hertzsprung-Russell diagram showing different classes of pulsating stars. Some of these are named after a particular member of the class. Others are acronyms, respectively, for: rapidly oscillating	Ap (roAp); Slowly Pulsating B (SPB); subdwarf B variables (sdBV). The group labelled GW Vir includes what has formerly been known as the PNNV stars (for Planetary Nebulae Nuclei Variables), and the variable hot DO white dwarfs (DOV); the DBV and DAV stars are variable DB (helium-rich) and DA (hydrogen-rich) white dwarfs. The parallel long-dashed lines indicate the Cepheid instability strip \citep{cunha_asteroseismology_2007}.
		Reprinted by permission from  Copyright Clearance Center: Springer Nature, Astron Astrophys Rev., \enquote{Asteroseismology and interferometry}, Cunha, \textcopyright 2007.}
	\label{fig:intro:puls-classes}
\end{figure}

Asteroseismology describes the study of the interiors of stars by using their oscillations as seismic waves. The first pulsating star was discovered in 1596 by David Fabricius. He noticed that the star \textomicron~Ceti (subsequently named \enquote{Mira}, the wonderful) vanished from the visible sky. Later on, it was realized that the star did so every eleven months and became the first known (semi) periodic variable star. Subsequently, more variable stars like \textdelta~Cephei (related stars are summarized as \enquote{Cepheids}) were discovered and their variations linked to radial pulsations. By now, a large number of groups of pulsating star is known to exist. Figure~\ref{fig:intro:puls-classes} shows their occurrences in a HRD. Historically, they have been classified on a phenomenological basis and mostly named after their prototype star. The physical reasons for the classification were uncovered later and are related to the different types of excited pulsation mode, mass and evolutionary state. Since this work relates mostly to sdB stars, the following describes their driving mechanism. Extensive overviews can be found, e.g., in \citet{cunha_asteroseismology_2007,aerts_asteroseismology_2010}.

\subsection{\textkappa~mechanism}\label{sec:kappa}
\citet{baker_pulsations_1962} proposed the \textkappa~mechanism to explain \textdelta~Cephei pulsations. 
For stellar matter than can be described as an ideal gas, an increase in compression of the atmosphere causes an increase in temperature and density. This usually results in a decrease of the opacity \textkappa, allowing energy to be transported more efficiently. This describes an equilibrium where temperature and pressure are balanced.
But in layers of the star where \textkappa~increases with increasing temperature, the incoming flux from inner layers can be temporally stored. These layers are associated with regions where (partial) ionization of chemical elements occur. 
When the layer expands in order to reach its equilibrium, the additionally accumulated energy is  released and the star can expand beyond its equilibrium radius.
When the material recedes, energy is again stored in the stellar interior, and the cycle can repeat as a periodic stellar oscillation.

Depending on the restoring force of the pulsations, modes are either of the nature of standing acoustic waves (referred to as pressure modes or p modes) or internal gravity waves (g modes). However, especially in evolved stars, both types of modes can be found.

Unlike solar oscillations, which are exited stochastically, pulsations excited by the \textkappa~mechanism are coherent in phase; their lifetime is very long, compared to observations spanning over years and change slowly due to stellar evolution. These modes are suitable for a pulsation timing analysis.

In variable sdB stars (sdBV) the bump in opacity is related to elements of the iron group \citep{charpinet_driving_1997}, which are accumulated in the driving layers by diffusion processes like gravitational settling and radiative levitation \citep[c.f.][]{heber_hot_2009-1,heber_hot_2016}. \citet{charpinet_potential_1996} predicted the nonradial pulsations theoretically almost simultaneously to their discovery by \citet{kilkenny_new_1997}. Effective temperatures $ T_{\text{eff}} $ typically range from \SIrange{28000}{35000}{\kelvin} and surface gravity $ \log g $ from 5.2 to 6.1. Observed pulsation amplitudes are usually smaller than \SI{50}{\milli\mag}. Hybrid pulsators, showing both p and g modes \citep[e.g.][]{schuh_hs_2006}, locate at the temperature boundary around \SI{28000}{\kelvin} between both classes of pulsators. About ten per cent of the sdB stars, located in the instability strip of the $ T_{\text{eff}} - \log g$ plane, exhibit pulsations above a few \si{\milli\mag}. This fraction might be biased due to detection limits of ground based observations. However, \textit{Kepler} observations confirmed the deficiency of p-mode pulsators in the instability strip \citep[e.g.][]{ostensen_at_2011}.

Due to the \ch{He} fusion into \ch{C}  and \ch{O} in the core of sdB stars, the internal structure varies slowly and thus modifies the condition of hydrostatic equilibrium. This results into slow, secular changes of the surface gravity and effective temperature, and slow changes in the pulsation period \citep{charpinet_adiabatic_2002}. Using long-term observations, these period changes can be measured.

\subsection{Asteroseismic observations}

Stellar oscillations generate motions and temperature variations on the surface and result in observable variability. These variations cause changes in brightness and colour, radial velocity and line profiles. Thus, pulsating stars can be studied using photometric and spectroscopic time series measurements, which can be used to generate frequency spectra - one of the most important tools in asteroseismology. The following focuses on time series photometry.

Since the observations are usually discrete and contain gaps, a harmonic analysis is performed using the Discrete Fourier Transformation
\begin{align*}
F(f) = \sum_{k=1}^{N} x(t_k) \exp\left( i 2 \pi f t_k \right),
\end{align*}
of the measurements $ x(t_k) $, which leads to an amplitude spectrum. The observations oppose limitations to this spectrum. Due to the discrete nature, the highest useful frequency to search for is called the Nyquist frequency 
\begin{align*}
f_{\text{Nyq}} = \frac{1}{2\Delta t},
\end{align*} 
with the time between two data points $ \Delta t $. In the case of unevenly spaced data, the Nyquist frequency does not exist, higher frequencies might be detectable. With a finite length of the observations, the spectrum has a finite frequency resolution proportional to the inverse of the total time span. Moreover, observations are likely to show gaps (e.g. day/night or seasonal observation windows, technical problems, poor observing conditions), some of them with a regular nature, which produce additional peaks in the Fourier space. These effects combined are known as the \enquote{window function} in the time domain or as \enquote{spectral window} in the Fourier space. The spectral window can show multiple peaks even if the underlying signal is monochromatic and complicate the interpretation of the data. Additionally, long gaps produce unwanted signals in the low frequency regime. In order to overcome the complex response in frequencies and amplitudes of the signal, even without noise, different forms of periodograms have been defined. 
A well-known and widely applied technique is the Lomb-Scargle periodogram \citep{scargle_studies_1982}. 
\citet{horne_prescription_1986,schwarzenberg-czerny_correct_1997} showed the equivalence of the variance reduction between a Lomb-Scargle periodogram and a least squares fit of sinusoids at test frequencies. Thus, fits of a harmonic series of sinusoids at test frequencies can be used to search for non-sinusoidal signals as well. The Lomb-Scargle periodogram does not require regular sampled data and is used for the analysis in this work (see \citet{press_numerical_2007,vanderplas_understanding_2017}; c.f. section~\ref{sec:analysis}).

\cleardoublepage

\part{The EXOTIME project} \label{sec:exotime-project}
\chapter{Targets} \label{sec:targets}
The EXOTIME program (EXOplanet search with the TIming MEthod) was born out of long-term monitoring of the object HS~2201+2610 in order to measure long-term drifts in the pulsation periods $ \dot{P} $. The large amount of data made it possible to not only constrain sdB star evolution from $ \dot{P} $ but resulted in the postulation of the substellar companion candidate V391~Peg\,b using the pulsation timing method \citep{silvotti_giant_2007}. With the puzzling formation history of sdB stars, the authors initiated the EXOTIME program in 2008. The project extends the long-term monitoring to the objects HS~0702+6043 (DW~Lyn), HS~0444+0458 (V1636~Ori), PG~1325+101 (QQ~Vir), EC~09582-1137 (V541~Hya) and HS~2201+2610 (V391~Peg). While the majority of exoplanet research is strongly focussed on planets (and moons) in the habitable zone and formation of planetary systems, the main goal of the EXOTIME project is to detect exoplanets orbiting evolved stars, using the stellar pulsations to conclude on timing variations (see section~\ref{sec:methods}). Transit and radial velocity searches are not efficient for stars with small radii and high gravities like sdB stars. Additionally, the project wants to characterize the target stars using asteroseismic methods and in particular the measurement of $ \dot{P} $.

The following sections describe the targets and their observations in detail. Their selection followed some basic properties: The targets must be observable from the Northern hemisphere and bright enough for a sufficient signal-to-noise ratio in the Johnson B band with \SIrange{1}{3}{m} class telescopes, since sdB stars emit the maximum energy in the near UV regime. A suitable target for the timing method should show two to four independent pulsation frequencies, coherent in phase and stable in amplitude ($ \SI{>1}{\milli\mag} \simeq \SI{0.1}{\percent} $). Each frequency leads to an independent timing measurement. But coherence and stability can only be verified with the long-term monitoring. To ensure the sdB star is not accompanied by a low-mass star in a binary system, there should be no strong IR excess compared to a black body spectrum. The observational parameters of the five EXOTIME targets are summarized in Table~\ref{tab:obs-params}.
Section~\ref{sec:observations} describes DW~Lyn, V1636~Ori, QQ~Vir, V541~Hya in detail and lists the atmospheric parameters in Table~\ref{tab:stellar}. 

\begin{table}[t]
\centering
\caption[Observational properties of the EXOTIME targets.]{Observational properties of the EXOTIME targets, taken from \mbox{\url{http://simbad.u-strasbg.fr/simbad/}} and amplitude of the main pulsation (measured in this work and by \citet{silvotti_giant_2007}). }
\begin{tabular}{lrrSSS}
\toprule
Target & {RA (J2000)} & {DE (J2000)} & {B mag} & {J$ - $H} & {ampl./\%} \\ 
\midrule
DW Lyn & \ra{07;07;9.80} & \ang[angle-symbol-over-decimal]{+60;38;50.16} & 14.7 & -0.123 & 2.19 \\ 
V1636 Ori & \ra{04;47;18.63} & \ang[angle-symbol-over-decimal]{+05;03;34.81} & 15.44 & 0.043 & 0.54
 \\ 
QQ Vir & \ra{13;27;48.56} & \ang[angle-symbol-over-decimal]{+09;54;51.05} & 13.73 & -0.14 & 2.6 \\ 
V541 Hya & \ra{10;00;41.82} & \ang[angle-symbol-over-decimal]{-11;51;34.59} & 15.01 & -0.25 & 0.31 \\ 
V391 Peg & \ra{22;04;12.10} & \ang[angle-symbol-over-decimal]{+26;25;7.82} & 14.41 & 0.01 & 0.85 \\ 
\bottomrule
\end{tabular}
\label{tab:obs-params}
\end{table}

Observations and results on a substellar companion for V391~Peg are published by \citet{silvotti_giant_2007}, alternative scenarios are discussed by \citet{silvotti_subdwarf_2008} and additional observations are published by \citet{silvotti_sdb_2018}. The latter publication placed the first interpretation of the $ O-C $ variations under discussion because of possible non-linear interactions between different pulsation modes that change arrival times. We proposed for continued observations at the Las Cumbres Observatory \citep[LCO][]{brown_cumbres_2013}, using \SI{2}{m} class telescopes in order to connect observations from \citet{silvotti_sdb_2018} with upcoming TESS observations in 2020. In 2017, we observed for about 52 hours, in 2019 for eight hours, each in the Johnson B band with exposure times of \SIrange{10}{20}{\second}. The observations are listed in Table~\ref{tab:obs-v391peg}.
\begin{table}[t]
	\centering
	\caption{Summary of the newly acquired observations for V391~Peg.}
	\begin{tabular}{ccSS}
		\toprule
		observatory & obs. date & {exp. time /\si{\second}} & {obs. time /\si{\hour}} \\
		\midrule
		LCO & 2017-09-12  & 10 & 1.22 \\ 
		\dots & 2017-09-14  & 10 & 0.68 \\ 
		& 2017-09-15  & 10 & 1.40 \\ 
		& 2017-09-16 & 10 & 4.72 \\ 
		& 2017-09-17  & 10 & 6.18 \\ 
		& 2017-09-18 & 10 & 1.22 \\ 
		& 2017-09-19  & 10 & 6.92 \\ 
		& 2017-09-20  & 10 & 0.72 \\ 
		& 2017-09-21  & 10 & 3.95 \\ 
		& 2017-09-22  & 10 & 2.70 \\ 
		& 2017-09-26  & 20 & 0.75 \\ 
		& 2017-09-28 & 20 & 1.75 \\ 
		& 2017-09-29  & 20 & 0.22 \\ 
		& 2017-09-30  & 20 & 0.75 \\ 
		& 2017-10-01 & 20 & 1.27 \\ 
		& 2017-10-02 & 20 & 0.75 \\ 
		& 2017-10-03 & 20 & 5.18 \\ 
		& 2017-10-05  & 20 & 0.23 \\ 
		& 2017-10-06  & 20 & 0.38 \\ 
		& 2017-10-07  & 20 & 0.77 \\ 
		& 2017-11-05 & 20 & 2.07 \\ 
		& 2017-11-06  & 20 & 0.75 \\ 
		& 2017-11-09  & 20 & 1.68 \\ 
		& 2017-11-10  & 20 & 4.22 \\ 
		& 2017-11-13 & 20 & 2.18 \\ 
		& 2019-09-07  & 20 & 1.10 \\ 
		& 2019-09-09  & 20 & 0.07 \\ 
		& 2019-09-10  & 20 & 2.48 \\ 
		& 2019-09-11  & 20 & 2.67 \\ 
		& 2019-09-12  & 20 & 1.30 \\ 
		\midrule
		$ \Sigma $ & & & 60.27 \\
		\bottomrule
	\end{tabular}
	\label{tab:obs-v391peg}
	\end{table}

\chapter{Multi-epoch photometric data}
\section{Time series photometry}
The timing method applied to the data and described in the next section imposes requirements on the photometric time series. Each observation run needs a high temporal resolution of less than about \SI{30}{\second} in order to sample the p-modes correctly and a high signal-to-noise ratio to detect their low amplitudes. A single phase measurement in the later analysis is required to reach an uncertainty in the order of one second. Therefore, the observations should cover about three consecutive nights, each with about three hours for one phase measurement. 

\subsection{Photometry}
In order to make use of the raw CCD images from the observations, the data need to be calibrated properly. This includes detector specific corrections (bias and dark field subtraction), as well as instrumental and sensitivity corrections (flat field division). The following photometric measurements were performed using the IDL package TRIPP (Time Resolved Imaging Photometry Package). Some data have been reduced using different software packages but the concept does not differ. A circular aperture is placed around the target star and measures the flux. An annulus around this aperture or a further field measures the background flux which is subtracted from the target flux. Since variations in the Earth's atmosphere on short and long term time scales (sky transparency, air mass) affect this measurement, the target flux is compared to the flux of multiple comparison stars. The field of view is assumed to be small enough that all atmospheric variations influence the flux of all stars equally. This procedure is applied to all images of an observation run and yields the light curve. Atmospheric extinction is correction by a second order polynomial in time.  

\subsection{Timing accuracy}
Accurate time stamps for each observation is crucial for the analysis of timing effects. Acquisition equipment is usually synced via network with a time server to the universal time (UTC). However, some of the first observations used for the project do not record the fraction of the second. Thus, the accuracy of individual observations is assumed to be better than \SI{\pm 0.5}{\second}.

Additional to the recording of precise time stamps, the observer needs to be positioned in a rest frame compared to the target. Without the necessary corrections, the result would be superimposed by the Earth's motion around the barycentre of the Solar system and discontinuous time standards. \citet{eastman_achieving_2010} describes the necessary formalism in detail. The Barycentric Coordinate Time (TDB) time frame including barycentric corrections for the Solar system is accurate down to \SI{3.4}{\milli\second} and applied to all observational time stamps (hereafter refereed as BJD(TDB)). 

\chapter{Methods} \label{sec:methods}
\section{$ O-C $ Pipeline}

Already in the 17th century, Ole R\oe mer made use of the travel time of light in order to measure the speed of light \citep[e.g.][]{soter_cosmic_2001}. An inversion of this scenario allows to measure changes of orbits using the finite speed of light. In astronomy, this idea is widely used for different applications, e.g., on pulsars, eclipsing binary systems, timing variations on planetary transits or stellar pulsations. In the context of stellar pulsations, this concept is referred to as the $ O-C $ method (\textit{O}bserved minus \textit{C}alculated). The approach compares the phase of a periodic function to a reference phase and thus allow conclusions on variations in the observed phase. While the following describes the method illustratively, the specific implementation of the $ O-C $ method in this work is explained in section~\ref{sec:analysis}.

\begin{figure}[tbp]
	\centering
	\input{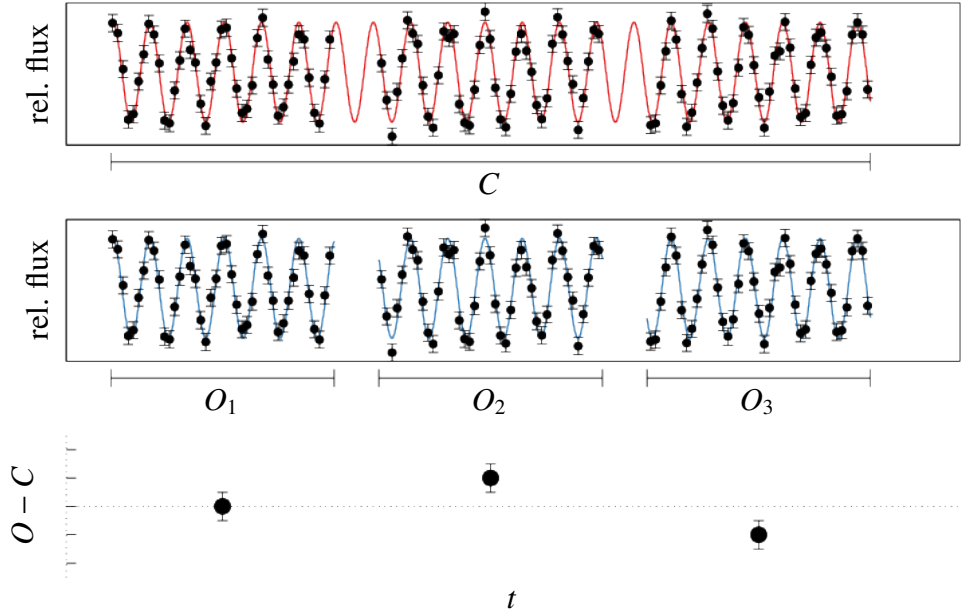} 
	\caption[Schematic picture of the $ O-C $ diagram construction.]{Schematic picture of the $ O-C $ diagram construction. \textit{Top:} A model to the full light curve yields the reference phase $ C $. \textit{Middle:} Light curve models to each epoch yield individual phase information $ O $. \textit{Bottom:} Construction of the $ O-C $ diagram from above measurements.}
	\label{fig:methods:oc-scheme}
\end{figure}

The schematic construction of an $ O-C $ diagram is pictured in Fig.~\ref{fig:methods:oc-scheme}. Requirement for a $ O-C $ analysis is a sufficiently long observational time span of the stellar pulsation. A light curve model fit to full data set provides the reference phase, referred to as $ C $. Similar fits to small sub-sets of the observation provides an independent light curve model. Thus, each sub-set, called epoch, provides a phase measurement, referred to as $ O $. Iterating through all epochs in time yield the differences between the average phase of the reference model and the phase measurements of the epochs, called $ O-C $ diagram.  

An $ O-C $ diagram with no variations implies an agreement between observed phases and the reference phase. Linear changes in the diagram suggest a wrong frequency for the reference model and thus a constant error added per cycle. More complex deviations are described below.	

\subsection{Linear change in period}

The following considerations can be found in \citet{kepler_detection_1991}. In order to make use of the timing method, a sufficiently stable timer in the observed system is required. \enquote{Sufficiently} refers to accuracy compared to the length of observations. In the case of stellar pulsations, this timer is a periodic brightness variation. The observed time of the $ E $th cycle can be expressed as
\begin{align*}
T_E = T_0 + PE,
\end{align*}
with the period $ P $ at $ T_0 $ and the epoch $ E $, as in the cycle number counted from $ T_0 $. 

Due to physical processes, the stellar pulsation period will change slowly over time (see section~\ref{sec:kappa}). If this change is small, $ T_E $ can be expand in a Taylor series:
\begin{align*}
T_E = T_0 + \diff{T}{E} \left( E-E_0 \right) + \frac{1}{2} \diff{^2T}{E^2} \left( E-E_0 \right)^2 + \mathcal{O}\left( \left( E-E_0 \right) \right)^2.
\end{align*}
Terms of higher order are neglected. Expanding the quadratic term, using $ \dif{T}/\dif{E} = P $, leads to
\begin{align*}
\diff{^2 T}{E^2} = \diff{P}{E} = \diff{t}{E} \diff{P}{t} = P \diff{P}{t}.
\end{align*}
Assuming $ 2E_0 \ll E $ finally yields
\begin{align*}
T_E = T_0 + PE + \frac{1}{2}P\dot{P}E.
\end{align*}

In order to compare the model of a slowly changing period with the null hypotheses (\enquote{the period is stable}), both identify with
\begin{align*}
O &\coloneqq T_0 + PE + \frac{1}{2} P \dot{P}E \\
C &\coloneqq T_0' + P'E,
\end{align*}
where $ O $ stands for \textit{observed} ephemerides with a constant period $ P' $ and $ C $ for the \textit{calculated} or \textit{computed} ephemerides. With $ \Delta T = T_0 - T_0' $ and $ \Delta P = P - P' $ and $ E = tP^{-1} $ this leads to
\begin{align*}
O-C = \Delta T_0 + \frac{\Delta P}{P} t + \frac{1}{2} \frac{\dot{P}}{P} t^2.
\end{align*}

In an $ O-C $ diagram, a fit with a parabolic function yields the evolutionary time scale $ \dot{P}/P $. 

\subsection{Light travel time effect}

\begin{figure}[tbp]
	\centering
	\begin{overpic}[scale=1]{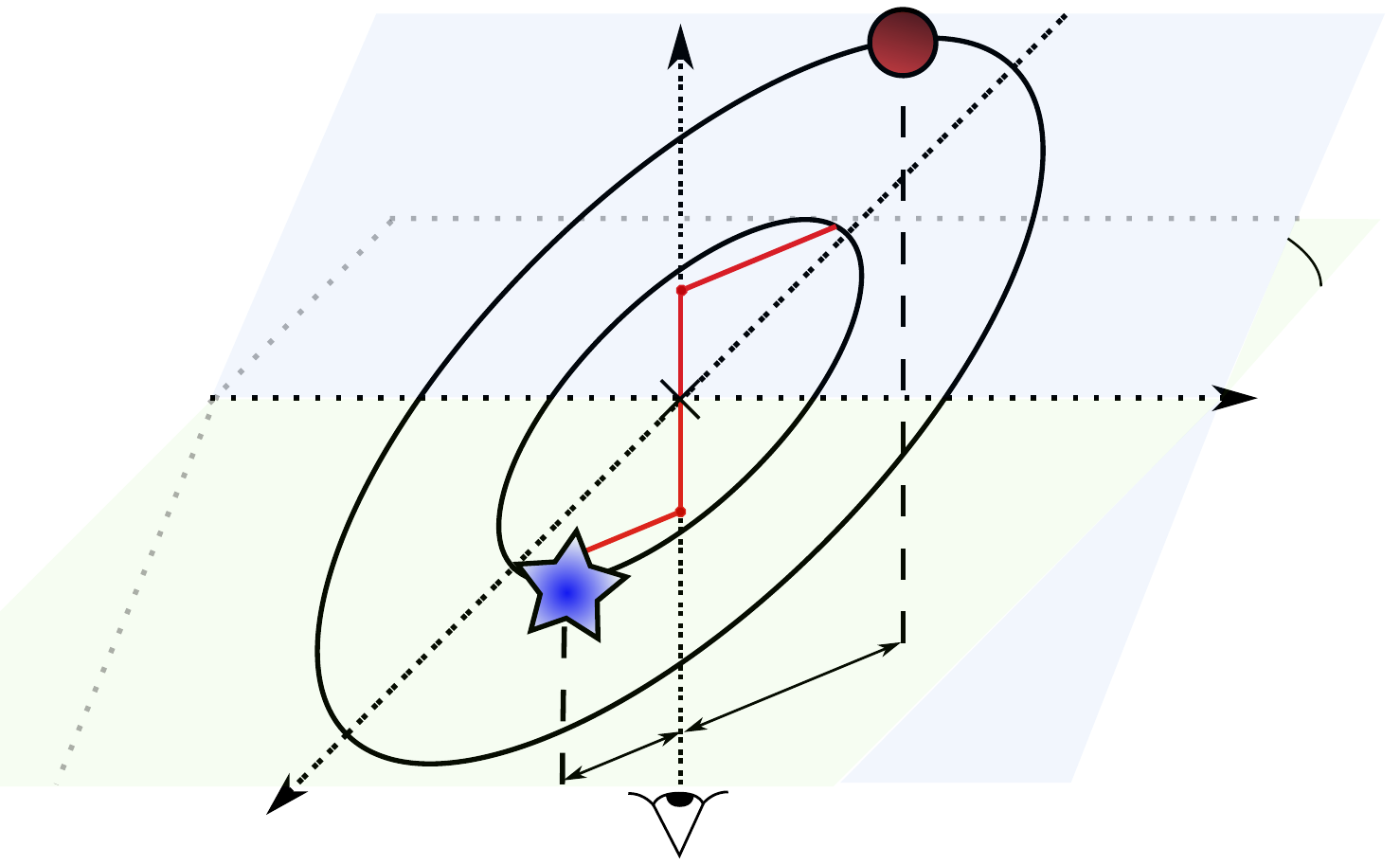}
		\put(18,6){$ x $}
		\put(45,57){$ z $}
		\put(88,30){$ y $}
		\put(44,9){$ a $}
		\put(54,15){$ a_{comp} $}
		\put(91.5,40){$ i $}
		\put(37,13){$ M $}
		\put(61,54){$ m $}
		\put(49,38){\textcolor{cb-red}{\boldmath$ +\Delta T $}}
		\put(49,29){\textcolor{cb-red}{\boldmath$ -\Delta T $}}
	\end{overpic}
	\caption[Schematic picture of the orbital configuration.]{Schematic picture of the orbital configuration.}
	\label{fig:methods:orbit}
\end{figure}
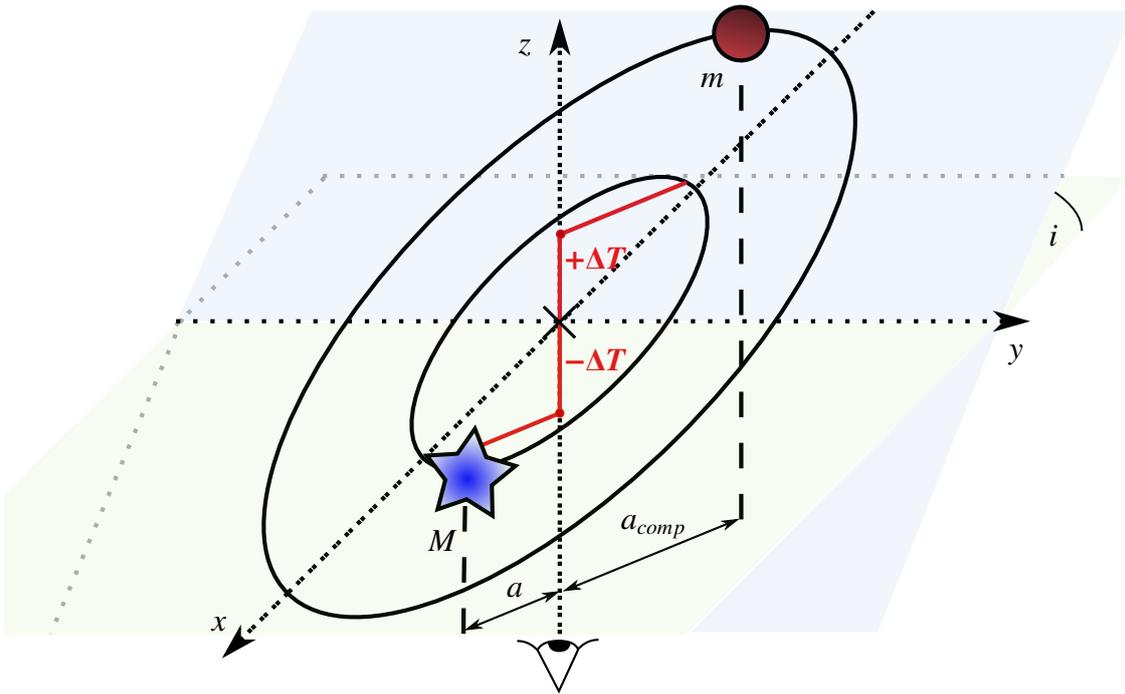

Assuming a two body system consisting of the pulsating star and a companion, both orbit their common centre of mass. The motion of pulsating star with mass $ M $ around the barycentre implies a variation in the projected line of sight, as show in Fig.~\ref{fig:methods:orbit}, which translates into a change in light travel time. The periodic pulsation of the star allow for measuring this change and hence, concluding on the minimum mass of the companion $ m $. Assuming a circular orbit due to the proposed formation history of companions of subdwarf B stars, the change in light travel time varies periodically and imposes a sinusoidal signal in the $ O-C $ diagram. The following describes the derivation of the expected amplitude of this light travel time effect \citep[e.g.][pp 29]{hilditch_introduction_2001}.

Let the line of sight be aligned with the $ z $ direction. With the inclination of the orbit $ i $, the argument of periapsis $ \omega $ and the true anomaly $ \vartheta $ of the Kepler ellipse, the $ z $ component of the radius vector $ r $ reads as
\begin{align*}
r_z = r \sin\left( i \right) \sin\left( \omega + \vartheta \right).
\end{align*}
Kepler's laws of planetary motion yield 
\begin{align*}
r = \frac{a \left( 1 - e^2 \right)}{1 + e \cos \left( \vartheta \right)}, \label{eq:rz1}
\end{align*}	
with the semi mayor axis $ a $ and the eccentricity $ e $. Thus,
\begin{align*}
r_z = \frac{a \left( 1 - e^2 \right)}{1 + e \cos \left( \vartheta \right)} \sin\left( i \right) \sin\left( \omega + \vartheta \right).
\end{align*}
With Kepler's Third Law follows for the semi major axis
\begin{align*}
a^3 = G \frac{m^3 P^2}{4 \pi^4 \left( M+m \right)^2}, 
\end{align*}	
where $ P $ is the orbital period. Assuming a circular orbit with $ e=0 $. The maximum displacement $ \Delta r_z $ in the orbit occurs at $ \sin\left( \omega + \vartheta \right) = \pm1 $:
\begin{align*}
\Delta r_z = 2 G^{1/3} m \left( \frac{P}{2\pi \left( M+m \right)} \right)^{2/3} \sin i .
\end{align*}
The light travel time follows from
\begin{align*}
\Delta T &= \frac{r_z}{c} \\
&= \frac{2 G^{1/3} m}{c} \left( \frac{P}{2\pi \left( M+m \right)} \right)^{2/3} \sin i,
\end{align*}
which is the semi-amplitude of the sinusoidal signal in the $ O-C $ diagram induced by a sub-stellar companion due to the light travel time effect.

\section{Artificial data testing}\label{sec:testing}
Once such a fast pipeline is deployed, it provides a convenient way to analyse artificial data with precisely known input parameters and test for the expected results. In order to avoid additional effects due to irregular observations and gaps in the data, the artificial data are mostly generated according to the PLATO scheme: cadence of \SI{25}{\second} for three years (i.e. $ 1.86\times10^6 $ measurements), and a signal to noise ratio of about 6, as expected for a \SI{14}{\mag} star in the B band observed with all 24 telescopes \citep{plato_esa/sre201113_plato_2011}. A different set of artificial data is created for a simplified ground-based observational scheme (three consecutive nights per month, each eight hours, for ten years, i.e. $ 1.24\times10^5 $). Since the results between these two sets are very similar, merely the PLATO-like scheme is discussed below. 

In total, about 50 different scenarios of frequency spacings and amplitudes were investigated. Some interesting cases are explained below.

\paragraph{Resonant frequencies} 
Pulsations with frequency-spacing in integer multiples might cause an unstable fitting process. To test for such scenarios, a set of several light curves was created. They simulate pulsation frequencies in 1:2, 1:3 and 1:5 resonances with a constant phase and amplitude. The resulting fits to individual modes tend to show oscillating phases but the simultaneous fit of all frequencies is stable and recovers successfully phase- and amplitude-information. Fig.~\ref{fig:methods:resonant} shows an example of resonant modes with a base frequency of $ \SI{240}{\per\day} $. Some modes tend to show an constant offset in the $ O-C $ diagram but are on average still close to the null hypothesis.
\begin{figure}[tbp]
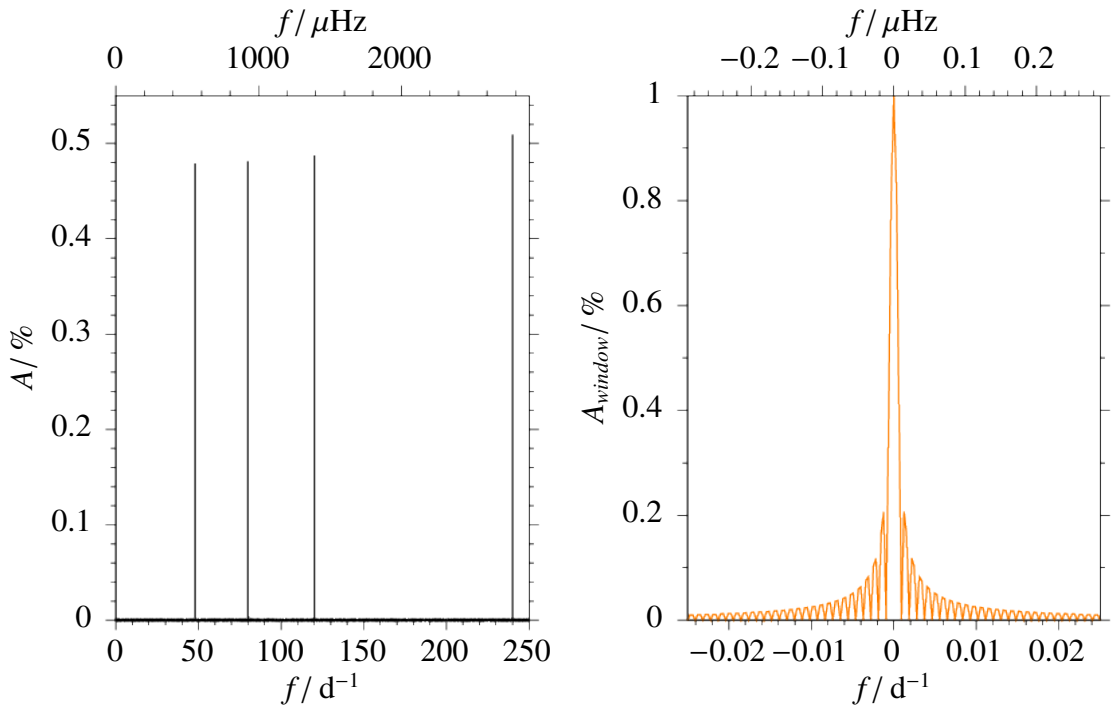

	\centering
	\input{figures/pipeline/resonant-02/powerspec.tex}	\input{figures/pipeline/resonant-02/window.tex}
	\caption[Amplitude spectrum for four resonant pulsations with frequencies.]{\textit{Left:} Amplitude spectrum for four resonant pulsations with frequencies of \SIlist{240;120;80;48}{\per\day} and amplitudes of \SI{0.5}{\percent}. \textit{Right:} Window function.}
	\label{fig:methods:resonant-window}
\end{figure}
\begin{figure}[tbp]
	\centering
	\input{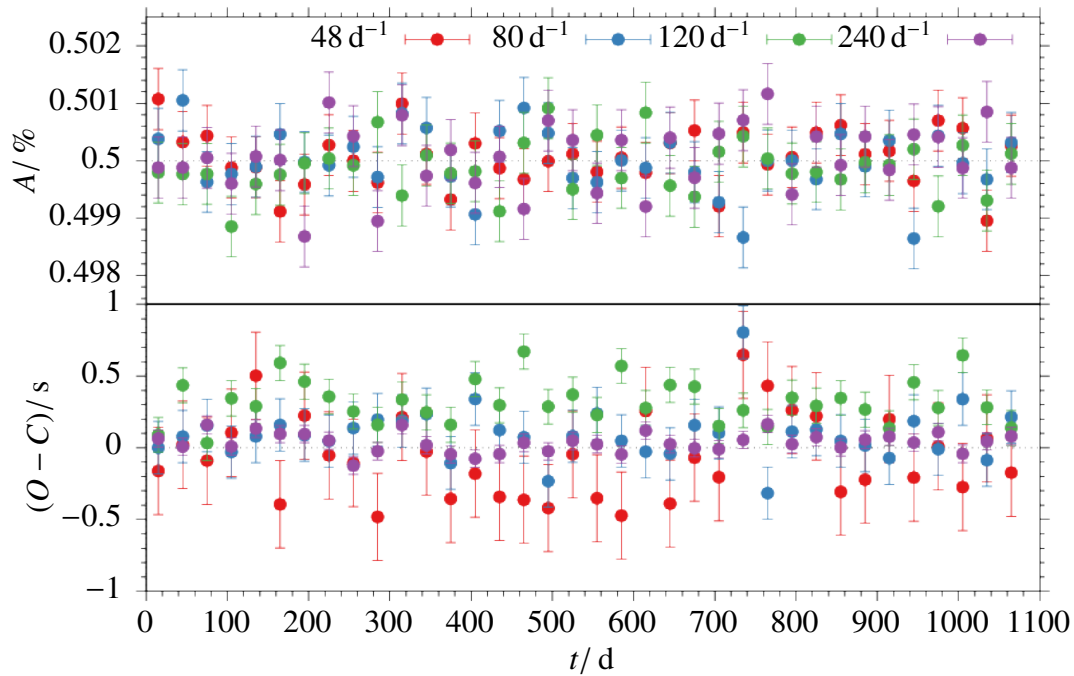}
	\caption[Results for four resonant pulsations.]{Results for four resonant pulsations with frequencies of \SIlist{240;120;80;48}{\per\day}. Each epoch contains data from one month. \textit{Top:} Reconstructed Amplitudes. \textit{Bottom:} $ O-C $ diagram.}
	\label{fig:methods:resonant}
\end{figure} 

\paragraph{Close frequencies}
Unresolved frequencies are a problem in the analysis of data and interpretation of their results. The frequency resolution is intrinsic to the observation pattern. Unfortunately, it is very difficult to conclude from an observational $ O-C $ pattern on unresolved frequencies. The vast range of possible combinations in number of frequencies, their spacing and amplitude ratios makes it almost impossible to conclude on a specific configuration to explain the observations. However, if there is a resolved multiplet in the observations, but the signal-to-noise ratio of individual amplitudes is not good enough to provide useful phase measurements, a simulated multiplet on the observational time grid can provide insights on the $ O-C $ behaviour of the main frequency. Additionally the window function can help to interpret amplitude spectra.

Artificial test cases can raise awareness of possible $ O-C $ patterns and caution the interpretation of observational results. The following tests were conducted with two pulsation frequencies and a constant phase which should result in a $ O-C $ diagram not different from zero.

\begin{figure}[tbp]
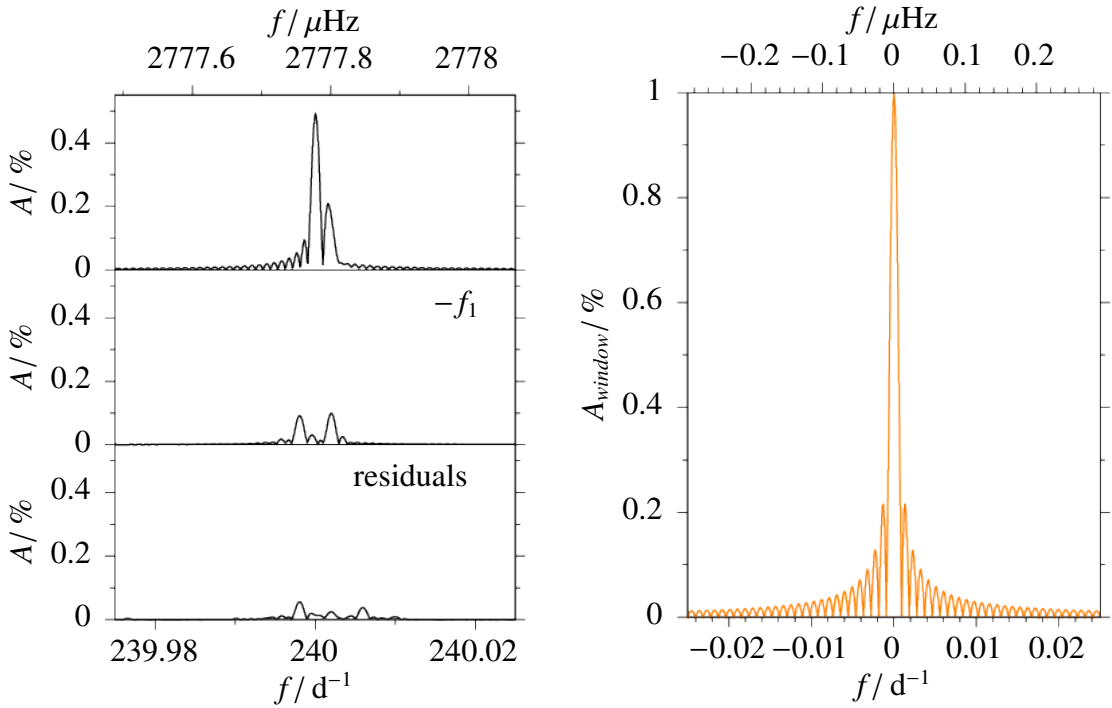

	\centering
	\input{figures/pipeline/close-03/powerspec.tex}	\input{figures/pipeline/close-03/window.tex}
	\caption[Amplitude spectrum for two close pulsations.]{\textit{Left:} Amplitude spectrum for two close pulsations with frequencies of \SIlist{240.000;240.002}{\per\day} and amplitudes of \SIlist{0.5;0.2}{\percent} (top), after subtraction of the mode at \SI{240.000}{\per\day} (middle) and residuals after subtraction of the mode at \SI{240.002}{\per\day} (bottom). \textit{Right:} Window function.}
	\label{fig:methods:close3-window}
\end{figure}

\begin{figure}[tbp]
	\centering
	\input{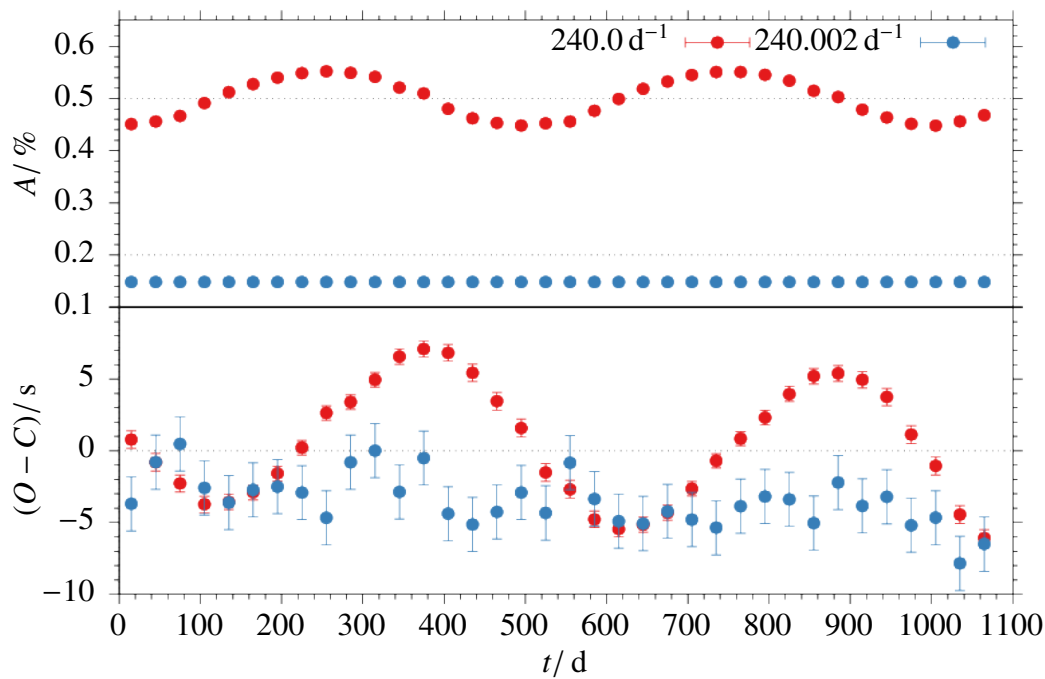}
	\caption[Results of the simultaneous fit of two close pulsations]{Results of the simultaneous fit of two close pulsations with frequencies of \SIlist{240.0;240.002}{\per\day} and similar amplitudes. Each epoch contains data from one month. \textit{Top:}Reconstructed  Amplitudes. \textit{Bottom:} $ O-C $ diagram.}
	\label{fig:methods:close3}
\end{figure}

With a sufficient frequency separation and signal-to-noise ratio for both pulsations (see Fig.~\ref{fig:methods:close3-window}), the $ O-C $ diagram might show mode-beating effects like in Fig.~\ref{fig:methods:close3}. The recovered amplitude and phase of the main pulsation mode show periodic variations, although they should appear constant. Thus, the simultaneous occurrence of amplitude- and phase-modulations in an $ O-C $ diagram should caution the interpretation in favour of a light travel time effect.

\begin{figure}[tbp]
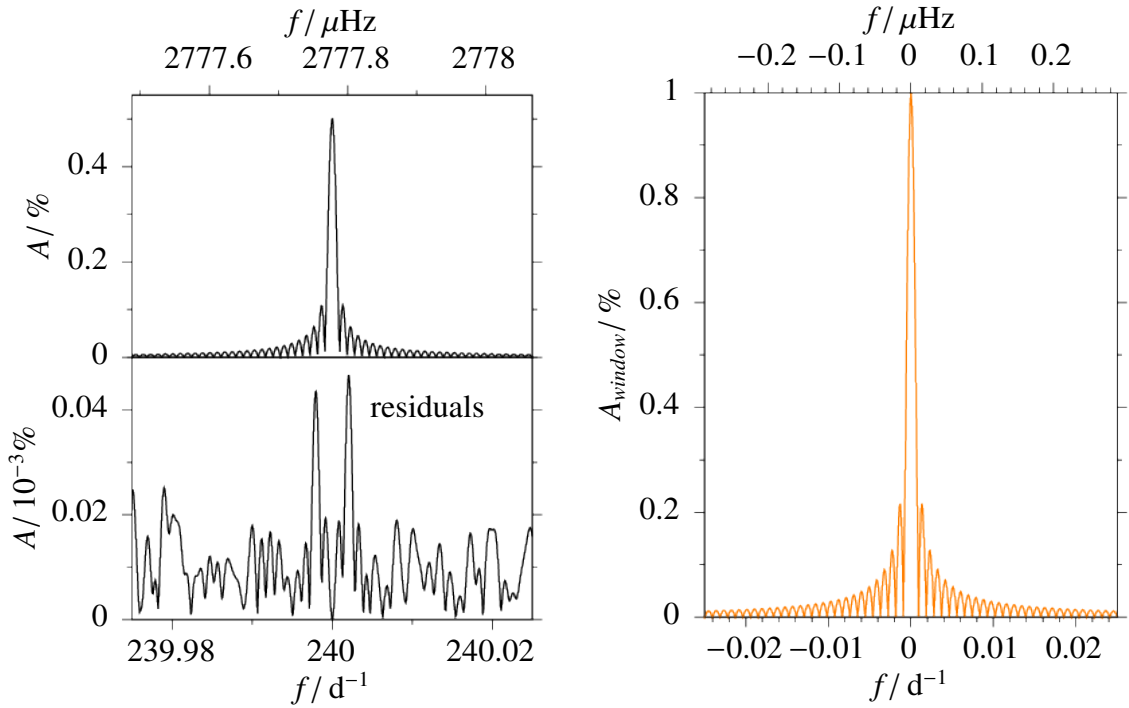

	\centering
	\input{figures/pipeline/close-04/powerspec.tex}	\input{figures/pipeline/close-04/window.tex}
	\caption[Amplitude spectrum for two close pulsations and very different amplitudes.]{\textit{Left:} Amplitude spectrum for two close pulsations with frequencies of \SIlist{240.0;240.002}{\per\day} and amplitudes of \SIlist{0.5;0.001}{\percent} (top), residuals after subtraction of the mode at \SI{240.0}{\per\day} (bottom). Note the different scale. The residual peaks are only slightly above the average noise level. \textit{Right:} Window function.}
	\label{fig:methods:close4-window}
\end{figure}

\begin{figure}[tbp]
	\centering
	\input{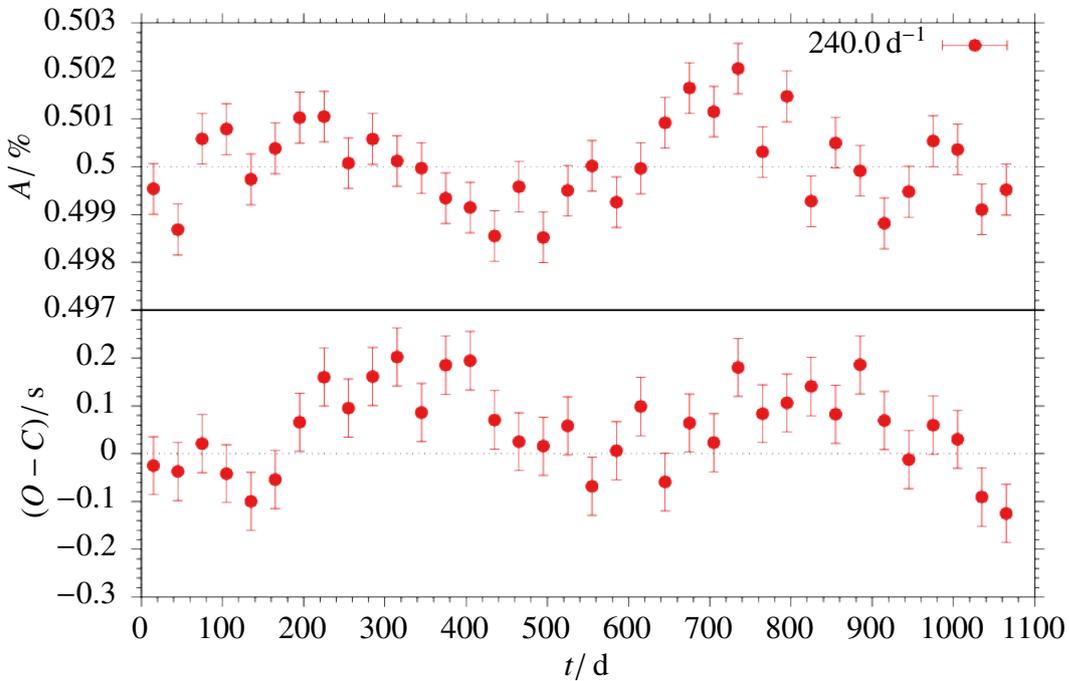}
	\caption[Results for two close pulsations and very different amplitudes.]{Results for two close pulsations with frequencies of \SIlist{240.0;240.002}{\per\day} and very different amplitudes. Each epoch contains data from one month. \textit{Top:} Reconstructed Amplitude. \textit{Bottom:} $ O-C $ diagram for the main pulsation.}
	\label{fig:methods:close4}
\end{figure}

When the signal-to-noise ratio for one frequency is barely over the noise level, it might be considered as noise after the pre-whitening of the dominant mode and neglected. 
Using the example above, but reducing the amplitude of the second pulsation from \SIlist{0.5}{\percent} to \SIlist{0.001}{\percent} (see Fig.~\ref{fig:methods:close4-window}), leads to the $ O-C $ diagram in Fig.~\ref{fig:methods:close4}. The second pulsation is not taken into consideration because the amplitude does not elevate above the noise level. The main pulsation shows periodic variations in the phase, as to be expected from a substellar companion. At the same time, the amplitude appears to be modulated as well, indicating a mode-beating effect. 
These scenarios strengthen the need for at least one additional phase variation measurement on an independent pulsation mode in order to conclude on reflex motion effects due to a substellar companion. 

\begin{figure}[tbp]
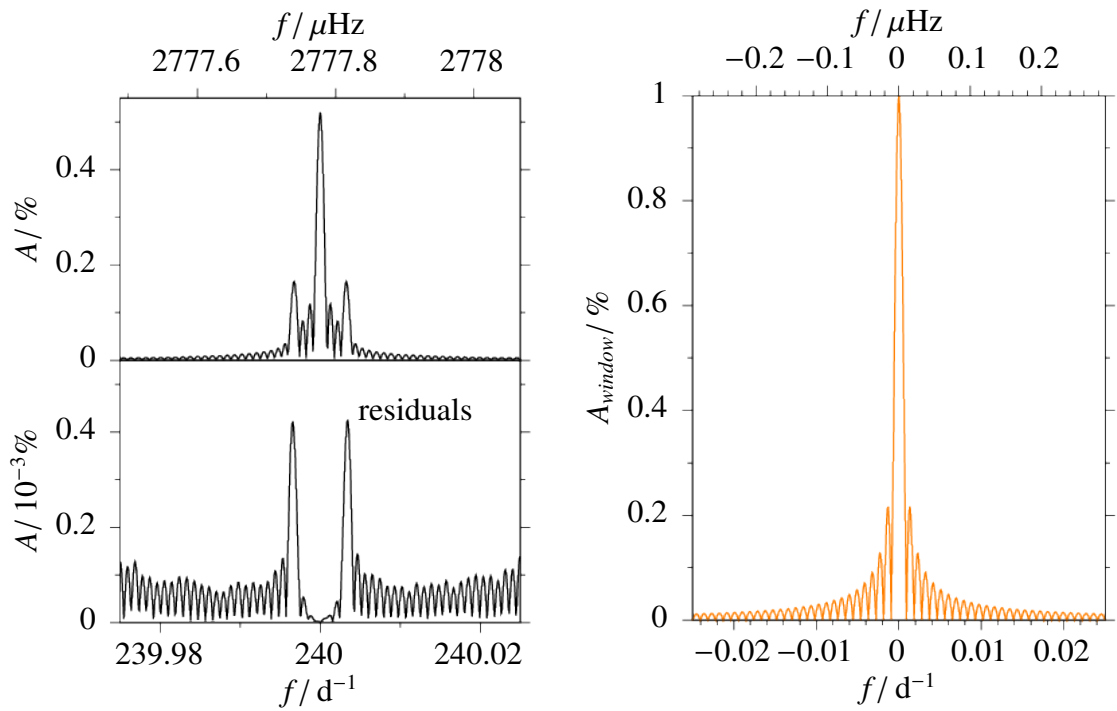
 
	\centering
	\input{figures/pipeline/amplitude-35/powerspec.tex}	\input{figures/pipeline/amplitude-35/window.tex}
	\caption[Amplitude spectrum for one pulsation of varying amplitude.]{\textit{Left:} Amplitude spectrum for one pulsation with a frequency of  \SIlist{240.0}{\per\day} and an amplitude of \SIlist{0.5}{\percent}, varying by \SI{50}{\percent} with a \SI{300}{\day} period (top), residuals after pre-whitening of the mode at \SI{240.0}{\per\day} (bottom). \textit{Right:} Window function.}
	\label{fig:methods:amplitude35-window}
\end{figure}

\begin{figure}[tbp]
	\centering
	\input{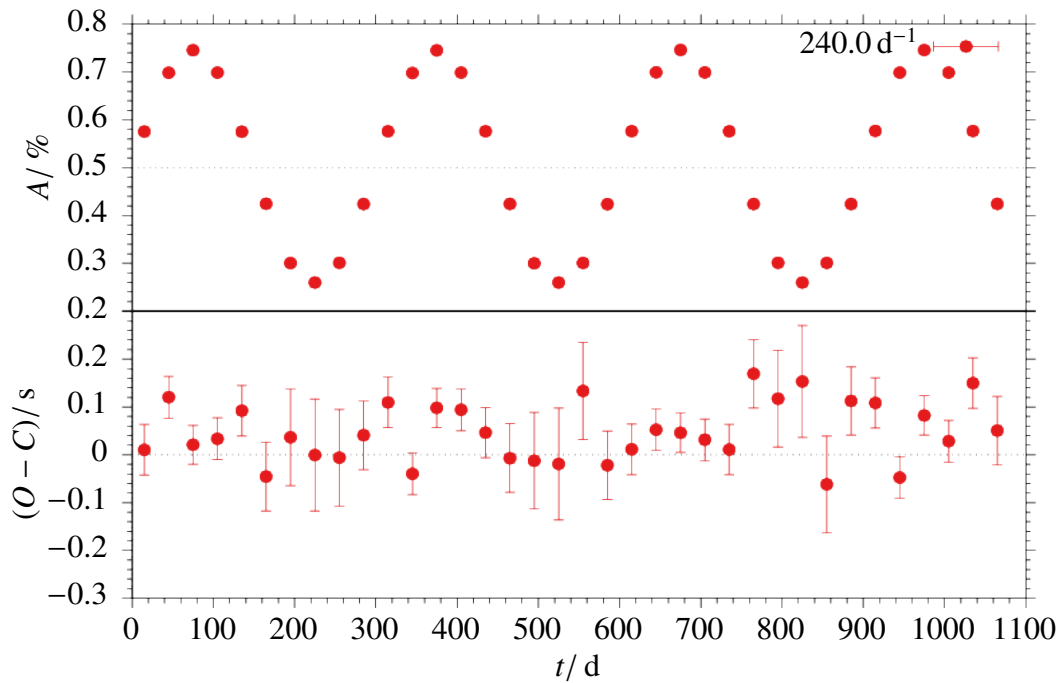}
	\caption[Results for one simulated pulsation with varying amplitude.]{Results for one simulated pulsation with a frequency of  \SIlist{240.0}{\per\day} and an amplitude of \SIlist{0.5}{\percent}. The amplitude is varying by \SI{50}{\percent} with a \SI{300}{\day} period. Each epoch contains data from one month. \textit{Top:} Reconstructed Amplitude. \textit{Bottom:} $ O-C $ diagram.}
	\label{fig:methods:amplitude35}
\end{figure}

\paragraph{Variable amplitude}
Previous observational studies show that amplitude modulations are a common phenomenon for sdB stars \citet{kilkenny_amplitude_2010}. But due to the relatively short observational baseline of such investigations, it is not clear whether amplitude modulations can appear on their own or together with phase modulations (as result of mode beating effects). Thus, the $ O-C $ method needs to be able to distinguish between the two scenarios, which can be tested by using a simulated mode with an amplitude modulation but no phase modulation. The example data set shown in Fig.~\ref{fig:methods:amplitude35-window} and \ref{fig:methods:amplitude35} exhibits an amplitude modulation of \SI{50}{\percent}. The $ O-C $ analysis recovers successful this amplitude variation. Also, the phase shows no simultaneous variation which enables the distinction between a mode beating and pure amplitude modulated scenario.

\section{Alternative approaches}
\subsection{Analytic signal}
The analytic signal (AS) is a complex-valued function, where real and imaginary part are related to each other by the Hilbert transform. In polar coordinates the AS can be expressed in terms of its amplitude and phase as a function of time, called envelope and instantaneous phase. The instantaneous frequency can be obtained by differentiating the instantaneous phase in respect to time. From the instantaneous phase, an $ O-C $ diagram can be inferred, which makes the AS a potentially useful tool to investigate phase, amplitude and frequency variations in stellar pulsations. See \citet{iii_mathematics_2007} for a comprehensive overview.

The AS $ \tilde{s}(t) $ of a signal $ s(t) $ is defined as 
\begin{align*}
\tilde{s}(t) = s(t) + i \mathcal{H}\left( s(t) \right),
\end{align*}
where the Hilbert transformation $ \mathcal{H} $ is defined over the convolution
\begin{align*}
\mathcal{H} = \frac{1}{\pi t} * s(t).
\end{align*}
While the original signal generates one amplitude value for each point in time, the analytical representation generates a rotating vector in the complex plane. Thus, the analytical representation provides instantaneous values for amplitude, frequency and phase, with only a single sample of a discrete signal. 		

The AS is implemented in the \verb|scipy| library \citep{jones_scipy_2011}. Since it is based on Fourier transforms, it requires equidistant steps in time. Thus, a space born observatory with a regular and continuous observing scheme is preferred over ground based observations by this method. Due to barycentric motion of the spacecraft the time steps will be distorted after correcting for the barycentric motion. Therefore, an interpolation in time is necessary which makes the analysis computationally very expensive. 

In order to test the implementation, the artificial light curves should represent this temporal behavior. The data were generated sampling one stellar pulsation mode according to the PLATO scheme (cadence of \SI{25}{\second} for three years) and an \enquote{inverse} barycentric correction was applied in order to create the irregular temporal spacing. 

\begin{figure}[tbp]
	\centering
	\input{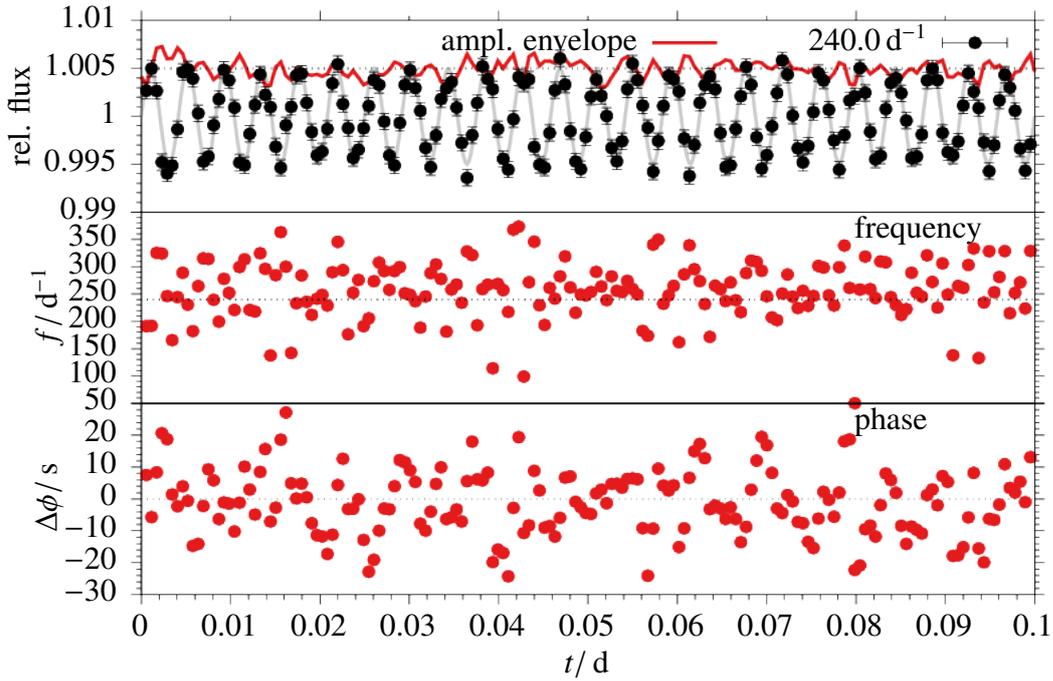}
	\caption[Results of the Hilbert transform for one pulsation.]{Results of the Hilbert transform for one pulsation with a frequency of  \SIlist{240.0}{\per\day} and an amplitude of \SIlist{0.5}{\percent} \textit{Top:} Flux measurements (black) sampled from the pulsation (grey) and amplitude envelope (red). \textit{Middle:} Instantaneous frequency \textit{Bottom:} Instantaneous phase minus reference phase.}
	\label{fig:methods:hilbert}
\end{figure}

For the analysis, the artificial data are interpolated on a grid with a spacing of the shortest time between two measurements, close to \SI{25}{\second}. From an amplitude spectrum, the frequency of the pulsation mode is selected, very similar to the classical $ O-C $ approach implemented for this work. In the presence of multiple pulsation frequencies, a frequency filtering is necessary. After the computation of the envelope, instantaneous phase and frequency, the expected linear cycle count is subtracted from the instantaneous phase to yield the $ O-C $ diagram (see Fig.~\ref{fig:methods:hilbert}). Since the AS is computed without a binning of data, the result appears to be scattering much more than the classical $ O-C $ result. Averaging the results over a sufficient amount of time smooths the measurements. Nevertheless, the process of interpolation before the actual analysis can start is computationally very expensive; about a factor 1000 more than the classical $ O-C $ calculation. This makes the Hilbert transform not practical for analysing irregular high cadence data. This approach has not been further pursued in this work.

\subsection{Independent confirmation via direct imaging}
Direct imaging with adaptive optics allow for observations near the refraction limit of the telescope. This might offer the possibility to confirm sub-stellar companion candidates where a clear detection using the pulsation timing is affected by non-linear effects of stellar pulsations.
However, possible companions are still not resolvable with current technology. V391~Peg and DW~Lyn have a distance of $ d = \SI{1238.7 \pm 82.2422}{pc} $ and $ d = \SI{1495.4389 \pm 130.3785}{pc} $, respectively \citep{gaia_collaboration_gaia_2018}. With semi major axis of about $ \SI{1.7}{\au} $ \citep{silvotti_giant_2007} and $ \SI{1.96}{\au} $, the hypothetical companions would be separated by $ \SI{1.37}{mas} $ and $ \SI{1.31}{mas} $ from the host star, respectively. This is about one order of magnitude smaller than the theoretical angular resolution of $ \SI{8}{\meter} $ class telescopes. The remaining EXOTIME targets are not significantly closer to Earth. Thus, the sub-stellar companion candidates in the EXOTIME sample can not be confirmed via direct imaging.

\part{The EXOTIME project: Signals in the $ O-C $ diagrams of the rapidly pulsating subdwarfs DW Lyn, V1636 Ori, QQ Vir, and V541 Hya} \label{sec:exotime}

This chapter is published as a journal article \enquote{The EXOTIME project: Signals in the $ O-C $ diagrams of the rapidly pulsating subdwarfs DW Lyn, V1636 Ori, QQ Vir and V541 Hya} in Astronomy and Astrophysics (A\&A, 638 (2020) A108, DOI: \url{https://doi.org/10.1051/0004-6361/201937172}). \textcopyright~F. Mackebrandt et al. 2020.\\
Authors: F. Mackebrandt\footnote{Max-Planck-Institut f\"ur Sonnensystemforschung, Justus-von-Liebig-Weg 3, 37077 G\"ottingen, Germany}\textsuperscript{,}\footnote{Institut f\"ur Astrophysik, Georg-August-Universit\"at G\"ottingen, Friedrich-Hund-Platz 1, 37077 G\"ottingen, Germany}
, S. Schuh\footnotemark[1]
, R. Silvotti\footnote{INAF -- Osservatorio Astrofisico di Torino, strada dell'Osservatorio 20, 10025 Pino Torinese, Italy}
, S.-L. Kim\footnote{Korea Astronomy and Space Science Institute, Daejeon 34055, South Korea}
, D. Kilkenny\footnote{Department of Physics and Astronomy, University of the Western Cape, Private Bag X17, Bellville 7535, South Africa}
, E. M. Green\footnote{Steward Observatory, University of Arizona, 933 N. Cherry Avenue, Tucson, AZ 85721, USA}
, R. Lutz\footnote{German Aerospace Center (DLR), Remote Sensing Technology Institute, M\"unchener Str. 20, 82234 We\ss ling, Germany}
, T. Nagel\footnote{Institute for Astronomy and Astrophysics, Kepler Center for Astro and Particle Physics, University of T\"ubingen, 72076 T\"ubingen, Germany}
, J. L. Provencal\footnote{University of Delaware, Department of Physics and Astronomy Newark, DE 19716, USA}\textsuperscript{,}\footnote{Delaware Asteroseismic Research Center, Mt. Cuba Observatory, Greenville, DE 19807, USA}
, T. Otani\footnote{Embry-Riddle Aeronautical University, Department of Physical Science and SARA, Daytona Beach, FL 32114, USA}\textsuperscript{,}\footnote{Florida Institute of Technology, Department of Physics \& Space Sciences, Melbourne, FL 32901, USA}
, T. D. Oswalt\footnotemark[11]\textsuperscript{,}\footnotemark[12]
, S. Benatti\footnote{INAF -- Osservatorio Astronomico di Padova, Vicolo dell’Osservatorio 5, 35122 Padova, Italy}
, L. Lanteri\footnotemark[3]
, A. Bonanno\footnote{INAF -- Osservatorio Astrofisico di Catania, Via S. Sofia 78, 95123 Catania, Italy}
, A. Frasca\footnotemark[14]
, R. Janulis\footnote{Institute of Theoretical Physics and Astronomy, Vilnius University, Gostauto 12, Vilnius 01108, Lithuania}
, M. Papar\'{o}\footnote{Konkoly Observatory, Research Centre for Astronomy and Earth Sciences, Konkoly-Thege M. \'ut 15-17, 1121 Budapest, Hungary}
, L. Moln\'ar\footnotemark[16]\textsuperscript{,}\footnote{MTA CSFK Lend\"ulet Near-Field Cosmology Research Group, Konkoly Observatory}
, R. Claudi\footnotemark[13]
, R. H. \O stensen\footnote{Department of Physics, Astronomy, and Materials Science, Missouri State University, Springfield, MO 65897, USA} \\
Contributions: F.M. analysed the data and wrote the manuscript; S.S. is CoI of the EXOTIME project, coordinated observations at CAHA and supervised the findings of this work; R.S. is PI of the EXOTIME project, coordinated observations at TNG and contributed Fig.~\ref{fig:rv-planet}, and the according description in section~\ref{sec:testing-hypothesis}. The following authors observed and reduced the data: S.-L.K. (LOAO); D.K. (SAAO); E.M.G. (MtB); R.L. (M/N); T.N. (Tue); J.L.P. (WET); T.O. (SARA-KP); T.D.O. (SARA-KP); S.B. (Asi); L.L. (Loi); A.B. (Serra la Nave); A.F. (Serra la Nave); R.J. (Mol); M.P. (Kon); L.M. (Kon); R.C. (Asi); R.H.\O{}. (NOT) \\
The manuscript has been read, discussed and approved by all named authors. \\

Photometric data of Fig.~\ref{fig:lc}, results in Fig.~\ref{fig:dwlyn-oc}, \ref{fig:v1636ori-oc}, \ref{fig:qqvir-oc} and \ref{fig:v541hya-oc} and figures in the appendix are only available at the CDS via anonymous ftp to \url{cdsarc.u-strasbg.fr} (\url{130.79.128.5}) or via \url{http://cdsarc.
u-strasbg.fr/viz-bin/cat/J/A+A/638/A108}.\\
Based on observations obtained at the 0.9 m SARA-KP telescope,
which is operated by the Southeastern Association for Research in
Astronomy (\url{saraobservatory.org}).

\chapter{Introduction}
Subdwarf B stars (sdBs) are sub-luminous stars with a mass of about $ \unit[0.5]{M_{\odot}} $ located at the blue end of the horizontal branch, which is the so-called extreme horizontal branch \citep[EHB,][]{heber_atmosphere_1986}. They maintain a helium burning core, but their thin hydrogen envelope ($ M_{env} < \unit[0.01]{M_{\odot}}$) cannot sustain hydrogen shell burning, which identifies sdBs with stripped cores of red giants \citep{heber_hot_2016}. Binary evolution with a common envelope (CE) is the favoured formation scenario for most sdBs. The sdB progenitor fills its Roche lobe near the tip of the RGB. A CE is formed when the mass transfer rate is sufficiently high and the companion star cannot accrete all the matter. 
For close binary systems with small initial mass ratios $ q < 1.2 \text{--} 1.5 $, two phases of mass transfer occur. The first Roche-lobe overflow is stable, whereas the second one is unstable, leading to the ejection of the CE. The resulting binary consists of an sdB star and a white dwarf in a short-period orbit. For initial mass ratios $ q > 1.2 \text{--} 1.5 $ the first mass-transfer phase is unstable and the CE is ejected, producing an sdB star with a non-degenerate (e.g. main sequence star) companion.
A more detailed review of formation and evolution of compact binary systems can be found in \citet{podsiadlowski_evolution_2008} and \citet{postnov_evolution_2014}. These formation scenarios cannot explain the observed additional occurrence of apparently single sdBs \citep{maxted_binary_2001}.
Among the proposed formation scenarios is the proposal by \citet{webbink_double_1984} that they could be formed by a merger of two helium white dwarfs. But such mergers are problematic, as they are expected to retain very little  hydrogen \citep{han_origin_2002} and be left with higher rotation rates than what has been observed \citep{charpinet_rotation_2018}.
Moreover, the overall observed mass distribution of single sdB stars is not consistent with that expected from the proposed formation scenario. Sub-stellar companions could resolve this disagreement between theory and observations. Planetary-mass companions like the candidates V391 Peg\,b \citep{silvotti_giant_2007}, KIC 05807616\,b,c \citep{charpinet_compact_2011}, KIC 10001893\,b,c,d \citep{silvotti_kepler_2014}, or brown dwarf companions like V2008-1753\,B \citep{schaffenroth_eclipsing_2015} or CS 1246 \citep{barlow_fortnightly_2011} indicate the existence of a previously undiscovered population of companions to apparently single sdBs.

Due to the high surface gravity and effective temperature (leading to few, strongly broadened spectral lines in the optical) and the small radii of sdBs, the detection efficiency for companions via methods like radial velocity variations or transits is small. The timing of stellar pulsations offers a complementary detection method, sensitive to large orbital separations. 

A small fraction of sdB stars shows pulsational variations in the $ p $- (pressure-) and $ g $- (gravity-) mode regimes.
Rapid $ p $-mode pulsators (sdBV\textsubscript{r}), discovered by \citet{kilkenny_new_1997}, show periods of the order of minutes and amplitudes of a few tens of $ \unit{mmag} $. Such pulsations were predicted by \citet[][et seq.]{charpinet_driving_1997} to be driven by the \textkappa-mechanism due to a $ Z $-opacity bump. For slow pulsators (sdBV\textsubscript{s}) the periods range from 30 to 80 $ \unit{min} $ with small amplitudes of a few $ \unit{mmag} $. This class was  discovered by \citet{green_discovery_2003} and the pulsations are explained by the \textkappa-mechanism as well \citep{fontaine_driving_2003}. Some sdB stars show both types of pulsation modes simultaneously (sdBV\textsubscript{rs}). These hybrid pulsators lie at the temperature boundary near $ \unit[28000]{K} $ between the two classes of pulsating stars, for example, the prototype for this class DW~Lyn \citep{schuh_hs_2006}, which is also addressed in this work, or Balloon~090100001 \citep{baran_multicolour_2005}.

Pulsations driven by the \textkappa-mechanism are coherent, which qualifies these objects for the timing method to search for sub-stellar companions. This method is based on the light-travel time effect, with the host star acting as a stable \enquote{clock} spatial movements of the star around the barycentre induced by a companion result in time delays of the stellar light measured by the observer. Examples of detections using this method are \enquote{pulsar-planets} \citep[e.g.][]{wolszczan_planetary_1992}, planets detected by transit timing variations \citep[e.g. Kepler 19\,c,][]{ballard_kepler-19_2011}, planets orbiting \textdelta Scuti stars \citep{murphy_planet_2016}, or eclipsing binaries \citep[e.g. V2051 Oph (AB)\,b,][]{qian_long-term_2015}. In particular, the detection of a late-type main sequence star companion to the sdB CS 1246 by \citet{barlow_fortnightly_2011}, subsequently confirmed with radial velocity data \citep{barlow_radial_2011}, or other studies like \citet{otani_orbital_2018}, demonstrate the viability of this method in sdB systems. On the other hand, the particular example of V391 Peg\,b is currently under discussion \citep{silvotti_sdb_2018} because of possible non-linear interactions between different pulsation modes, that change arrival times
(see \citealt{zong_oscillation_2018} for a detailed study of amplitude/frequency variations related to nonlinear effects). Stochastically driven pulsations, are suspected by \citet{reed_followup_2007,kilkenny_amplitude_2010} and their nature confirmed by \citet{ostensen_stochastic_2014}. 
Also the candidate detections of KIC 05807616 and KIC 10001893 are uncertain, since other sdBs observed within the \textit{Kepler K2}- mission exhibit $ g $-modes with long periods up to few hours. They question the interpretation of the low-frequency variations for KIC 05807616 and KIC 10001893 \citep[e.g.][]{krzesinski_planetary_2015,blokesz_analysis_2019}.

In order to detect sub-stellar companions orbiting rapidly pulsating sdB stars, the \textit{EXOTIME} observational programme (EXOplanet search with the TIming MEthod) has been taking long term data since 1999. \textit{EXOTIME} conducted a long-term monitoring programme of five rapidly pulsating sdB stars. V391 Peg has been discussed by \citet{silvotti_giant_2007,silvotti_sdb_2018}. In this paper, we present the observations of DW Lyn and V1636 Ori, previously discussed in \citet{lutz_light_2008,schuh_exotime_2010,lutz_search_2011,lutz_exotime_2011}, and re-evaluate their findings using an extended set of observations. In addition, the observations of QQ Vir and V541 Hya are presented and analysed. In the beginning of the programme, the mode stability was tested for all targets over a timespan of months in order to ensure the pulsation modes were coherent.

For the DW Lyn observations, \citet{lutz_exotime_2011} found no significant signals in a periodogram of the $ O-C $ data of the two analysed pulsation frequencies, which would indicate sub-stellar companions. A tentative signal in the second frequency (in this work labelled $ f_2 $, as well), formally corresponding to an 80-day companion orbit, is concluded to arise from mode beating of an unresolved frequency doublet. The analysis of V1636 Ori revealed a signal at $ \unit[160]{d} $ in the periodogram of the main frequency $ O-C $ data \citep{lutz_exotime_2011}. Although this periodicity showed a significance of only $ \unit[1]{\sigma} $, \citet{lutz_exotime_2011} predicted an increase of significance with follow-up observations. We are using this extended data set in our work, now incorporating observations up to 2015.

This paper is organised as follows. Section~\ref{sec:observations} describes the observational aspects within the \textit{EXOTIME} programme and the data reduction, followed by a description of our analysis in Sect.~\ref{sec:analysis}. Our results are presented in Sect.~\ref{sec:results}, together with a discussion.

\chapter{Observations and data reduction} \label{sec:observations}
The observational data necessary for the analysis are comprised of many individual data sets gathered over the course of up to two decades. The detection method demands the observation of a target for a total time base at least as long as one orbit of a potential companion, which can span several years. This requires coordinated campaigns with observatories using \textasciitilde 1 -- 4 $ \unit{m} $ telescopes.
In order to derive sufficient accuracy for the analysis, observations with at least three to four consecutive nights, each with a minimum of two to three hours per target are required. To resolve the short-period p-modes the cadence must be shorter than about $ \unit[30]{s} $ but still with a sufficient signal-to-noise ratio (S/N). All observations used the Johnson-Bessel \textit{B} band. The correct time stamps for each observation are of most importance for the timing analysis. Most observatories of this study contributed already successfully to the work of \citet[][Table 2]{silvotti_giant_2007,silvotti_subdwarf_2008}. The following lists some references where telescopes used for this study contributed successfully to other timing-relevant observations.
Konkoly RCC $ \unit[1.0]{m} $ Telescope 
: \citet{provencal_2006_2009,stello_multisite_2006},
Mt. Lemmon Optical Astronomy Observatory: \citet{bischoff-kim_gd358_2019,lee_pulsating_2014},
Serra la Nave $ \unit[0.9]{m} $: \citet{bonanno_pg_2003,bonanno_asteroseismology_2003},
SARA-KP $ \unit[0.9]{m} $ telescope: \citet{kilkenny_orbital_2014,baran_pulsations_2018}.

Table~\ref{tab:stellar} lists the atmospheric parameters of the stars, and Table~\ref{tab:photometric} summarises the photometric observations obtained at multiple medium-class telescopes. Fig.~\ref{fig:lc} summarises the observational coverage.
\begin{table}
		\centering
		\caption{Atmospheric parameters of the targets.}
		\label{tab:stellar}
\begin{threeparttable}[htbp!]
	\begin{tabular}{lr@{$ \pm $}lr@{$ \pm $}lr@{$ \pm $}lc}
		\toprule
		Target & \multicolumn{2}{c}{$ \unit[T_{eff}/]{K} $} & \multicolumn{2}{c}{$ \log \left( g/\unitfrac{cm}{s} \right) $} & \multicolumn{2}{c}{$ \log \left( \frac{N \left(\ch{He}\right)}{N \left(\ch{H}\right)} \right) $} & Ref. \\
		\midrule
		DW Lyn & 28\,400 & 600 & 5.35 & 0.1 & -2.7 & 0.1 & 1 \\
		V1636 Ori & 33\,800 & 1\,000 & 5.60 & 0.15 & -1.85 & 0.20 & 2 \\
		QQ Vir & 34\,800 & 610 & 5.81 & 0.05 & -1.65 & 0.05 & 3 \\
		V541 Hya & 34\,806 & 230 & 5.794 & 0.044 & -1.680 & 0.056 & 4 \\
		\bottomrule
	\end{tabular}
	\begin{tablenotes}
	\item[1] \citet{dreizler_hs0702+6043_2002}; \item[2] \citet{ostensen_four_2001}; \item[3] \citet{telting_radial-velocity_2004}; \item[4] \citet{randall_observations_2009}
	\end{tablenotes}
\end{threeparttable}
\end{table}

\begin{figure}[htbp!]
\begin{leftfullpage}
	\centering
	\begin{subfigure}[b]{\textwidth}
		\centering
		\input{figures/exotime/dwlyn/lc_clipped.tex}
		\caption{}
		\label{fig:dwlyn-lc}
	\end{subfigure}
	\hfil
	\begin{subfigure}[b]{\textwidth}
		\centering
		\input{figures/exotime/v1636ori/lc_clipped.tex}
		\caption{}
		\label{fig:v1636ori-lc}
	\end{subfigure}
\end{leftfullpage}
\end{figure}
\begin{figure}[htbp!]\ContinuedFloat
\begin{fullpage}
	\begin{subfigure}[b]{\textwidth}
		\centering
		\input{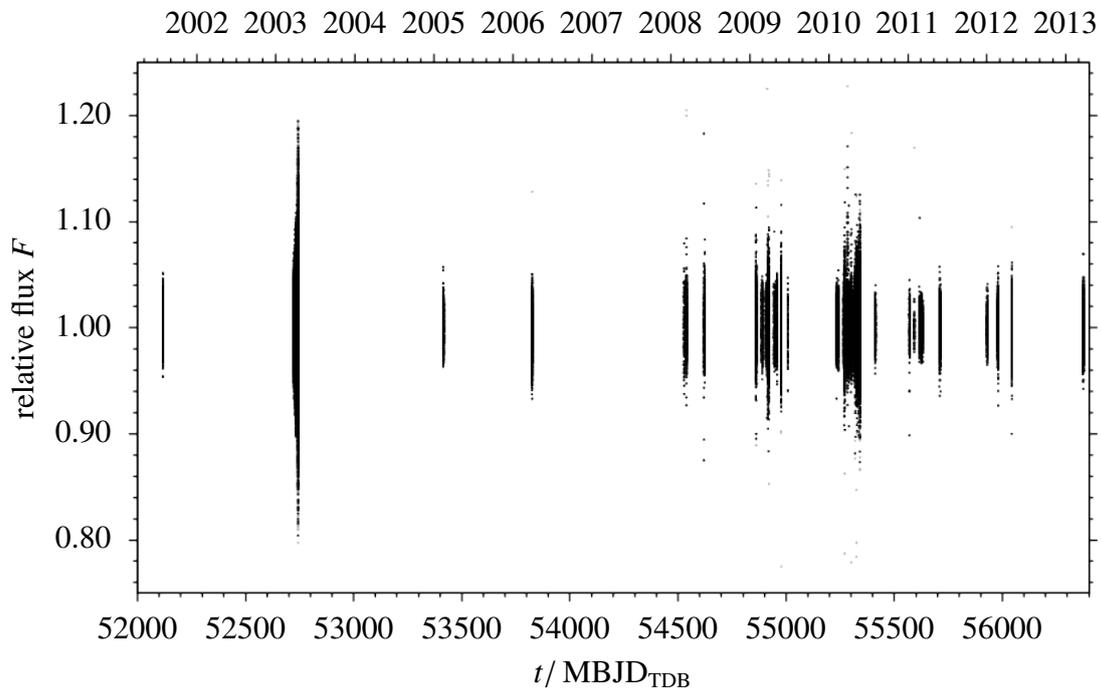}
		\caption{}
		\label{fig:qqvir-lc}
	\end{subfigure}
	\hfil
	\begin{subfigure}[b]{\textwidth}
		\centering
		\input{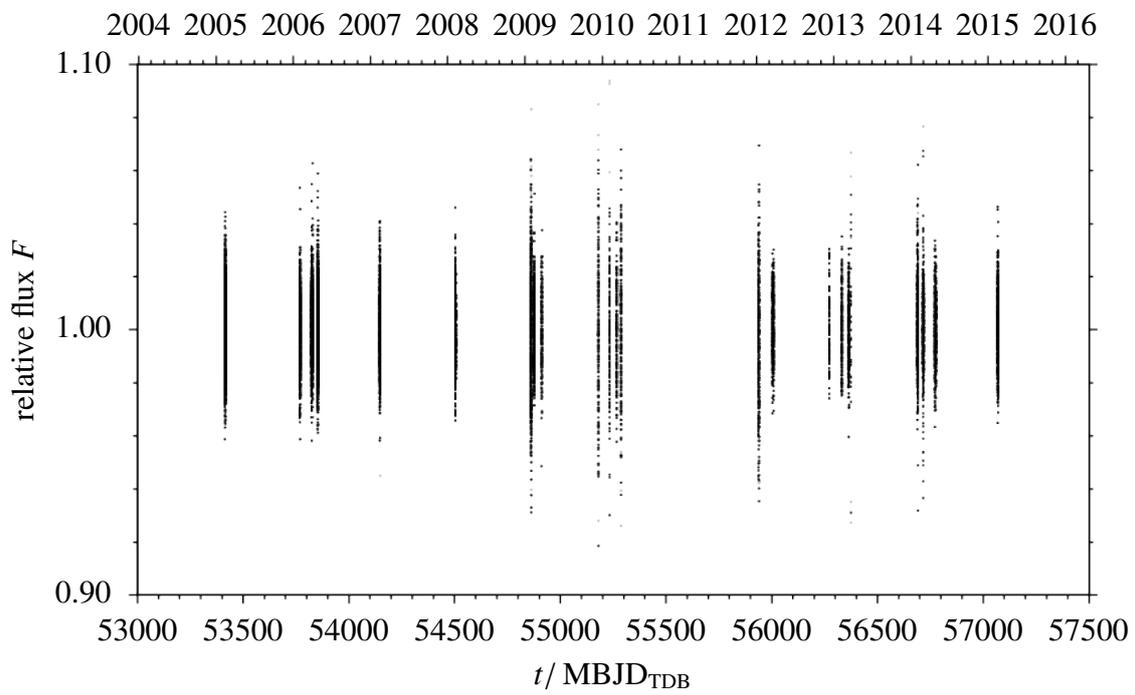}
		\caption{}
		\label{fig:v541hya-lc}
	\end{subfigure}
	\caption[Light curves.]{Light curves. Grey points are considered outliers and partially exceeding the plotting range. (a) DW Lyn. (b) V1636 Ori. (c) QQ Vir. (d) V541 Hya.}
	\label{fig:lc}
\end{fullpage}
\end{figure}

\begin{sidewaystable}
	\centering
	\caption[Summary of the observing time per target, per site.]{Summary of the observing time per target, per site in hours.}
	\label{tab:photometric}
	\begin{threeparttable}
	\begin{tabular}{lrrrr}
		\toprule
		Site & DW Lyn & V1636 Ori & QQ Vir & V541 Hya \\
		\midrule
		Asiago $ \unit[1.8]{m} $ Copernico Telescope (Asi) & 20.25 &  &  &  \\
		Calar Alto Observatory $ \unit[2.2]{m} $ (CAHA) & 52.38 & 32.49 & 48.73 & 10.19 \\
		Baker $ \unit[0.4]{m} $ &  &  & 41.10 &  \\
		BAO $ \unit[0.85]{m} $ &  &  & 47.70 &  \\
		BOAO $ \unit[1.85]{m} $ &  &  & 25.60 &  \\
		G\"ottingen IAG $ \unit[0.5]{m} $ Telescope (Goe) & 52.53 &  &  &  \\
		Konkoly RCC $ \unit[1.0]{m} $ Telescope (Kon) & 14.27 &  & 4.76 &  \\
		La Palma $ \unit[0.6]{m} $ &  &  & 37.30 &  \\
		Mt. Lemmon Optical Astronomy Observatory $ \unit[1.0]{m} $ (LOAO) & 167.76 & 40.12 & 126.11 & 24.15 \\
		Loiano $ \unit[1.5]{m} $ Telescope (Loi) &  & 2.66 & 78.40 &  \\
		Lulin Observatory $ \unit[1]{m} $ Telescope (Lul) & 9.30 &  &  &  \\
		Moletai $ \unit[1.6]{m} $ Telescope (Mol) & 13.47 &  & 2.95 &  \\
		MONET/North Telescope $ \unit[1.2]{m} $ (M/N) & 138.90 & 41.29 & 34.24 &  \\
		Mt. Bigelow Kuiper Telescope $ \unit[1.5]{m} $ (MtB) & 440.85 &  &  &  \\
		Nordic Optical Telescope $ \unit[2.5]{m} $ (NOT) & &  & 3.86 &  \\
		SARA-KP $ \unit[0.9]{m} $ telescope  & 66.26 &  & 1.80 &  \\
		Serra la Nave $ \unit[0.9]{m} $ &  &  & 26.40 &  \\
		South African Astronomical Observatory $ \unit[1]{m} $ (SAAO) &  & 64.04 & 36.93 & 166.31 \\
		Steward Observatory Bok Telescope $ \unit[2.2]{m} $ (StB) & 12.00 &  &  &  \\
		Telescopio Nazionale Galileo $ \unit[3.6]{m} $ (TNG) &  & 7.00 & 3.37 &  \\
		T\"ubingen $ \unit[0.8]{m} $ Telescope (Tue) & 23.05 &  &  &  \\
		Whole Earth Telescope (WET) & & & 40.00 & \\
		Wise $ \unit[1]{m} $ &  &  & 9.00 &  \\
		\midrule
		$ \Sigma $ & 998.21 & 187.80 & 568.25 & 200.67 \\
		\bottomrule
	\end{tabular}
	\begin{tablenotes}
		\textbf{Notes.}Detailed Tables, including observing dates and times per observatory is available online at the CDS, as well as Tables listing the allocation into the epochs. Observations at Baker Observatory, Mt. Bigelow Kuiper Telescope, Nordic Optical Telescope, Steward Observatory Bok Telescope were initially collected for other project(s) but also used for this work.
	\end{tablenotes}
\end{threeparttable}
\end{sidewaystable}

\section{DW Lyn}
\citet{dreizler_hs0702+6043_2002} identified DW Lyn (HS~0702+6043) as a $ p $-mode pulsator. \citet{schuh_hs_2006} discovered additional $ g $-mode pulsations making this star the prototype of hybrid sdB pulsators.

There are photometric data available from 1999. Large gaps make a consistent $ O-C $ analysis difficult. Regular monitoring within the \textit{EXOTIME} programme ran from 2007 until the beginning of 2010. Further observations cover a period up to the end of 2010. These multi-site observations are described in \citet{lutz_light_2008,lutz_long-term_2008,lutz_exotime_2011}. Here, we add observations made with the SARA-KP $ \unit[0.9]{m} $ telescope at Kitt Peak National Observatory in Arizona
, that used exposure times of $ \unit[30]{s} $.

\section{V1636 Ori}\label{sec:obs-v1636ori}
\citet{ostensen_four_2001} discovered V1636 Ori (HS~0444+0458) as a pulsating sdB star. \citet{reed_resolving_2007} conducted a frequency analysis, reporting one small and two large amplitude $ p $-modes.

V1636 Ori was observed between August 2008 and January 2015 for the EXOTIME project. About a third of the data was obtained using the $ \unit[1]{m} $ South African Astronomical Observatory (SAAO) with the UCT and STE3 CCD instruments.
Observations at the $ \unit[1.2]{m} $ MONET/North telescope, equipped with an Apogee 1k$ \times $1k E2V CCD camera, were taken in 2x2 binning, using $ \unit[20]{s} $ exposure times.
Observations at the $ \unit[2.2]{m} $ Calar Alto Observatory (CAHA) used the CAFOS instrument with $ \unit[10]{s} $ exposure time.
Two nights were obtained at the $ \unit[1.5]{m} $ telescope at Loiano observatory, using the BFOSC (Bologna Faint Object Spectrograph \& Camera) instrument and $ \unit[15]{s} $ exposure times.
Between October 2008 and December 2009, observations at the $ \unit[1]{m} $ Mt. Lemmon Optical Astronomy Observatory (LOAO) were conducted with a 2k$ \times $2k CCD camera with exposure times of $ \unit[12]{s} $ and $ \unit[20]{s} $.
The observations at the $ \unit[3.6]{m} $ Telescopio Nazionale \textit{Galileo} (TNG) in August 2008 and 2010 were performed with the DOLORES instrument and $ \unit[5]{s} $ exposure times.

\section{QQ Vir}
The discovery of QQ Vir (PG~1325+101) as a multi-period pulsator was reported in \citet{silvotti_pg_2002}, followed by a frequency analysis and asteroseismological modelling by \citet{silvotti_rapidly_2006} and \citet{charpinet_rapidly_2006}, respectively.

Observations of QQ Vir in 2001 and 2003 are described in \citet{silvotti_pg_2002} and \citet{silvotti_rapidly_2006}, respectively. Between March 2008 and April 2010, the object was observed as part of the EXOTIME project \citep{benatti_exotime_2010}.
Additionally, one observation run in February 2005 was performed at the $ \unit[1.5]{m} $ telescope at Loiano observatory, using the BFOSC instrument.
Most of the observations were obtained in 2009, 2010, 2011 and 2012 at the LOAO, using an exposure time of $ \unit[10]{s} $.
The CAHA and MONET/North observations were conducted with $ \unit[10]{s} $ and $ \unit[20]{s} $ exposure times, respectively.
The Loiano observatory performed additional observations in 2009, 2010 and 2011 with the $ \unit[1.5]{m} $ telescope, using BFOSC and an exposure times of $ \unit[12]{s} $, $ \unit[15]{s} $ and $ \unit[20]{s} $.
Observations at the Mol\.{e}tai Astronomical Observatory (Mol) in 2008 were performed using the $ \unit[1.6]{m} $ telescope and an Apogee 1k$ \times $1k E2V CCD camera using $ \unit[17.5]{s} $ of exposure time.
Observations at the SAAO used the same instrumental set up as described in section~\ref{sec:obs-v1636ori}.
The TNG observed in 2010 and 2011.
A DARC-WET campaign on QQ Vir was performed in May 2010.

\section{V541 Hya}
V541 Hya (EC~09582-1137) was discovered by \citet{kilkenny_three_2006}. \citet{randall_observations_2009} conducted an asteroseismological analysis of this target.

Between 2005 and 2015, a large number of observations were obtained at the SAAO, using the same instrumentation noted in Sect.~\ref{sec:obs-v1636ori} and exposure times of $ \unit[10]{s} $. 
The LOAO conducted observations in 2009, 2012 and 2013 with exposure times of $ \unit[20]{s} $.
During March 2009 and February and March 2010, V541 Hya was observed at the CAHA, using an exposure time of $ \unit[10]{s} $.

\section{TESS observations}
The primary goal of the NASA Transiting Exoplanet Survey Satellite (TESS) space telescope is to detect exoplanets transiting bright nearby stars \citep{ricker_transiting_2015}. However, the extensive time series photometry is valuable for asteroseimology and the TESS Asteroseismic Science Consortium (TASC) coordinates short cadence observations of pulsating evolved stars. TESS observed V1636 Ori and V541 Hya with a cadence of $ \unit[120]{s} $ between November 15, 2018 and December 11, 2018, and February 2, 2019 and February 27, 2019, respectively. We used the light curves provided by the MAST archive\footnote{\url{https://archive.stsci.edu/}} that had common instrumental trends removed by the Pre-Search Data Conditioning Pipeline \citep[PDC, ][]{stumpe_kepler_2012}. Light curves and amplitude spectra are presented in Fig.~\ref{fig:lc-tess} and \ref{fig:tess-power}. The two minutes (\enquote{short}) cadence undersamples the $ p $-modes at about $ \unit[140]{s} $. In combination with the large photometric scatter, the amplitude spectra show no evidence of the $ p $-modes. Thus, we do not make use of the TESS observations in our study.

\section{Data reduction}
For the EXOTIME observations, the data reduction was carried out using the IDL software TRIPP \citep[Time Resolved Imaging Photometry Package, see][]{schuh_ccd_2000}. TRIPP performs bias-, dark-, flat-field corrections, and differential aperture photometry to calculate the relative flux of a target with respect to one or more comparison stars and extinction-corrections (second order polynomial in time). In the presence of sub-stellar companions, we might expect variations in the arrival times of stellar pulsations on the order of seconds to tens of seconds. The corresponding uncertainties are expected to be about one second. These uncertainties rise from observational constrains, such as smearing and sampling effects due to the integration time. The accuracy of individual time-stamps is better than $ \pm \unit[0.5]{s} $.
All time stamps were converted from GJD(UTC) to BJD(TDB), according to \citet{eastman_achieving_2010}, with an accuracy well below the expected observational uncertainty.

Typical S/N for our ground-based observations range from 60 for large amplitude pulsations, to 3 for the smallest pulsation-amplitudes we investigate in this work. The amplitude spectra in Sect.~\ref{sec:results} show also pulsations with smaller S/N, but these are not suitable for timing analysis because the uncertainties are too large (see Table~\ref{tab:add-freq}).
\chapter{Analysis} \label{sec:analysis}

In order to detect variations in the arrival time of stellar pulsations, we developed a pipeline to process the reduced data. A schematic flowchart of our pipeline is presented in Fig.~\ref{fig:flowchart}. The input consists of the light curve (time series) and the dates of the observational epochs. In a light curve, typically spanning several years at a very low duty cycle of 0.2 to 1.7 per cent, an epoch consists of a few roughly consecutive nights of observation.

\tikzset{
	ultra thick/.style={line width=3pt}
}

\tikzstyle{input} = [rectangle, thick, minimum width=0.1cm, minimum height=0.1cm, text centered, text width=0.2\textwidth, draw=cb-green, fill=cb-green!20]
\tikzstyle{process} = [rectangle, rounded corners, thick, minimum width=0.1cm, minimum height=0.1cm, text centered, text width=0.25\textwidth, draw=cb-blue, fill=cb-blue!20]
\tikzstyle{decision} = [diamond,  aspect=4, rounded corners, thick, minimum width=0.2\textwidth, minimum height=1cm, text centered, draw=cb-orange, fill=cb-orange!20]
\tikzstyle{arrow} = [thick,->,>=stealth]

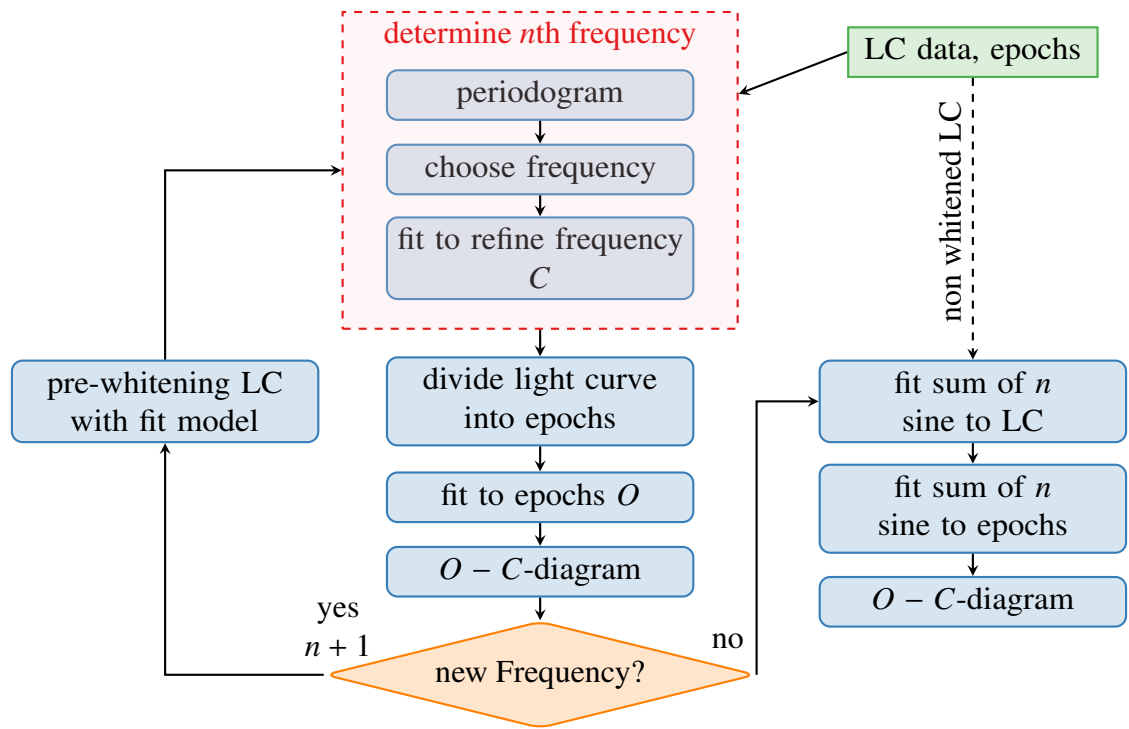
\begin{figure*}[t]
	\centering
	\begin{tikzpicture}[align=center, node distance=5ex]
	\node (input) [input] {LC data, epochs};
	
	\node (entropy) [process, left of=input, xshift=-25ex, yshift=-3ex] {periodogram};
	
	\node (choose) [process, below of=entropy, yshift=-0.2ex] {choose frequency};
	
	\node (refine) [process, below of=choose, yshift=-1.2ex] {fit to refine frequency\\ $ C $};
	
	\node (period) [draw=cb-red, yshift=1ex, fit= (entropy) (choose) (refine), inner sep=3ex, dashed, thick,fill=cb-red!20, fill opacity=0.2] {};
	
	\node [yshift=-1.75ex, cb-red] at (period.north) {determine $n$th frequency};
	\node (chunks) [process, below of = refine, yshift=-5ex] {divide light curve into epochs};
	
	\node (phaseshift) [process, below of = chunks, yshift=-1.7ex] {fit to epochs $ O $};
	
	\node (oc) [process, below of = phaseshift, yshift=-0.2ex] {$ O-C $-diagram};
	
	\node (repeat) [decision, below of = oc, yshift=-2.2ex] {new Frequency?};
	
	\node (whitening) [process, left of = chunks, xshift=-21ex] {pre-whitening LC with fit model};
	
	\node (multiple-all) [process, right of = chunks, xshift=25ex] {fit sum of $n$ sine to LC};
	\node (multiple-fixed) [process, below of = multiple-all, yshift=-2.5ex] {fit sum of $n$ sine to epochs};
	\node (oc-multi) [process, below of = multiple-fixed, yshift=-1.5ex] {$ O-C $-diagram};
	
	
	\draw [arrow] (input.west) -- (period);
	\draw [arrow] (entropy) -- (choose);
	\draw [arrow] (choose) -- (refine);
	\draw [arrow] (period) -- (chunks);
	\draw [arrow] (chunks) -- (phaseshift);
	\draw [arrow] (phaseshift) -- (oc);
	\draw [arrow] (oc) -- (repeat);
	\draw [arrow] (repeat.west) -- node[anchor=south, xshift=1ex, yshift=0.5ex] {yes\\$n+1$} ++(0,0) -| (whitening.south) ;
	\draw [arrow] (whitening.north) -- ++(0,0) |- (period.180) ;
	\draw [arrow] (repeat.east) -- node[anchor=south, xshift=-2ex, yshift=1ex] {no} ++(0,0) |- (multiple-all.west);
	\draw [arrow] (multiple-all) -- (multiple-fixed);
	\draw [arrow] (multiple-fixed) -- (oc-multi);
	\draw [arrow, dashed] (input.south) -- node[anchor=south, rotate=90, xshift=0.5ex] {non whitened LC} (multiple-all.north);
	
	\end{tikzpicture}
	\caption[Flow chart representing the analysis of the time of arrival]{Flow chart representing the analysis of the time of arrival. Light curve (LC) and start/ end time of each observational epoch are provided as input. Each frequency is analysed, leading to an intermediate $ O-C $-diagram, and subtracted from the LC by itself before the sum of all sinusoidal functions is fitted simultaneously to the LC, resulting in the final $ O-C $-diagram.} 
	\label{fig:flowchart}
\end{figure*}

\paragraph{Outlier removal:}In case no uncertainty in the flux measurement $ F $ is provided, the root-mean-square of each observation is used as an approximate photometric error for the later analysis. We have used a running median filter to exclude $ \unit[5]{\sigma} $ flux-outliers. The length of the window size depends on the cadence of the observations. We constrained it to be not longer than half of the period of the main frequency.
The analysis is performed for each frequency individually before all frequencies were analysed simultaneously.

\paragraph{Full data fit:}For the individual fitting of pulsations, we first determined the frequency of the main signal. For this, we used the \verb|astropy| package to calculate the Lomb-Scargle periodogram \citep{astropy_collaboration_astropy_2013,astropy_collaboration_astropy_2018}. From this periodogram, we selected the frequency with the largest amplitude to continue. In the next step, we performed the fit of a sinusoidal function to the light curve, using
\begin{align}
	F(t) = A \sin \left( f t + \phi \right) + o
\end{align}
with amplitude $ A $, frequency $ f $, phase $ \phi $ and offset $ o $. The minimization problem is solved using the \textit{scipy} implementation of the Trust Region Reflective algorithm \citep{jones_scipy_2011}. The selected frequency from the amplitude spectrum serves as initial value, the full-width-at-half-maximum of the corresponding peak in the periodogram is used as a 	boundary. The amplitude-guess is taken from the amplitude spectrum. In case of highly varying amplitudes, the initial value can be set manually. The fitting routine returns the parameters and their variance. 

\paragraph{Epoch fit:}Frequency and offset are all kept fixed for the following analysis of the individual observational epochs. The starting value of the current phase fit is determined by the average of the previous $ j $ phase values (or the global fit value from above in case there are no $ j $ previous values yet) in order to keep the fitting process stable and avoid \enquote{phase-jumps}. For our target sample, a value of $ j = 3 $ has proven to be reasonable, except when observational gaps span over several years. 

The uncertainty in the phase measurement scales inverse with the length of the epochs. Thus, this length is chosen in a way to minimize the uncertainties of the fit but at the same time keep the epochs as short as possible to maximize the temporal resolution of the final $ O-C $ diagram. Often, the observations themselves constrain the length of the epochs (e.g. three consecutive nights of observations and a gap of several weeks before the next block of observations). If possible, we aim for an epoch length such that the timing uncertainties are of the order of one second. The phase information of the global and the epoch fit result in a intermediate $ O-C $ diagram. 

As a last step in the single-frequency-analysis, the fitted model is subtracted from the light curve. We noticed significant amplitude variations for some of our targets. Thus we subtract the model using the amplitude of the individual epochs. This pre-whitening procedure is repeated for every relevant pulsation in the data.

\paragraph{Multi frequency fit:}Close frequencies are likely to introduce artificial trends in the arrival times in such a step-by-step analysis. Thus, the sum of all sinusoidal functions,
\begin{align}
	F(t) = A_n \sin \left( f_n t + \phi_n \right) + o
\end{align}
is fitted to the non-whitened light curve, where $ n $ is the number of investigated frequencies. The previously retrieved values for amplitude, frequency, phase and offset are used as initial values. 
We are using the phase information $ \phi_n $ of the light curve as reference phase, namely calculated phase $ C $ in the final $ O-C $ diagram.
Similar to the single-frequency analysis, the observational epochs are now fitted individually using the sum of sinusoidal functions to yield the observed phase information $ O $.

\begin{figure}[htbp!]
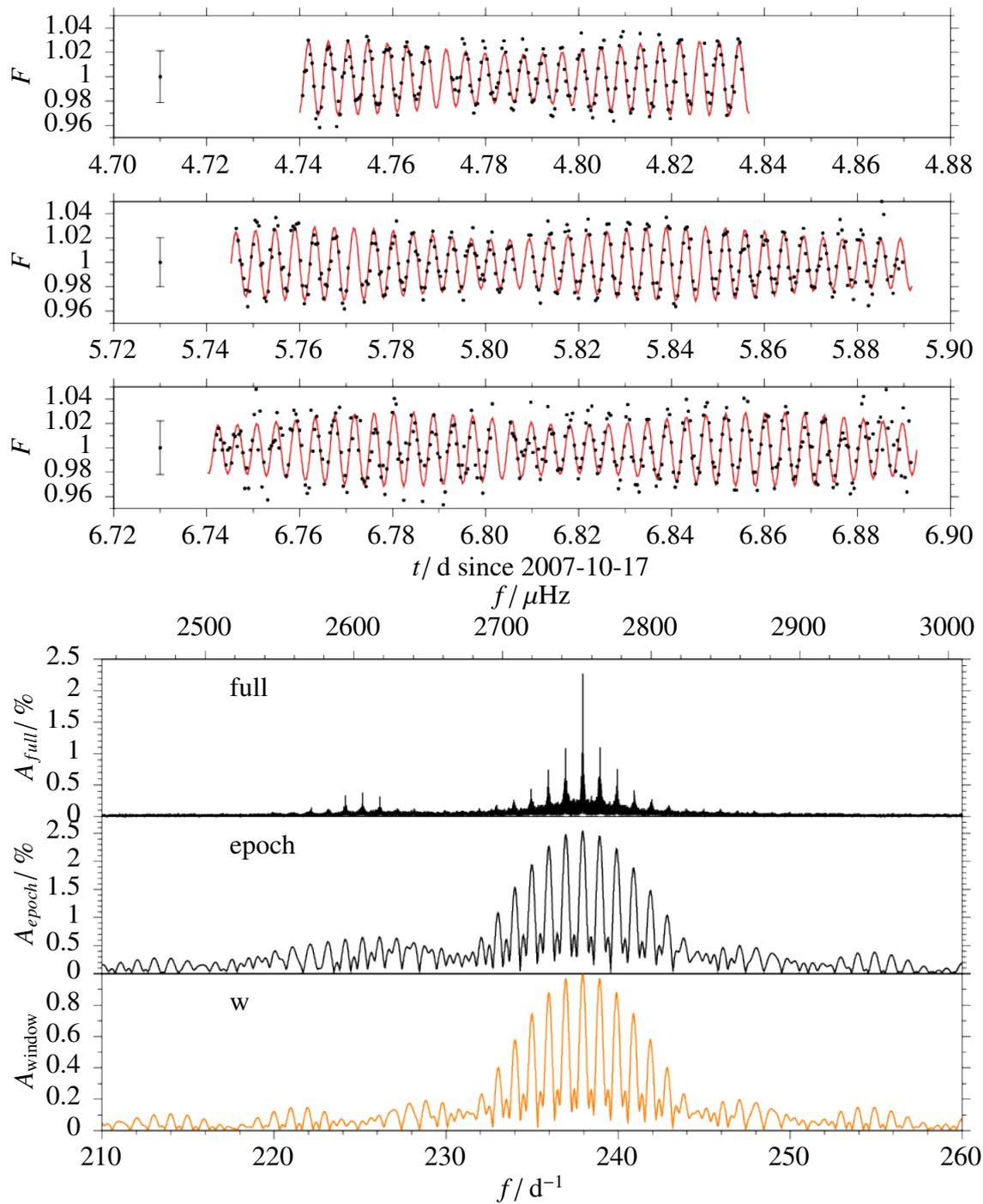

	\centering
	\begin{subfigure}[b]{\textwidth}
		\centering
		\input{figures/exotime/dwlyn/epoch/lc.tex}
	\end{subfigure}
	\begin{subfigure}[b]{\textwidth}
		\centering
		\input{figures/exotime/dwlyn/epoch/window.tex}
	\end{subfigure}
	\caption[Example observations of DW~Lyn.]{Example observations of DW~Lyn from October 21.,22. and 23., 2010 at the Lulin observatory (\textit{top}, from top to bottom), combined used as one $ O-C $ measurement. The errorbar on the left of each plot represents the photometric uncertainty. The red line shows the simultaneous two frequency fit to the epoch data. The amplitude spectra in the lower panel show the spectrum of the full data set (\textit{top}), the spectrum of this epoch (\textit{middle}), and the respective window function computed at $ f_1 $ (\textit{bottom}).}
	\label{fig:dwlyn-lc-crop}
\end{figure}

The results of the simultaneous fit for each target in this paper are summarised in Table~\ref{tab:fit}. We list pulsation modes not used for the timing analysis in Table~\ref{tab:add-freq}. Figures~\ref{fig:dwlyn-lc-crop} -- \ref{fig:qqvir-v541hya-epoch-power} show example light curves of the targets for one epoch each, including their multi frequency fit and the respective amplitude spectrum.

\begin{table*}[t]
\centering
	\caption[Parameters of the simultaneously fitted pulsations per target.]{Parameters of the simultaneously fitted pulsations per target over their full observational time span and the pulsation period $ P $.}
	\label{tab:fit}
	\begin{threeparttable}[htbp!]
	\begin{tabular}{llr@{.}lr@{.}lr@{.}lr@{.}l}
		\toprule
		Target & & \multicolumn{2}{c}{$ \unit[f/]{d^{-1}} $} & \multicolumn{2}{c}{$ \unit[P/]{s} $} & \multicolumn{2}{c}{$ \unit[A/]{\%} $} & \multicolumn{2}{c}{$ \unit[\phi/]{d} $} \\
		\midrule
		DW Lyn& $ f_1 $ & 237 & 941160\,(8) & {363} & {114982(12)} & 2 & 19\,(9) & 54394 & 741\,(5) \\
		& $ f_2 $ & 225 & 15898\,(5) & {383} & {72887(9)} & 0 & 35\,(9) & 54394 & 74\,(3) \\
		V1636 Ori & $ f_{1} $ & 631 & 7346\,(2) & {136} & {76629(5)} & 0 & 54\,(3) & 54698 & 72\,(3) \\
		& $ f_2 $ & 509 & 9780\,(3) & {169} & {4191(1)} & 0 & 24\,(3) & 54698 & 72\,(7) \\
		QQ Vir & $ f_{1} $ & 626 & 877627\,(3) & {137} & {8259429(7)} & {2} & {6\,(1)} & 52117 & 924\,(6) \\
		 & $ f_{2} $ & 552 & 00714\,(9) & {156} & {51971(3)} & {0} & {10\,(9)} & 52117 & 9\,(1) \\
		 & $ f_{3} $ & 642 & 0515\,(1) & {134} & {56864(3)} & {0} & {07\,(1)} & 52117 & 9\,(2) \\
		V541 Hya & $ f_{1} $ & 635 & 32218\,(5) & {135} & {993993(11)} & {0} & {31\,(8)} & 53413 & 88\,(3) \\
		& $ f_2 $ & 571 & 28556\,(3) & {151} & {237850(8)} & 0 & 21\,(7) &  53413 & 88\,(3) \\
		\bottomrule
	\end{tabular}
	\begin{tablenotes}
		\textbf{Notes.} The phase $ \phi $ refers to the time corresponding to the first zero-crossing of the function after the first measurement $ t_0 $ in $ \unit{MBJD} $.
	\end{tablenotes}
\end{threeparttable}
\end{table*}

\begin{figure}[htbp!]
	\centering
	\begin{subfigure}[b]{\textwidth}
		\centering
		\input{figures/exotime/v1636ori/epoch/lc.tex}
	\end{subfigure}
	\begin{subfigure}[b]{\textwidth}
		\centering
		\input{figures/exotime/v1636ori/epoch/window.tex}
	\end{subfigure}	
	\caption[Example observations of V1636~Ori.]{Example observations of V1636~Ori from March 20., 21. and 22., 2009 at the CAHA (\textit{top}, from top to bottom), combined used as one $ O-C $ measurement. The errorbar on the left of each plot represents the photometric uncertainty. The red line shows the simultaneuos two frequency fit to the epoch data. The amplitude spectra in the lower panel show the spectrum of the full data set (\textit{top}), the spectrum of this epoch (\textit{middle}), and the respective window function computed at $ f_1 $ (\textit{bottom}).}
	\label{fig:v1636-lc-crop}
\end{figure}

\begin{figure}[htbp!]
	\centering
	\begin{subfigure}[b]{\textwidth}
		\centering
		\input{figures/exotime/qqvir/epoch/lc.tex}
	\end{subfigure}
	\begin{subfigure}[b]{\textwidth}
		\centering
		\input{figures/exotime/qqvir/epoch/window.tex}
	\end{subfigure}	
	\caption[Example observations of QQ Vir.]{Example observations of QQ Vir from February 20., 21. and 22., 2012 at the Monet telescope (\textit{top}, from top to bottom), combined used as one $ O-C $ measurement. The errorbar on the left of each plot represents the photometric uncertainty. The red line shows the simultaneuos two frequency fit to the epoch data. The amplitude spectra in the lower panel show the spectrum of the full data set (\textit{top}), the spectrum of this epoch (\textit{middle}), and the respective window function computed at $ f_1 $ (\textit{bottom}).}
	\label{fig:qqvir-v541hya-lc-crop}
\end{figure}

\begin{figure}[htbp!]
	\centering
	\begin{subfigure}[b]{\textwidth}
		\centering
		\input{figures/exotime/v541hya/epoch/lc.tex}
	\end{subfigure}
	\begin{subfigure}[b]{\textwidth}
		\centering
		\input{figures/exotime/v541hya/epoch/window.tex}
	\end{subfigure}	
	\caption[Example observations of V541~Hya.]{Example observations of V541~Hya from February 6., 7. and 12., 2008 at the SAAO (\textit{top}, from top to bottom), combined used as one $ O-C $ measurement. The errorbar on the left of each plot represents the photometric uncertainty. The red line shows the simultaneuos two frequency fit to the epoch data. The amplitude spectra in the lower panel show the spectrum of the full data set (\textit{top}), the spectrum of this epoch (\textit{middle}) and the respective window function computed at $ f_1 $ (\textit{bottom}).}
	\label{fig:qqvir-v541hya-epoch-power}
\end{figure}

\chapter{Results and discussion} \label{sec:results}
In the following, we discuss the implications of the obtained amplitude spectra and $ O-C $ measurements on the evolutionary state and presence of sub-stellar companions to the targets.

\section{DW Lyn} \label{sec:results-dwlyn}
\begin{figure}[bp!]
	\centering
	\input{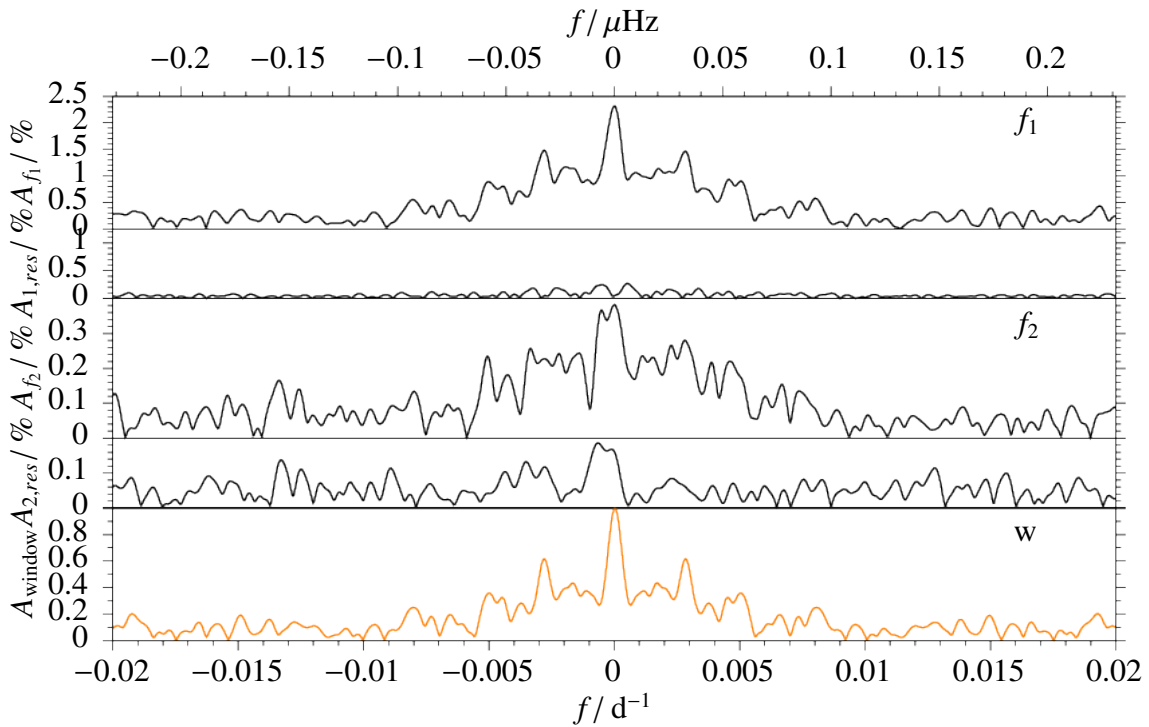}
	\caption[Amplitude spectrum of DW Lyn.]{Amplitude spectrum of DW Lyn of the main pulsation frequency $ f_1 =  \unit[237.941160]{d^{-1}} $ (\textit{top}), $ f_2 =  \unit[225.15898]{d^{-1}} $ (\textit{middle}) with the respective residuals after the pre-whitening below, and the normalised window-function (\textit{bottom}).}
	\label{fig:dwlyn-window}
\end{figure}

The amplitude spectrum of DW Lyn in Fig.~\ref{fig:dwlyn-power} reveals two strong pulsation modes at $ f_1 = \unit[237.941160]{d^{-1}} $ and $ f_2 = \unit[225.15898]{d^{-1}} $. A closer look to the amplitude spectrum in Fig.~\ref{fig:dwlyn-window} reveals small asymmetries compared to the window function. The pre-whitening of both frequencies leaves residuals well above noise level in the amplitude spectrum, indicating unresolved multiplets or mode splitting, especially for $ f_2 $. 

\begin{figure*}[tp!]
	\centering
	\input{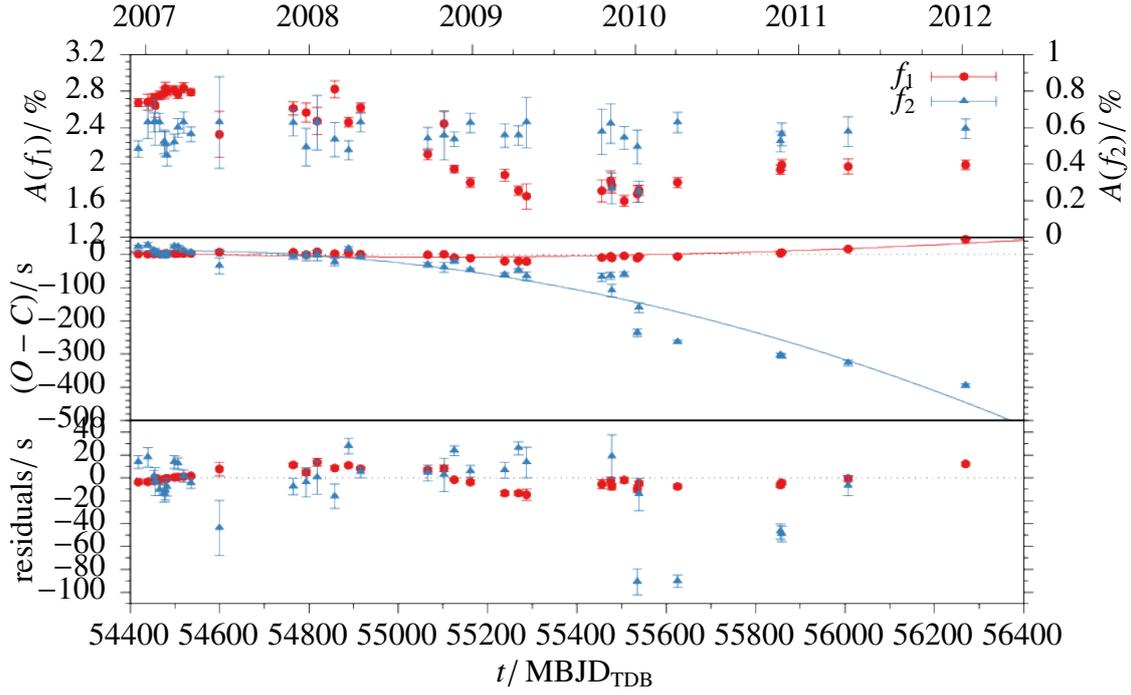}
	\caption[Results for the two main pulsations of DW Lyn.]{Results for the two main pulsations of DW Lyn. \textit{Top panel:} Amplitudes. \textit{Middle panel:} Fits of the $ O-C $ data with second order polynomials in time. \textit{Lower panel:} Residuals.}
	\label{fig:dwlyn-oc}
\end{figure*}

The S/N of modes at higher frequencies, for example, at about $ \unit[320]{d^{-1}} $ and $ \unit[480]{d^{-1}} $, are too small for a stable $ O-C $ analysis (see Table~\ref{tab:add-freq}). 
Therefore, the $ O-C $ diagram in Fig.~\ref{fig:dwlyn-oc} shows the analysis of the two main pulsation modes, with the time-dependent variation of the pulsation amplitudes. 

In order to determine evolutionary times-cales of the pulsations, we investigate the long-term evolution in the $ O-C $ data. A constant change in period results in a second-order term as a function of time \citep{sterken_o-c_2005}, which allows us to derive a value for the secular change of the period $ \dot{P} $, and hence the evolutionary time-scale. Results of the fits of the second order polynomial are included in Fig.~\ref{fig:dwlyn-oc}, which are $ \dot{P}/P_{f1} = \left( 5.8 \pm 0.2 \right) \times \unit[10^{-5}]{d^{-1}} $ and $ \dot{P}/P_{f2} = \left( -29.3 \pm 0.8 \right) \times \unit[10^{-5}]{d^{-1}} $. 
Assuming $ \dot{P} $ is based on stellar evolution, stellar model calculations show that the sign of the rate of period change indicates the phase of the sdB after the zero-age extreme horizontal branch \citep[ZAEHB;][]{charpinet_adiabatic_2002}. For $ p $-modes, a positive $ \dot{P} $ relates to the first evolutionary phase of the ZAEHB, in which the surface gravity decreases due to \ch{He} burning in the core. A negative $ \dot{P} $ would correspond to the second evolutionary phase, in which the sdB contracts because the depletion of \ch{He} in its core, and this happens before the post-EHB evolution. The turning point between these two states occurs between 87 and 91 Myr after the ZAEHB. According to our measurement of a positive $ \dot{P} $ for $ f_1 $, DW Lyn would still be in its first evolutionary phase. With the lack of a mode identification from an asteroseismic model for DW~Lyn, we can not directly compare the measured $ \dot{P} $ with theoretical predictions from \citet{charpinet_adiabatic_2002}. 
However, stellar models with pulsation periods of around $ \unit[360]{s} $ show values for $ \dot{P} $ with a comparable order of magnitude as our measurement $ \dot{P} = \left( 4.3 \pm 0.15 \right) \times \unit[10^{-1}]{s \, Myr^{-1}} $, for example, $ \dot{P} = 1.62 \unit{s \, Myr^{-1}} $ for a model with a mode of $ l=0,k=0 $ at the age of $ \unit[67.83]{Myr} $ \citep[appendix C]{charpinet_adiabatic_2002}.
The large $ \dot{P} $ of $ f_2 $ is consistent with the apparent mode splitting seen in the amplitude spectra in Fig.~\ref{fig:dwlyn-window}, and thus does not reflect the evolutionary phase of DW Lyn.

After subtracting the long term-trend,
small time scale features are evident. For example, the $ O-C $ data for $ f_2 $ show an oscillating behaviour with a significance of $ \unit[3]{\sigma} $ within the first 200 days, while the arrival times for $ f_1 $ remain constant during the same period of time. In later epochs, the $ O-C $ data for both frequencies agree mostly within $ \unit[2]{\sigma} $. During the second half of the observations, the phase of $ f_2 $ jumps by about $ \unit[100]{s} $. This behavior lacks an explanation.

Additionally, the evolution of the pulsation-amplitudes in Fig.~\ref{fig:dwlyn-oc} shows a comparable oscillating behaviour for $ f_2 $ within the first epochs similar to the change in arrival times. Although the periodic variations in amplitude are not as significant as for the phase, the occurrence of simultaneous phase- and amplitude-modulations indicate a mode-beating of two close, unresolved frequencies. The residuals in the amplitude spectrum support this explanation. In later observations, the amplitude remains almost constant within the uncertainties. The beating mode might lose energy or shift frequency over time.
The amplitude for the $ f_1 $ pulsation drops by about 1 per cent (amplitude) or about 35 per cent (relative) to the second half of the observation campaign with a similar quasi-periodic variation as the phase. The residuals in the amplitude spectrum show no indication of an unresolved frequency leading to mode-beating. Besides stochastically driven pulsation modes, \citet{kilkenny_amplitude_2010} suggest energy transfer between modes as possible explanation for amplitude variations. For both frequencies, a possible interaction between amplitude and phase of pulsations is not well understood.

\section{V1636 Ori} 
\begin{figure}[tp!]
	\centering
	\input{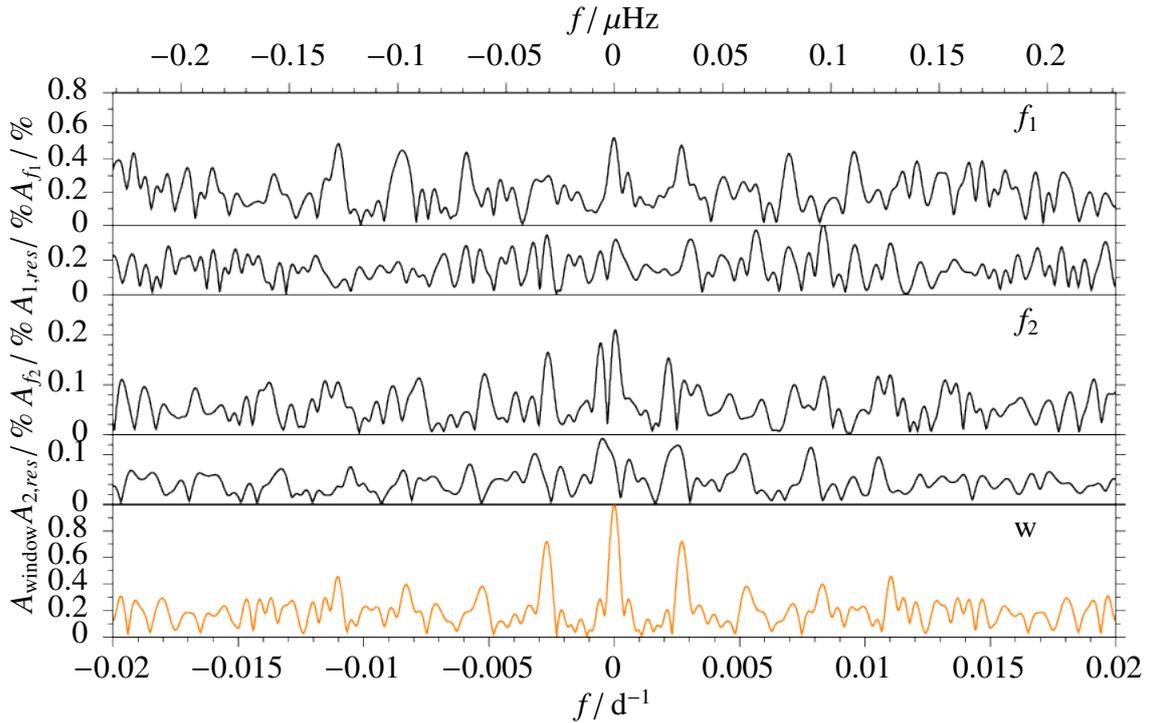}
	\caption[Amplitude spectrum of V1636 Ori.]{Amplitude spectrum of V1636 Ori of the main pulsation frequency $ f_1 =  \unit[631.7346]{d^{-1}} $ (\textit{top}), $ f_2 =  \unit[509.9780]{d^{-1}} $ (\textit{middle}) with the respective residuals after the pre-whitening below and the normalised window-function (\textit{bottom}).}
	\label{fig:v1636ori-window}
\end{figure}

The amplitude spectrum of V1636 Ori in Fig.~\ref{fig:v1636ori-power} shows two main pulsation modes with frequencies at $ f_1 = \unit[631.7346]{d^{-1}} $ and $ f_2 =  \unit[509.9780]{d^{-1}} $. The S/N is not sufficient to use a third pulsation mode at $ \unit[566.2]{d^{-1}}$ ($ \unit[6553]{\mu Hz} $, \citealt{reed_resolving_2007}). The amplitude spectrum of TESS data in Fig.~\ref{fig:tess-power} shows no evidence for $ g $-mode pulsations with amplitudes greater than 0.4 per cent.
A detailed look at the spectra of the two main frequencies in Fig.~\ref{fig:v1636ori-window} shows mode splitting, likely due to a change in frequency over the long time of observation.

\begin{figure*}[tp!]
	\centering
	\input{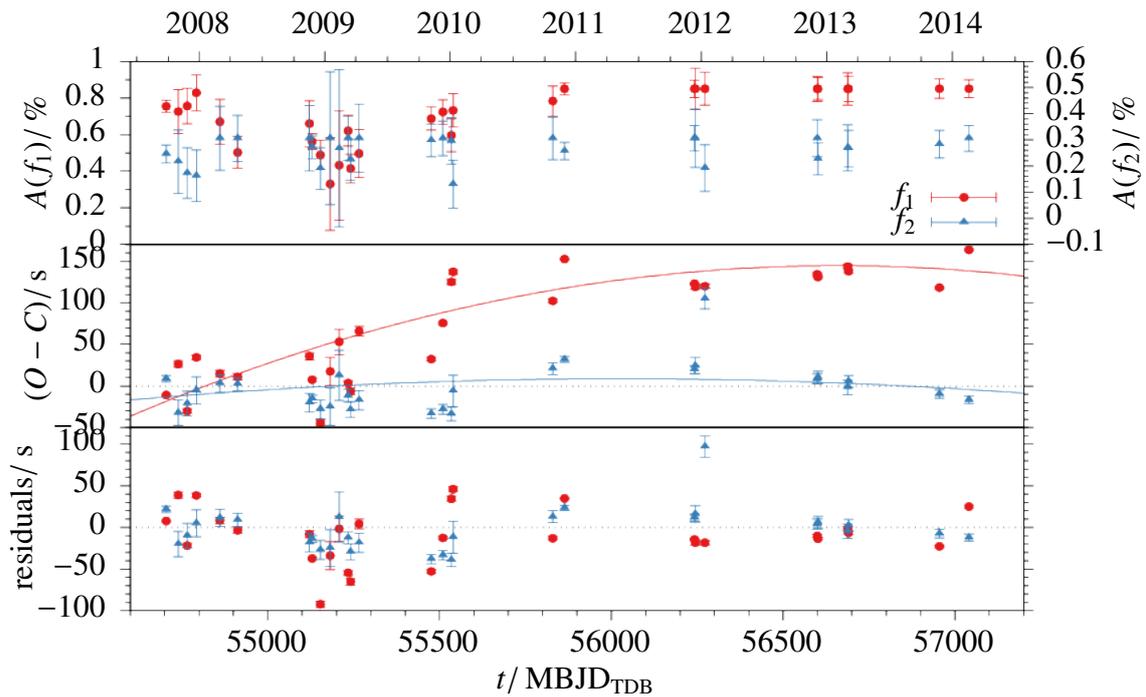}
	\caption[Results for the two main pulsations of V1636 Ori.]{Results for the two main pulsations of V1636 Ori. \textit{Top panel:} Amplitudes. \textit{Middle panel:} Fits of the $ O-C $ data with second order polynomials in time. \textit{Lower panel:} Residuals.}
	\label{fig:v1636ori-oc}
\end{figure*}

The $ O-C $ diagram in Fig.~\ref{fig:v1636ori-oc} shows the two main pulsation modes and the variation of the pulsation amplitudes.

From the second order fit in time, we derive the changes in period $ \dot{P}/P_{f1} = \left( -8.54 \pm \right. $  $ \left. 0.14 \right) \times \unit[10^{-5}]{d^{-1}} $ and $ \dot{P}/P_{f2} = \left( -2.5 \pm 0.5 \right)\times \unit[10^{-5}]{d^{-1}} $. We caution the interpretation of these values as evolutionary time scales since the  apparent mode splitting seen in Fig.~\ref{fig:v1636ori-window} could explain these trends as well.

The residuals after subtracting the long term trend show a large variation. They change by up to about $ \unit[\pm50]{s} $ for $ f_1 $ ($ \sim \unit[14]{\sigma} $ significance) and up to about $ \unit[\pm30]{s} $ for $ f_2 $ ($ \sim \unit[3]{\sigma} $ significance). The amplitude for $ f_1 $ drops by about 0.25 per cent (amplitude) or about 33 per cent (relative)  in the time between $ \unit{MBJD} = \unit[55100]{d} \text{ and } \unit[55300]{d}$, and returns to its previous level afterwards, while the amplitude for $ f_2 $ remains constant within the uncertainties. This decrease in amplitude coincides with earlier arrival times in the $ O-C $ diagram. As already discussed in the previous section, a possible amplitude- and phase-interaction is not well understood. The $ f_1 $ pulsation mode may not be coherent on such long time-scales but of a short-term stochastic nature not resolvable by our data set \citep[e.g.  KIC~2991276,][]{ostensen_stochastic_2014}.

\section{QQ Vir}
\begin{figure}[tp!]
	\centering
	\input{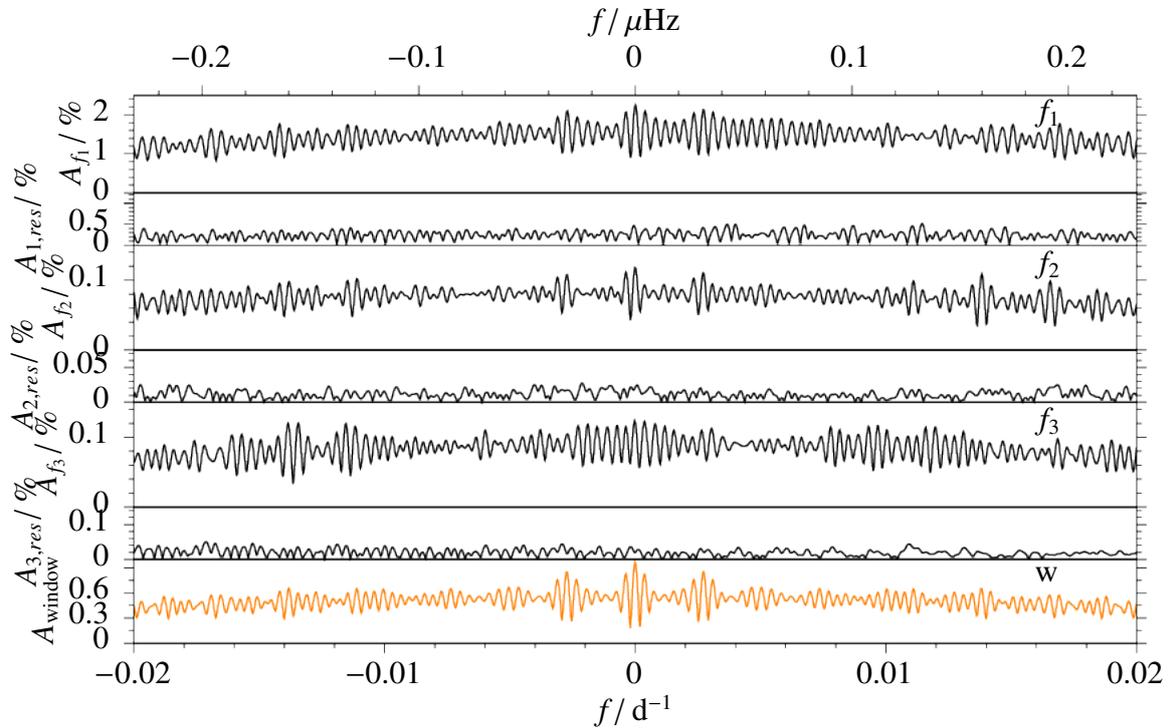}
	\caption[Amplitude spectrum of QQ Vir.]{Amplitude spectrum of QQ Vir of the main pulsation frequency $ f_1 =  \unit[626.877628]{d^{-1}} $ (\textit{top}), $ f_2 =  \unit[552.00713]{d^{-1}} $ (\textit{top middle}),  $ f_3 =  \unit[642.0516]{d^{-1}} $ (\textit{bottom middle}) with the respective residuals after the pre-whitening below and the normalised window-function (\textit{bottom}).}
	\label{fig:qqvir-window}
\end{figure}

Fig.~\ref{fig:qqvir-power} shows the amplitude spectrum for the QQ Vir observations. The main frequency at about $ f_1 = \unit[626.877628]{d^{-1}} $ is presented in Fig.~\ref{fig:qqvir-window} in detail and shows asymmetries compared to the window function. After the pre-whitening process, a close frequency at about $ \unit[626.881270]{d^{-1}} $ remains but attempts to model this pulsation fail with uncertainties too large for the timing analysis. There appear two more frequencies suitable for our study. The amplitude spectra around $ f_2 =  \unit[552.00713]{d^{-1}} $  and $ f_3 = \unit[642.0516]{d^{-1}} $ are presented next to $ f_1 $ in Fig.~\ref{fig:qqvir-window}. Another peak at about $ \unit[665]{d^{-1}} $ consists of at least two frequencies at $ \unit[664.488549]{d^{-1}} $ and $ \unit[665.478133]{d^{-1}} $, but they are not sufficiently resolvable within the individual epochs, and lead to too large uncertainties in the $ O-C $ analysis.

\begin{figure*}[tp!]
	\centering
	\input{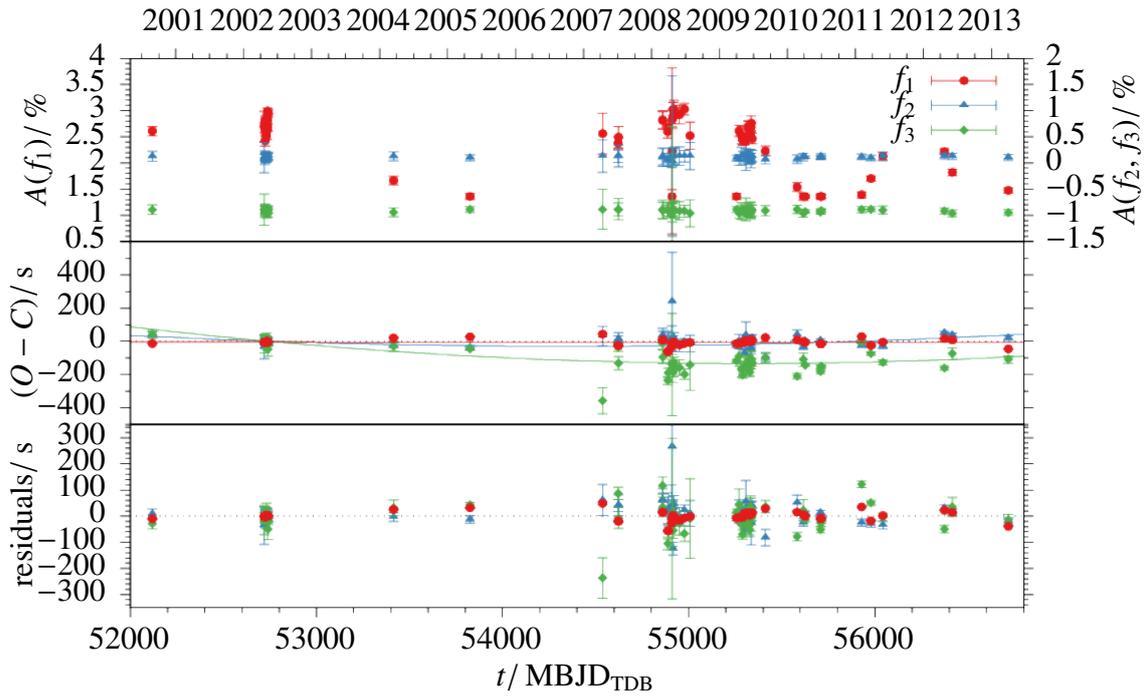}
	\caption[Results for the three main pulsations of QQ Vir.]{Results for the three main pulsations of QQ Vir. \textit{Top panel:} Amplitudes. $ f_3 $ has a vertical offset of -1 for clarity. \textit{Middle panel:} Fits of the $ O-C $ data with second order polynomials in time. \textit{Lower panel:} Residuals.}
	\label{fig:qqvir-oc}
\end{figure*}

Fig.~\ref{fig:qqvir-oc} shows the resulting $ O-C $ diagram and the amplitudes at different epochs. Due to the large observational gap from 2003 to 2008 with only one block of observations in between, we had difficulties avoiding errors in cycle count. In order to avoid a phase jump, we increased the averaging window for initial phase values to $ q = 6 $. With this set up, the changes in pulsation frequencies read as follows: $ \dot{P}/P_{f1} = \left( 1.7 \pm 1.6 \right)\times \unit[10^{-7}]{d^{-1}} $, $ \dot{P}/P_{f2} = \left( 2.4 \pm 0.4 \right)\times \unit[10^{-5}]{d^{-1}} $ and $ \dot{P}/P_{f3} = \left( 4.0 \pm 0.5 \right)\times \unit[10^{-6}]{d^{-1}} $. While $ f_2 $ and $ f_3 $ show no significant variation of pulsation amplitude, $ f_1 $ varies by 1.5 per cent (amplitude) or 50 per cent (relative). Thus, the corresponding phase changes should be interpreted with caution. \citet{charpinet_rapidly_2006} identify the radial order $ k $ and degree $ l $ from asteroseismic modelling to be $ f_1 $: $ l=2,k=2 $; $ f_2 $: $ l=4,k=1 $; $ f_3 $: $ l=3,k=2 $. These combinations do not allow a direct comparison of our $ \dot{P} $ measurements to the model calculations from \citet{charpinet_adiabatic_2002}, but the sign of $ \dot{P} $ indicates QQ~Vir to be in the stage of \ch{He} burning.

\section{V541 Hya}
\begin{figure}[tp!]
	\centering
	\input{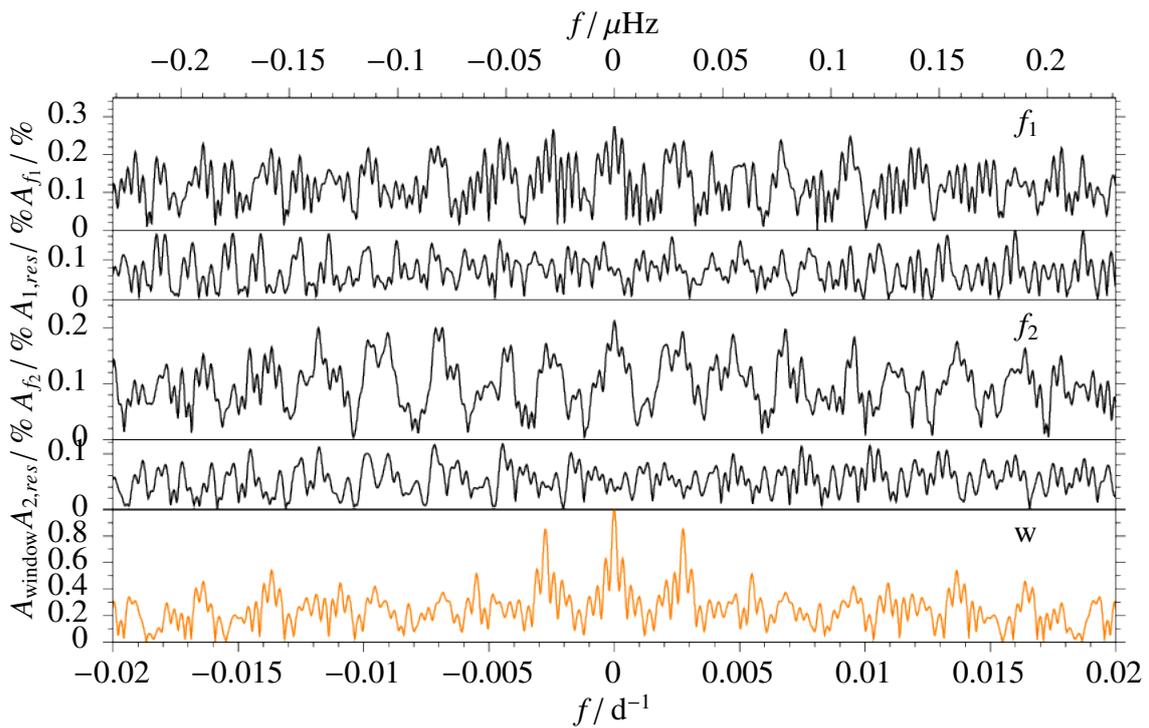}
	\caption[Amplitude spectrum of V541 Hya.]{Amplitude spectrum of V541 Hya of the main pulsation frequency $ f_1 =  \unit[635.32218]{d^{-1}} $ (\textit{top}), $ f_2 =  \unit[571.28556]{d^{-1}} $ (\textit{middle}) with the respective residuals after the pre-whitening below and the normalised window-function (\textit{bottom}).}
	\label{fig:v541hya-window}
\end{figure}

The amplitude spectrum in Fig.~\ref{fig:v541hya-power} shows two pulsation modes with frequencies at $ f_1 = \unit[635.32218]{d^{-1}} $ and at $ f_2 = \unit[571.28556]{d^{-1}} $.
Both of them show a complex behavior (Fig.~\ref{fig:v541hya-window}), indicating unresolved multiplets and/or frequency changes that we see also in the O-C diagrams (Fig.~\ref{fig:v541hya-oc}).
The S/N for a third frequency at $ \unit[603.88741]{d^{-1}} $ is not sufficient for the $ O-C $ analysis. Similar to V1636 Ori, the amplitude spectrum obtained from the TESS light curve in Fig.~\ref{fig:tess-power} shows no evidence for $ g $-mode pulsations with amplitudes greater than 0.4 per cent.

\begin{figure}[tp!]
	\centering
	\input{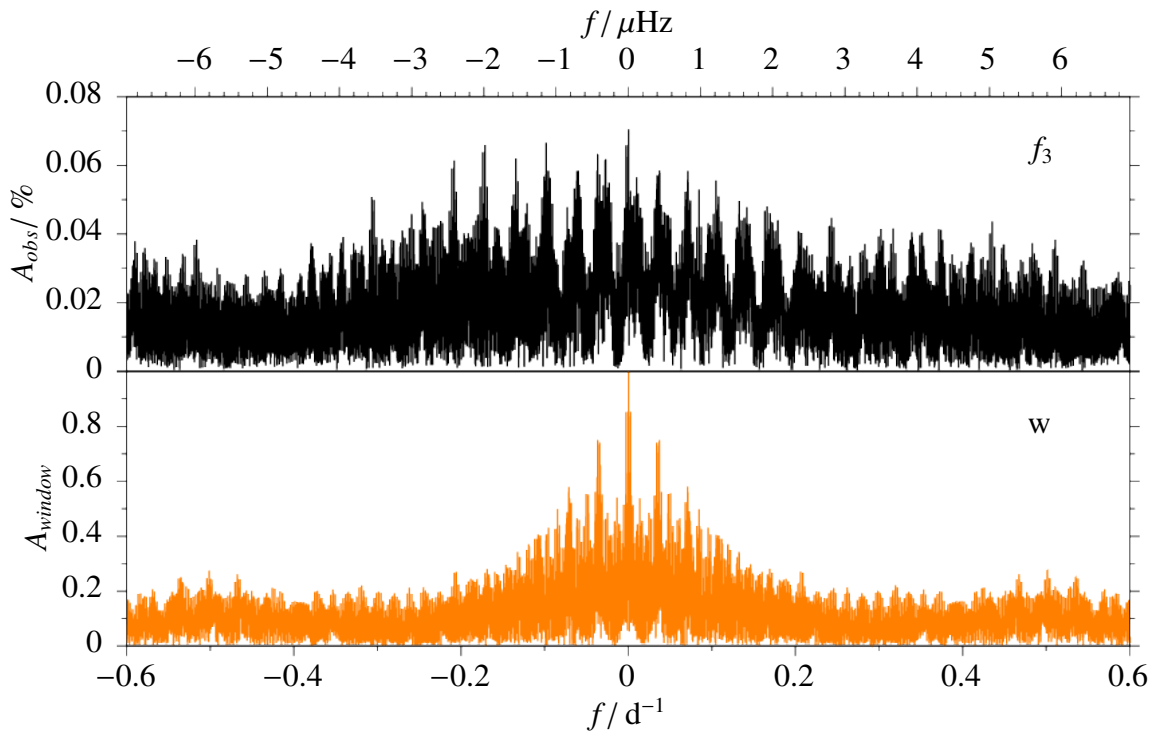}
	\caption[Amplitude spectrum with respect to the pulsation frequency $ f_3  $ of V541 Hya.]{Amplitude spectrum with respect to the pulsation frequency $ f_3 =  \unit[603.88741]{d^{-1}} $ of V541 Hya (\textit{top}) and the normalised window-function (\textit{bottom}).}
	\label{fig:v541hya-windowf3}
\end{figure}

\begin{figure*}[tp!]
	\centering
	\input{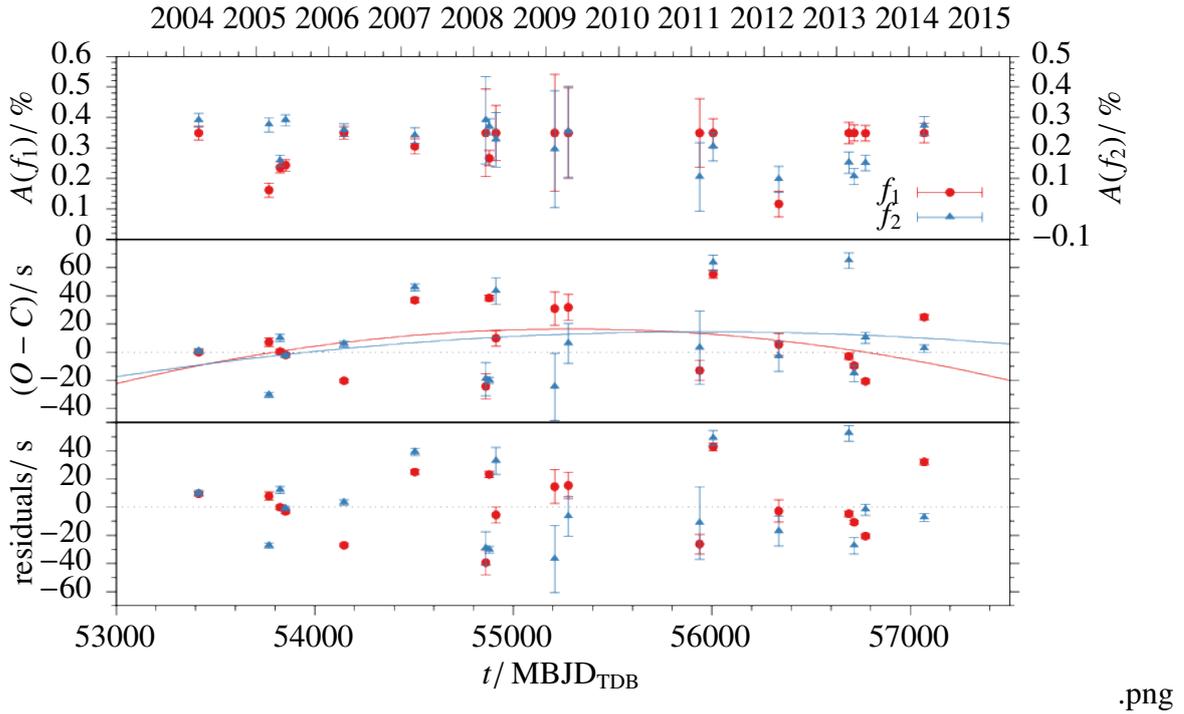}
	\caption[Results for the two main pulsations of V541 Hya.]{Results for the two main pulsations of V541 Hya. \textit{Top panel:} Amplitudes. \textit{Middle panel:} Fits of the $ O-C $ data with second order polynomials in time. \textit{Lower panel:} Residuals.}
	\label{fig:v541hya-oc}
\end{figure*}

\citet{randall_observations_2009} speculated about rotational mode splitting for $ f_3 $ with $ \Delta f_{3,-}  = \unit[5.12]{\mu Hz} $ and $ \Delta f_{3,+}  = \unit[3.68]{\mu Hz} $.
The asteroseismic modelling associates $ f_1 $ with a $ l=0 $ mode and $ f_2 $ with $ l=0\text{ or }1 $ mode (depending on the favoured model). $ f_3 $ corresponds to a $ l=2 $ mode. They caution this interpretation due to their limited resolution in frequency space, the mode splitting could be an unresolved quintuplet. Our data set shows no clear evidence for a mode splitting with $ \Delta f_{3,-}  = \unit[5.12]{\mu Hz} $ or $ \Delta f_{3,+}  = \unit[3.68]{\mu Hz} $ (see Fig.~\ref{fig:v541hya-windowf3}) but rather a mode splitting for $ f_1 $ and $ f_2 $ with about $ \Delta f = \unit[0.08]{\mu Hz} $ (Fig.~\ref{fig:v541hya-window}). Assuming these modes are of degree $ l=1 $, this could be interpreted as a triplet. But \citet{randall_observations_2009} model these modes with a degree of $ l=0 $ which does not support a mode splitting into triplets.

The $ O-C $ diagram in Fig.~\ref{fig:v541hya-oc} shows the analysis of the two main pulsation modes and the variation of the pulsation amplitudes. The second order fits in time indicated in the $ O-C $ diagram correspond to changes in period of $ \dot{P}/P_{f1} = \left( -1.49 \pm 0.11 \right)\times \unit[10^{-5}]{d^{-1}} $ and $ \dot{P}/P_{f2} = \left( -0.7 \pm 1.5 \right)\times \unit[10^{-5}]{d^{-1}} $. For $ f_2 $, the change in period does not significantly differ from the null hypothesis. Assuming these changes origin from stellar evolution, V541~Hya might just have passed the point of sign change in $ \dot{P} $ and at the beginning of the contraction phase. While the arrival times scatter widely, the amplitudes of both pulsations remain almost constant within the uncertainties. If V541~Hya is in its evolution close to starting the contraction phase, as indicated by a $ \dot{P} $ close to zero, the changes in stellar structure may cancel the strict phase coherence.

\section{Testing the sub-stellar companion hypothesis} \label{sec:testing-hypothesis}

In order to set upper limits to the mass of a companion, we computed a series of synthetic $ O-C $ curves for different orbital periods and companion masses, assuming circular orbits, and compared these curves with the $ O-C $ measurements after subtracting the long-term variations.

For each synthetic $ O-C $ curve, we selected the phase that gives the best fit to  the data using a weighted least squares algorithm. For each observational point, we computed the difference, in absolute value and in $\sigma$ units (where $\sigma$ is the $ O-C $ error), between $ O-C $ and the synthetic value. The greyscale in Fig.~\ref{fig:rv-planet} corresponds to the mean value of this difference in $\sigma$ units, which means that the presence of a companion is indicated by a minimum (bright areas) of this parameter. We see that in V1636~Ori, QQ~Vir and V541~Hya the mean difference for $ f_1 $ is always very high, implying that the data are not compatible with a companion.  However these results are limited by the fact that the $ O-C $ diagrams of 
these stars are \enquote{contaminated} by other irregular variations, presumably due 
to other reasons like non-linear interactions between different pulsation 
modes, for example, and therefore these constraints to the orbital period and mass of a companion must be taken with some caution. For the f2 and f3 measurements, the mean difference to the synthetic data is smaller in sigma units (because of the larger uncertainties) and very uniform. The uncertainties of the O-C measurements are not small enough to favour a set of models in the period-mass parameter space.

For $ f_1 $ of DW~Lyn, there is a significant minimum at about $ \unit[1450]{d} $ ($\sim\unit[4]{years}$) and $\sim\unit[5]{M_{\jupiter}} $\footnote{$ \unit[1]{M_{\jupiter}} $ (Jupiter mass) $ = 1.899 \cdot \unit[10^{27}]{kg} $}, which is well visible also in the $ O-C $ diagram of Fig.~\ref{fig:dwlyn-oc}. This periodicity is not visible in the second frequency $f_2$ which, however, has much larger error bars due to the much lower amplitude of $f_2$ with respect to $f_1$.

\citet{lutz_exotime_2011} described a periodicity at 80 days, detected for $ f_2 $. We can recover this signal, however with a low significance. This would correspond to a light travel time amplitude of $ \unit[4]{s} $ (for $ m\,\sin i \approx \unit[15]{M_{\jupiter}} $), which is smaller than $ \unit[15]{s} $ measured by \citet{lutz_exotime_2011}. Nevertheless, this signal is not confirmed by $ f_1 $. Thus, we rule out a companion induced signal in the arrival times due to the lack of simultaneous signals in $ f_1 $ and $ f_2 $ with similar amplitude. The tentative signal in $ f_2 $ is better explained by mode beating, as already described in Sect.~\ref{sec:results-dwlyn}. The variations seen in the first 200 days of the $ O-C $ diagram in Fig.~\ref{fig:dwlyn-oc} correspond to a periodicity of about 80 days and are accompanied by variations in the amplitude of the pulsation.

For V1636 Ori, \citet{lutz_search_2011} predicted a period at $ \unit[160]{d} $ and amplitude of $ \unit[12]{s} $. This can not be confirmed as a companion-induced signal. A periodic signal with an amplitude of $ \unit[6.5]{s} $ (for $ m\,\sin i \approx \unit[15]{M_{\jupiter}} $) is indicated in the analysis of $ f_1 $ but at a low significance and accompanied by many other signals of similar significance. This periodicity is not confirmed by a significant signal in the measurements of $ f_2 $.

\begin{figure}[tp!]
	\centering
	\begin{subfigure}[b]{\textwidth}
		\centering
		\input{figures/exotime/dwlyn/rv/oc-model.tex}
		\caption{}
		\label{fig:dwlyn-rv-planet}
	\end{subfigure}
	\begin{subfigure}[b]{\textwidth}
		\centering
		\input{figures/exotime/v1636ori/rv/oc-model.tex}
		\caption{}
		\label{fig:v1636ori-rv-planet}
	\end{subfigure}
\end{figure}
	\begin{figure}[tp!]
		\ContinuedFloat
		\centering
		\begin{subfigure}[b]{\textwidth}
		\centering
		\input{figures/exotime/qqvir/rv/oc-model.tex}
		\caption{}
		\label{fig:qqvir-rv-planet}
	\end{subfigure}
\end{figure}
\begin{figure}[tp!]
\ContinuedFloat
	\begin{subfigure}[b]{\textwidth}
		\centering
		\input{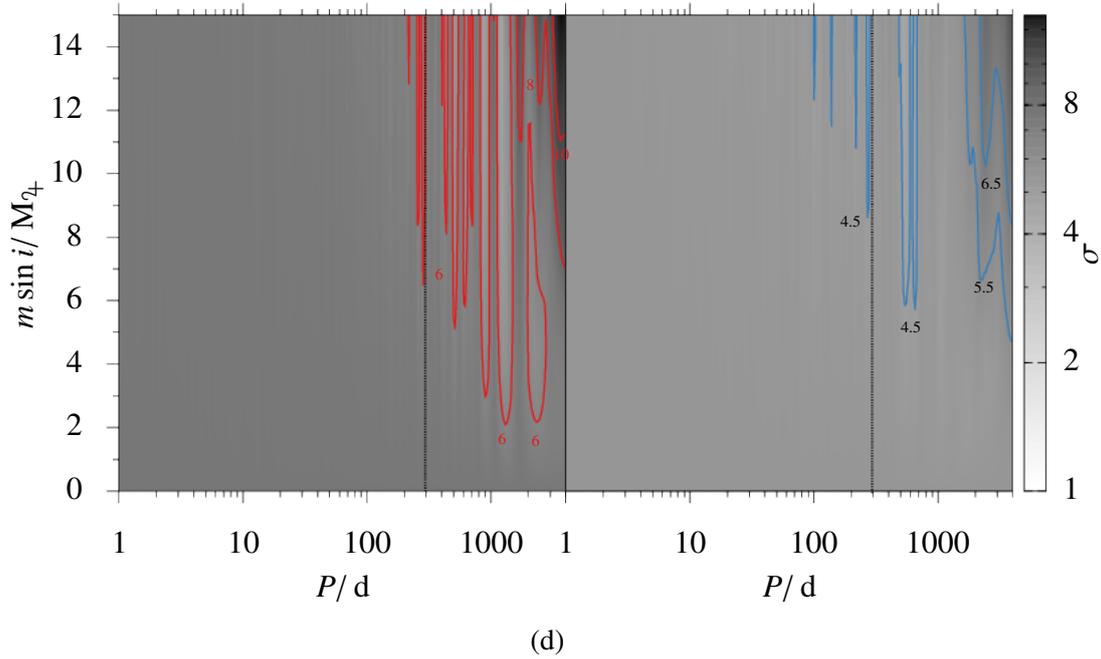}
		\caption{}
		\label{fig:v541hya-rv-planet}
	\end{subfigure}
	\caption[Minimum companion mass as a function of orbital period.]{Minimum companion mass as a function of orbital period. Greyscale shows the difference between the $ O-C $ measurements and artificial $ O-C $ data generated for a given combination of companion mass and orbit. We note that at this stage, the phase optimization of the artificial data is done independently for each pulsation frequency. The median of gaps in between the epochs is indicated by a vertical dotted line. See text for more details. (a) DW~Lyn. Contour lines for $ f_1 $ are placed at 2, 3, 4 and $ \unit[5]{\sigma} $ (\textit{left panel}) and for $ f_2 $ at 3.25 and $ \unit[3.5]{\sigma} $ (\textit{right panel}), as indicated by their labels. The planetary signal proposed by \citet{lutz_exotime_2011} at a period of $ \unit[80]{d} $ is indicated as dashed line. (b) V1636 Ori. Contour lines for $ f_1 $ are placed at 9 and $ \unit[10]{\sigma} $ (\textit{left panel}) and for $ f_2 $ at 2 and $ \unit[2.2]{\sigma} $ (\textit{right panel}), as indicated by their labels. The planetary signal proposed by \citet{lutz_exotime_2011} at a period of $ \unit[160]{d} $ is indicated as dashed line. (c) QQ Vir. Contour lines for $ f_1 $ are placed at 15, 20 and $ \unit[25]{\sigma} $ (\textit{left panel}), for $ f_2 $ at 1.1 and $ \unit[1.3]{\sigma} $ (\textit{middle panel}) and for $ f_3 $ at 1.3 and $ \unit[1.5]{\sigma} $ (\textit{right panel}), as indicated by their labels. (c) V541 Hya. Contour lines for $ f_1 $ are placed at 6, 8 and $ \unit[10]{\sigma} $ (\textit{left panel}) and for $ f_2 $ at 4.5, 5.5 and $ \unit[6.5]{\sigma} $ (\textit{right panel}), as indicated by their labels.}
\label{fig:rv-planet}
\end{figure}

\chapter{Summary and Conclusion}

In this work, we present ground-based multi-site observations for the four sdBs, DW Lyn, V1636 Ori, QQ Vir and V541 Hya. We investigate variations in the arrival times of their dominant stellar pulsation modes to draw conclusions about secular period drifts and possible sub-stellar companions. All light curves are analysed homogeneously. 

From the $ O-C $ measurements, we derive an evolutionary time scale from the change in period $ \dot{P} $. 
Comparing to model calculations from \citet{charpinet_adiabatic_2002}, we infer the evolutionary phase of the target. Although some $ \dot{P} $ measurements are influenced by mode splitting, we can tell from the sign of $ \dot{P}_1 $ of DW~Lyn that the star is likely still in the stage of central \ch{He}  burning. We can draw a similar conclusion from the sign of $ \dot{P} $ of QQ~Vir. The $ \dot{P}$ measurements of V1636~Ori are likely affected by mode splitting, making it difficult to interpret the results in the context of stellar evolution. V541~Hya shows $ \dot{P} $ measurements close to zero, which indicates the star being at the transition phase between \ch{He} burning and contraction due to the depletion of \ch{He} in the core.

Comparing atmospheric properties from Table~\ref{tab:stellar} with the evolutionary tracks for different models from Fig.~1 in \citet{charpinet_adiabatic_2002}, we can confirm the hypothesis that DW~Lyn and QQ~Vir are in their \ch{He} burning phase. V541~Hya agrees within $ \unit[2]{\sigma} $ of the $ \log g $ measurement with one model at the turning point between the two evolutionary stages.

However, we can not exclude frequency and amplitude variations on smaller time scales than resolvable by our data set. Using temporally higher resolved \textit{Kepler}-data of KIC 3527751, \citet{zong_oscillation_2018} caution about long-term frequency or phase evolutions ascribing to non-linear amplitude and frequency modulations in pulsating sdBs. We see such effects already in our data set, even with a low temporal resolution compared to the \textit{Kepler} sampling with a duty cycle of more than 90 per cent.

Observations on DW Lyn and V1636 Ori were published by \citet{lutz_light_2008,schuh_exotime_2010,lutz_search_2011,lutz_exotime_2011}. Our analysis of these observations, including extended data sets, do not confirm the tentative companion periods of 80 and 160 days, respectively. These signals more likely arise due to mode beating indicated by partly unresolved frequency multiplets and amplitude modulations. 

Almost all analysed pulsation modes show formal significant changes in arrival times but the amplitudes of these periodic signals do not correlate with frequencies, excluding the light-travel time effect due to orbital reflex motions for such variations and thus giving upper limits on companion masses. Only DW~Lyn might have a planetary companion on a long orbital period, as indicated by one arrival time measurement. But this can not be confirmed with a second measurement, due to larger uncertainties. Additionally, more studies question the presence of already proposed companions, for example,  \citet{krzesinski_planetary_2015,hutchens_new_2017}. Our unique sample of long-term observations shows a complex behavior of mode- and amplitude interactions in sdBs which should be addressed in further studies. Until this has been addressed, caution is advised when interpreting $ O-C $ pulse arrival times in terms of companions.

\chapter{Appendix}
\section{TESS data}
\begin{figure*}[ht!]
	\centering
	\begin{subfigure}[b]{\textwidth}
		\centering
		\input{figures/exotime/v1636ori/lc_clipped_tess.tex}
		\caption{}
		\label{fig:v1636ori-lc-tess}
	\end{subfigure}
\end{figure*}
	\begin{figure*}[ht!]\ContinuedFloat
	\begin{subfigure}[b]{\textwidth}
		\centering
		\input{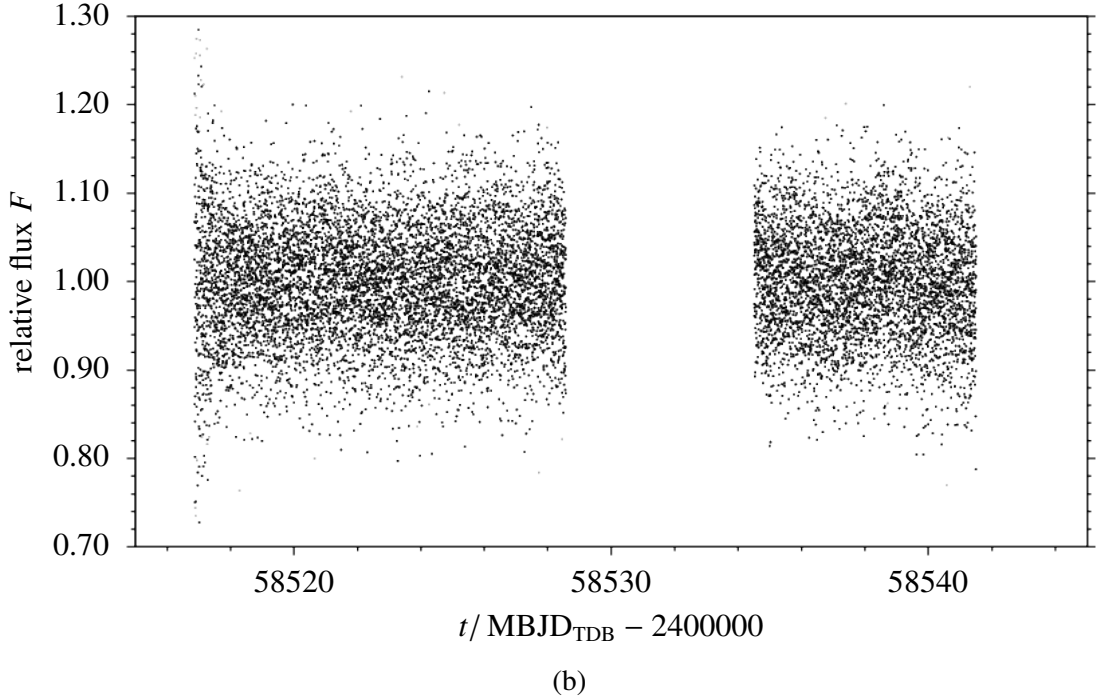}
		\caption{}
		\label{fig:v541hya-lc-tess}
	\end{subfigure}
	\caption[Light curves of the TESS observations.]{Light curves of the TESS observations. Grey points are considered outliers and partially exceeding the plotting range. (a) V1636~Ori. (b) V541~Hya.}
	\label{fig:lc-tess}
\end{figure*}

\begin{figure*}[ht!]
	\centering
	\input{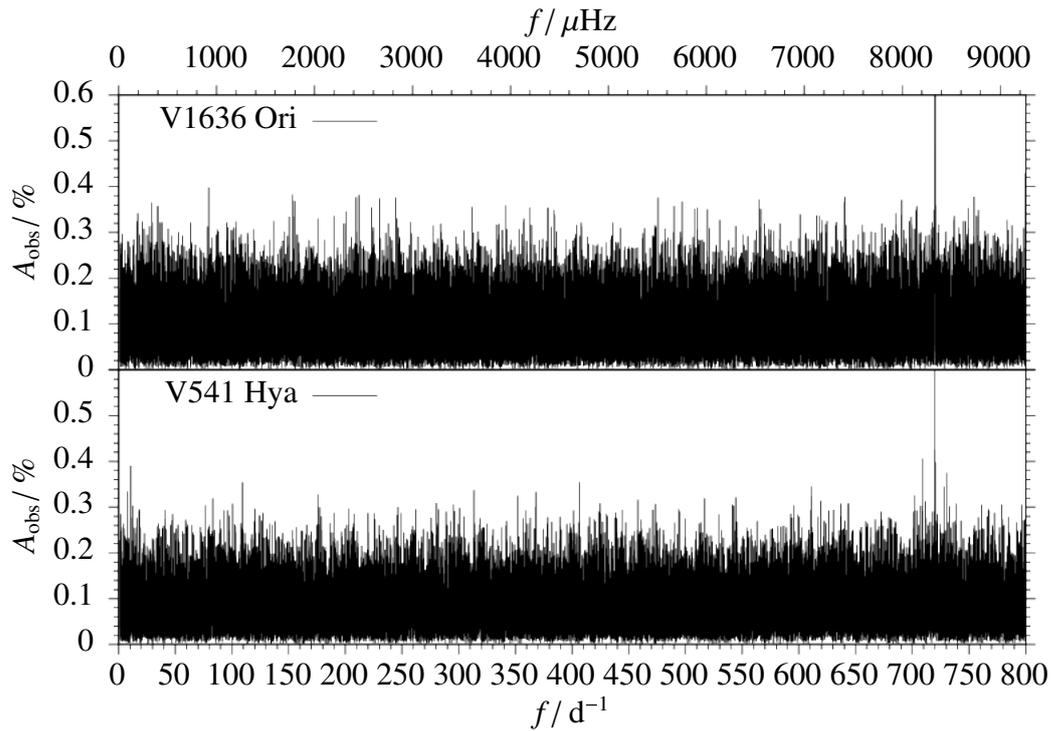}
	\caption[Amplitude spectrum of the TESS observations.]{Amplitude spectrum of the TESS observations. \textit{Upper panel} spectrum of V1636 Ori. \textit{Lower panel} spectrum of V541 Hya. The only peak above the noise level is the $ \unit[120]{s} $ alias due to the cadence of the observations.}
	\label{fig:tess-power}
\end{figure*}

\newpage
\section{Amplitude spectra}

\begin{table}[h]
	\centering
	\caption[Additional identified pulsation modes.]{Additional pulsation modes identified for our targets not used in the $ O-C $ analysis due to their low signal-to-noise-ratio.}
	\label{tab:add-freq}
	\begin{tabular}{lr@{.}lr@{.}l}
		\toprule
		Target & \multicolumn{2}{c}{$ \unit[f/]{d^{-1}} $} & \multicolumn{2}{c}{$ \unit[A/]{\%} $}  \\
		\midrule
		DW Lyn & 475 & 8231(2) & 0 & 09(18) \\
		 & 319 & 4042(3) & 0 & 06(12) \\
		 & 463 & 0100(6) & 0 & 03(18) \\
		V1636 Ori & 566 & 24031(3) & 0 & 6(3) \\
		QQ Vir & 733 & 0704(1) & 0 & 3(1) \\
		 & 664 & 4886(1) & 0 & 2(1) \\
		 & 572 & 73611(5) & 0 & 19(9) \\
		 & 664 & 7122(1) & 0 & 1(1) \\
		 & 434 & 1522(6) & 0 & 01(7) \\
		 & 502 & 410(2) & 0 & 01(9) \\
		V541 Hya  & 531 & 16759(16) & 0 & 03(7) \\
		 & 603 & 88741(6) & 0 & 03(8) \\
		\bottomrule
	\end{tabular}
\end{table}
\begin{figure*}[h]
	\centering
	\input{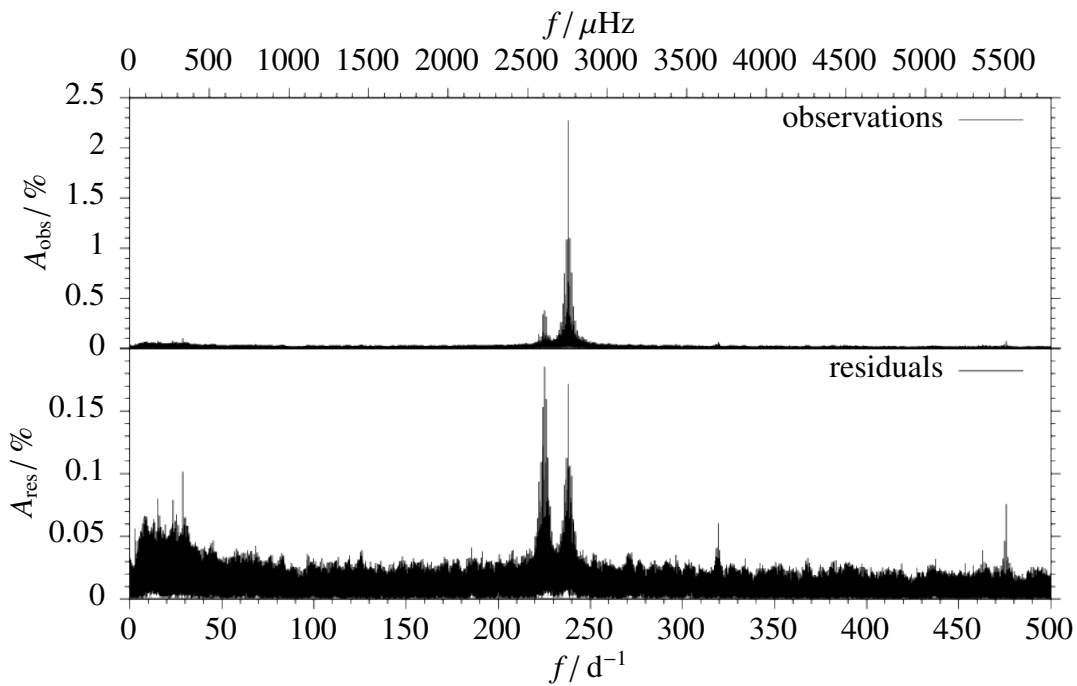}
	\caption[Amplitude spectrum of DW Lyn.]{Amplitude spectrum of DW Lyn. \textit{Upper panel:} observations $ A_{obs} $. \textit{Lower panel} residuals $ A_{res} $ after subtracting the light curve models from the observations.}
	\label{fig:dwlyn-power}
\end{figure*}
\begin{figure*}[h]
	\centering
	\input{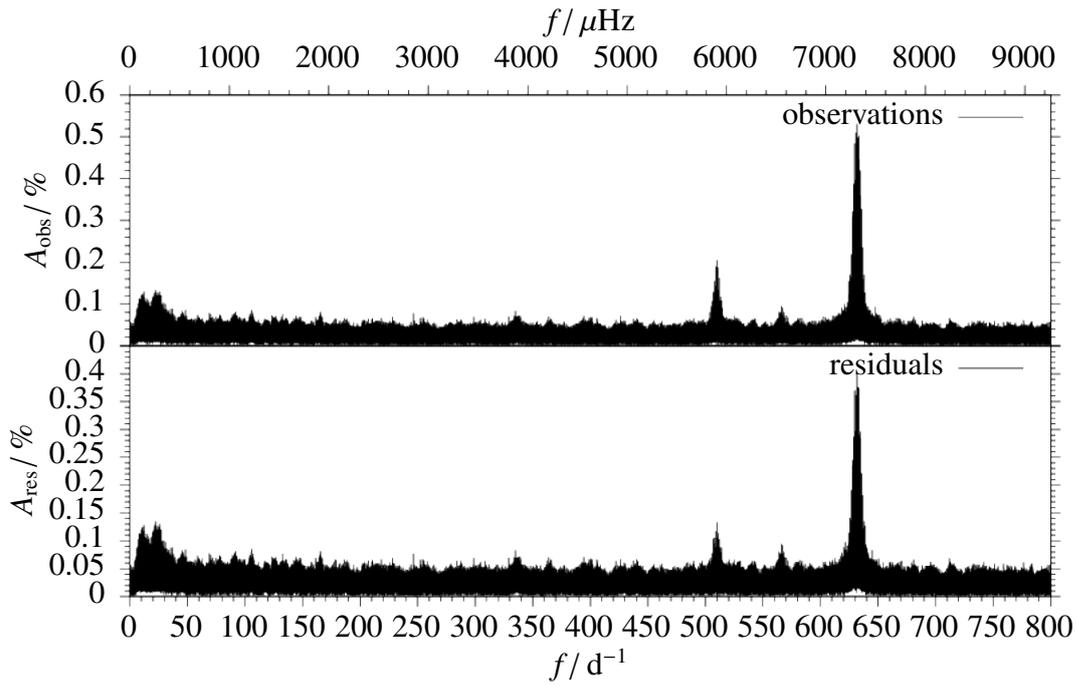}
	\caption[Amplitude spectrum of V1636 Ori.]{Same as Fig.~\ref{fig:dwlyn-power} but for V1636 Ori.}
	\label{fig:v1636ori-power}
\end{figure*}
\begin{figure*}[h]
	\centering
	\input{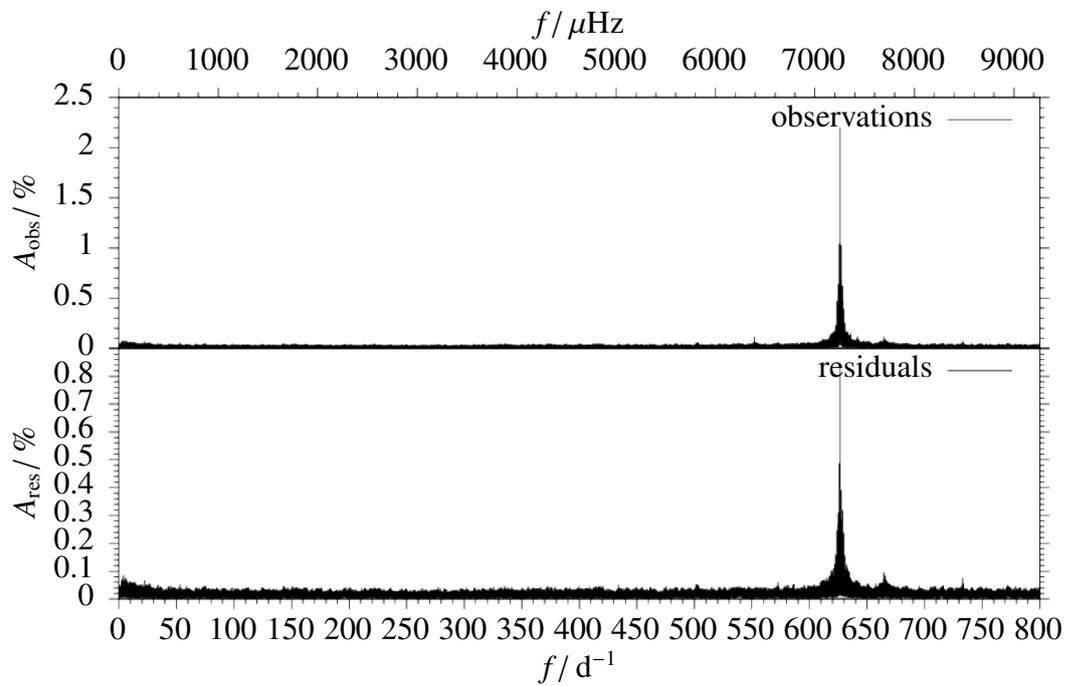}
	\caption[Amplitude spectrum of QQ Vir.]{Same as Fig.~\ref{fig:dwlyn-power} but for QQ Vir.}
	\label{fig:qqvir-power}
\end{figure*}
\begin{figure*}[h]
	\centering
	\input{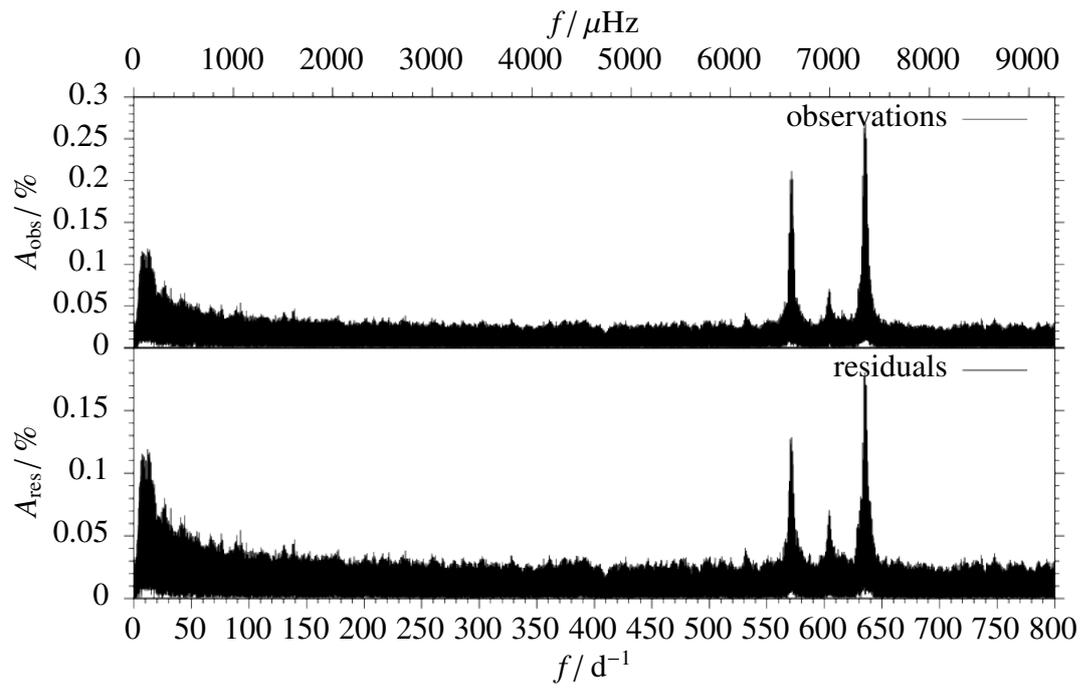}
	\caption[Amplitude spectrum of V541 Hya.]{Same as Fig.~\ref{fig:dwlyn-power} but for V541 Hya.}
	\label{fig:v541hya-power}
\end{figure*}

\clearpage
\paragraph*{\footnotesize Acknowledgements.}
{\footnotesize
	We thank Wen-Shan Hsiao for observing at the Lulin Observatory and Mike D. Reed for observing at the Baker Observatory and Elia Leibowitz for observing at the WISE observatory.
	F.M. conducted the work in this paper in the framework of the International Max-Planck Research School (IMPRS) for Solar System Science at the University of G\"ottingen (Volkswagen Foundation project grant number VWZN3020).
	DK thanks the SAAO for generous allocations of telescope time and the National Research Foundation of South Africa and the University of the Western Cape for financial support.
	TDO gratefully acknowledges support from the U.S. National Science Foundation grant AST-0807919.
	L.M.\ was supported by the Premium Postdoctoral Research Program of the Hungarian Academy of Sciences. This project has been supported by the Lend\"ulet Program  of the Hungarian Academy of Sciences, project No. LP2018-7/2019.
	Based on observations made with the Italian Telescopio Nazionale Galileo (TNG) operated on the island of La Palma by the Fundaci\'{o}n Galileo Galilei of the INAF (Istituto Nazionale di Astrofisica) at the Spanish Observatorio del Roque de los Muchachos of the Instituto de Astrofisica de Canarias.
	Based on observations collected at the Centro Astron\'{o}mico Hisp\'{a}nico en Andaluc\'{i}a (CAHA) at Calar Alto, operated jointly by the Andalusian Universities and the Instituto de Astrof\'{i}sica de Andaluc\'{i}a (CSIC).
	Based on observations collected at Copernico telescope (Asiago, Italy) of the INAF - Osservatorio Astronomico di Padova.
	Based on observations made with the Nordic Optical Telescope, operated by the Nordic Optical Telescope Scientific Association at the Observatorio del Roque de los Muchachos, La Palma, Spain, of the Instituto de Astrofisica de Canarias.
	This paper includes data collected by the TESS mission. Funding for the TESS mission is provided by the NASA Explorer Program.
}
\cleardoublepage

\part{Application to larger target pool} \label{sec:other-targets}
	\chapter{Further asteroseismic targets}
	The search for sub-stellar companions using the pulsation timing method is not only limited to sdB stars but can also be applied to a greater variety of pulsating stars, as long as the pulsations are driven by a \textkappa~mechanism, i.e., coherent in phase. However, modes of long periods decrease the precision of timing measurements. Additionally, large pulsation amplitudes tend to show non-linear effects, which excludes the application of this technique. A case study of \textbeta~Cephei, RR~Lyrae and \textdelta~Scuti (\textdelta~Sct) stars can be found in \citet{silvotti_potential_2011}. Among these, \textdelta~Sct stars are the most promising pulsators to follow-up. The following describes \textdelta~Sct stars and the application of the pulsation timing method in more detail.

\section{\textdelta~Scuti stars}

Planet occurrence rates predict a maximum for host stars with masses of \asymmnumerr{1.9}{0.1}{0.5}$\SI{}{M_\odot}$ \citep{reffert_precise_2015}, close to the mass of main sequence A stars in the instability strip where \textdelta~Sct pulsators are common (see Fig.~\ref{fig:intro:puls-classes}). However, actual planet detections are affected by observational selection effects. Planetary transits are difficult to observe because of the pulsational luminosity variations and due to wider orbits of planets around A stars \citep{johnson_retired_2011,lloyd_retired_2011}, resulting in lower transit probabilities. For the radial velocity method, A stars are not well suited because of their fast rotation, with the mean of the equatorial rotational velocity distribution exceeding \SI{100}{\kilo\meter\per\second} \citep{abt_relation_1995,royer_rotational_2007}. Their high effective temperatures compared to solar-like stars leads to fewer and shallower absorption lines, which can additionally be distorted by pulsations. This leaves the pulsation timing method as a promising alternative to search for companions. \textdelta~Sct stars show radial and non-radial p-modes with amplitudes up to \SI{0.5}{\percent} and with pulsation periods between about \SI{20}{\minute} and \SI{8}{\hour} \citep[e.g.][]{qian_lamost_2017,balona_kepler_2012}. The pulsations are driven by the \textkappa~mechanism due to partial ionization of \ch{He} II.

The mostly uninterrupted observations of space missions like \textit{Kepler} and TESS provide a large sample of \textdelta~Sct observations for this purpose. The two missions and some of their applications are described in the following sections.

	\chapter{Kepler}
	Sections of this chapter are part of a paper in preparation to be submitted as an article to Astronomy and Astrophysics. 

\section{\textit{Kepler} mission}
The \textit{Kepler} mission \citep{borucki_kepler_2010} was launched in 2009 to observe a fixed patch of sky in the Cygnus constellation with more than \SI{100}{\deg^2} field of view to discover transiting Earth-like exoplanets in the habitable zone. The telescope's mirror measures \SI{0.95}{m} in diameter and the camera is sensitive for a wavelength range of \SIrange{430}{890}{\nano\meter}. Asteroseismic observations of pulsating stars were organized by the Kepler Asteroseismic Science Consortium (KASC). Some were observed in a high cadence mode of \SI{60}{\second} (\textit{short cadence}), the nominal cadence measures \SI{30}{\minute} (\textit{long cadence}). 

\subsection{sdB stars}

The \textit{Kepler} field contained a small number of sdB stars \citep[e.g.][]{mcnamara_classification_2012}. None of them showed indications for a planetary transit. However, KIC~05807616 and KIC~10001893, two pulsating sdB stars in the \textit{Kepler} field, show some evidence for the presence of planetary companions \citep{charpinet_compact_2011,silvotti_kepler_2014}. However, interpretation of these signals is questioned, as already discussed in chapter~\ref{sec:exotime}.

\subsection{\textdelta~Scuti stars}

\textit{Kepler} observed around one thousand \textdelta~Sct stars in its field of view. The primary goal of their observations are conclusions about the stellar structure and evolution via asteroseismology, as the transition from deep to shallow convective envelopes takes place in the region of the HRD where \textdelta~Sct are located (see Fig.~\ref{fig:intro:puls-classes}). This opens the opportunity to use the extensive \textit{Kepler} data set and apply the pulsation timing method to the observations. The following describes one particular system as a benchmark for which a planetary companion was already discovered.

\subsubsection{KIC~7917485}

\citet{murphy_finding_2014,murphy_finding_2016} used the \textit{Kepler} data to search for binary systems using the light travel time effect. \citet{murphy_planet_2016} (hereafter M16) detected a planet orbiting its host star KIC~7917485 in a \SI{840}{\day} orbit. The following validates this finding using the $ O-C $ pipeline implemented for this work.

\begin{figure}[ht!]
	\centering
	\begin{subfigure}[b]{\textwidth}
		\centering
		\input{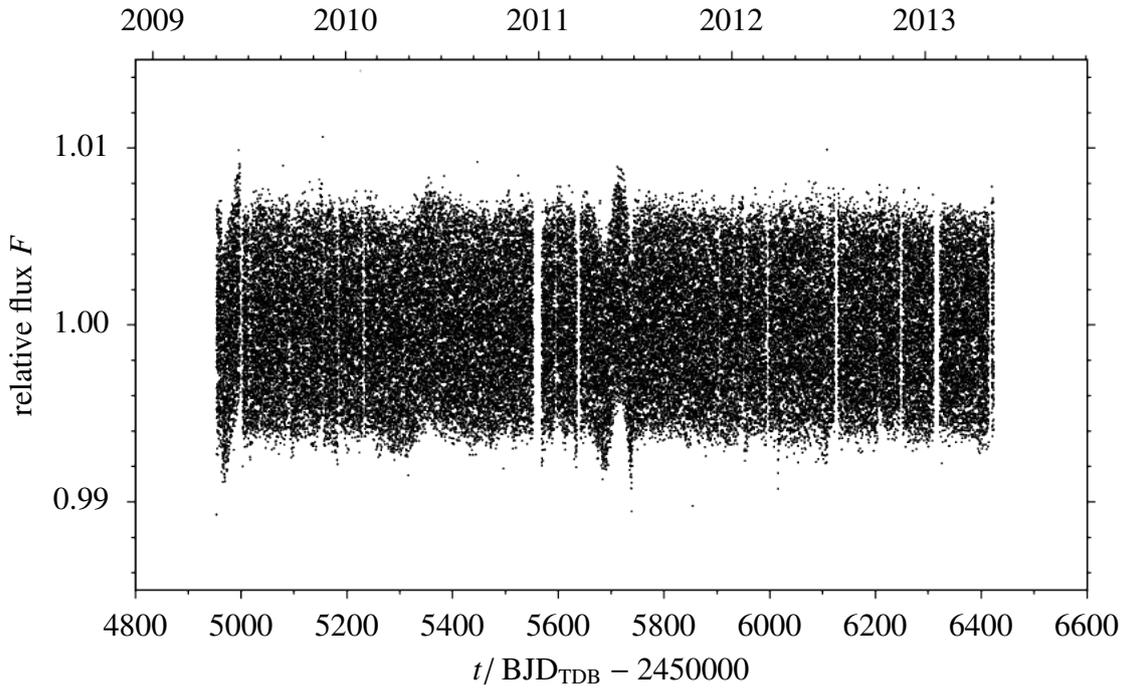}
		\caption{KIC~7917485 light curve.}
		\label{fig:KIC7917485-lc-kepler}
	\end{subfigure}
	\vspace{10pt}
	\begin{subfigure}[b]{\textwidth}
		\centering
		\input{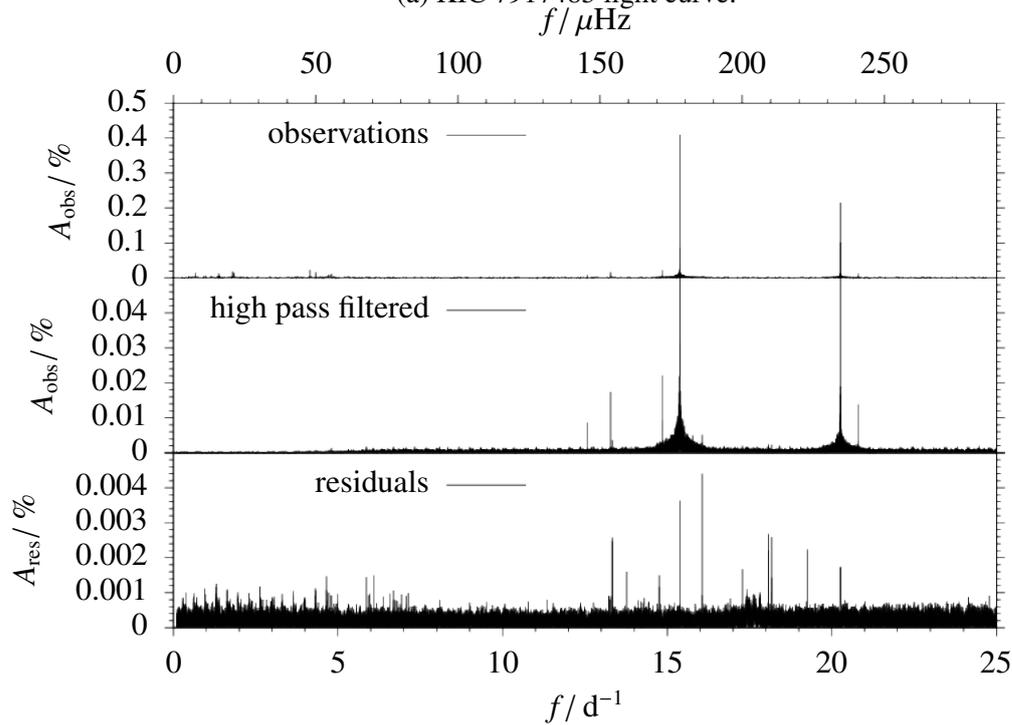}
		\caption{KIC~7917485 amplitude spectrum.}
		\label{fig:KIC7917485-power}
	\end{subfigure}
	\caption[KIC~7917485 data.]{KIC~7917485 data. \textit{Top:} Light curve. Grey points are considered outliers and partially exceeding the plotting range. \textit{Bottom:} Amplitude spectrum. The top panel shows the observations. The middle panel the amplitude spectrum after applying the high pass filter. Note the different scale on the y-axis in order to make small amplitudes visible. The bottom panel shows the residuals after modelling six pulsation modes.}
	\label{fig:KIC7917485-lc}
\end{figure}

KIC~7917485 was observed between May 2009 and May 2013 by \textit{Kepler}. We use the light curves provided by the MAST archive\footnote{\url{https://archive.stsci.edu/}} which had common instrumental trends removed by the Pre-Search Data Conditioning Pipeline \citep[PDC, ][]{stumpe_kepler_2012}. In order to remove remaining long term trends, we make use of the \verb|kepflatten| routine of the PyKE package \citep{still_pyke_2012,ze_vinicius_keplergo/pyke_2018}. The light curve and the corresponding amplitude spectrum are shown in Fig.~\ref{fig:KIC7917485-lc}. 

In order to reproduce the results of M16, we apply a high-pass frequency filter. This removes frequencies below $ \unit[5]{d^{-1}} $, which are most likely  artefacts due to instrumental effect. We apply this filter forward and backward in time to compensate for phase delays.

\begin{figure}[ht!]
	\centering
	\input{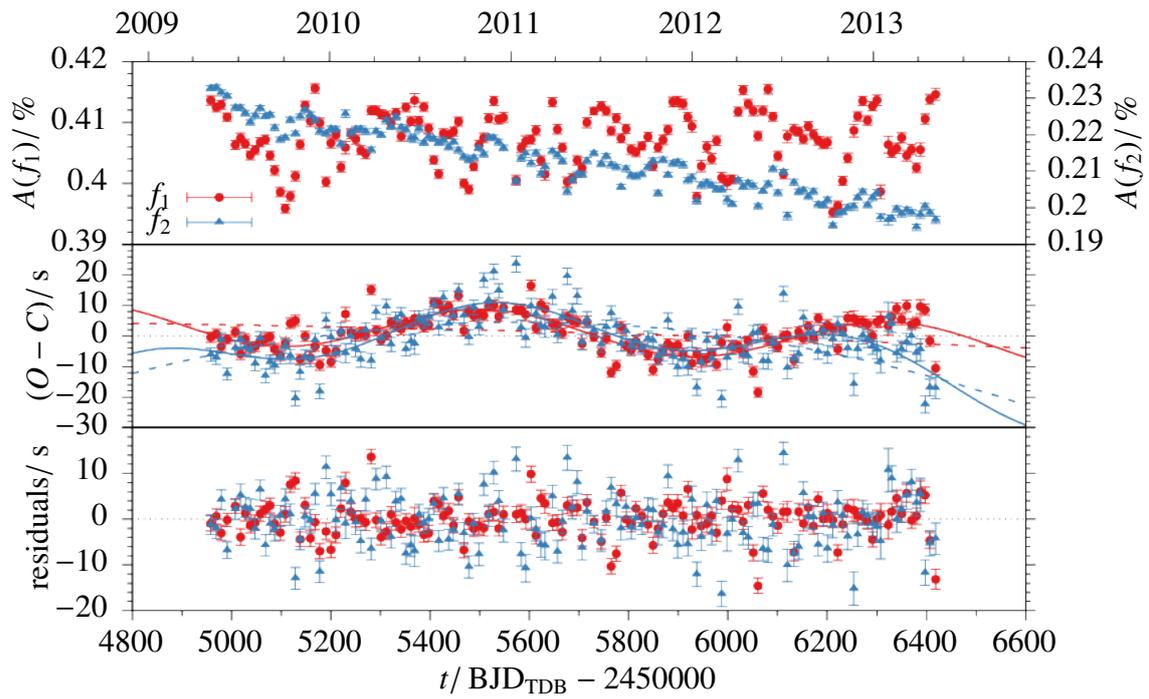}
	\caption[Results for the two main pulsations of KIC~7917485.]{Results for the two main pulsations of KIC~7917485. \textit{Top panel:} Amplitudes. \textit{Middle panel:} Fits of the $ O-C $ data with second order polynomials only (dashed lines) and full fit including the light travel time effect (solid lines). \textit{Lower panel:} Residuals.}
	\label{fig:KIC7917485-oc}
\end{figure}

For the $ O-C $ analysis, we applied our pipeline and fitted all pulsations with amplitudes above \SI{0.005}{\percent}. Although the signal-to-noise ratio is sufficient for only two modes, the simultaneous fitting process of four more pulsations prevents any unaccounted interactions between the modes. The time dependent variation in pulsation amplitude and phase is shown in Fig.~\ref{fig:KIC7917485-oc}. Each epoch measures about ten days. The main pulsation frequency $ f_1 = \SI[separate-uncertainty = false]{15.3830074\pm0.0000004}{\per\day} $ exhibits large semi-periodic variations which correlate with the observational quarters of the \textit{Kepler} mission. Thus, they are likely to be instrumental systematics. The amplitude of $ f_2 = \SI[separate-uncertainty = false]{20.2629053\pm0.0000008}{\per\day} $ shows a similar behaviour on top of a general trend of decreasing amplitude with time. The phase variations for both pulsations show clearly a sinusoidal modulation with an amplitude of about \SI{10}{\second} and a period of about \SI{800}{\day}. We fit simultaneously a quadratic long term trend and a sinusoidal model for a circular companion orbit to the data. The results are presented in Table~\ref{tab:KIC7917485}. 
\begin{table}[t]
	\centering
	\caption[Orbital parameters for the KIC~7917485 system.]{Orbital parameters for the KIC~7917485 system. Values for M16 are taken from Table 2 in \citet{murphy_planet_2016}. M16 did not account for a change in pulsation period $ \dot{P} $. The model of this work uses a circular orbit model for the companion, i.e. $ e=0 $.}
	\sisetup{table-text-alignment = center,
		table-align-uncertainty
	}
	\begin{tabular}{lSSS}
		\toprule
		& {M16} & {this work $ f_1 $}  & {this work $ f_2 $} \\ 
		\midrule
		$ \Delta T /\si{\second} $ & \asymmnumerr{7.1}{0.5}{0.4} & 6.8 \pm 0.3 & 6.6 \pm 0.4 \\
		$ P_{\text{orbit}}/\si{\day} $ & \asymmnumerr{840}{22}{20} & 821.50 \pm 0.08 & 726.72 \pm 0.08 \\
		$ e $& \asymmnumerr{0.15}{0.13}{0.10} & 0 & 0 \\
		$ \dot{P}/\si{\second\,Myr^{-1}}  $ &  & -0.001 \pm 0.001 & -0.035 \pm 0.002  \\
		\bottomrule
	\end{tabular}
	\label{tab:KIC7917485}
\end{table}
While a change in period $ \dot{P} $ for $ f_1 $ is almost not detectable, $ f_2 $ shows a significant $ \dot{P} $, which coincides with the decrease in pulsation amplitude of 0.04 per cent (amplitude) or about 18 per cent (relative). M16 did not account for a change in period, thus we cannot compare our measurements.

For the orbital fit of the companion, M16 used the weighted mean of both phase measurements. Since we want to confirm the nature of the companion with two independent measurements, we do not follow this procedure. Our sinusoidal orbit fit for $ f_1 $ agrees within the uncertainties with the measurement of M16. However, the orbital period measured for $ f_2 $ is about 100 days significantly shorter than for $ f_1 $. This might be attributed to our additional fit for the change in period, combined with larger $ O-C $ uncertainties towards the end of the observations (approximately two times larger), due to the decreasing pulsation amplitude. 

In order to compare the M16 results better with the measurements of this work, we will be implementing a full orbit fit on both measurements simultaneously. Additionally, the uncertainties appear rather small in comparison to M16. A different treatment of uncertainties might give more reliable error bars.
	\chapter{TESS}
	\section{TESS mission}

The Transiting Exoplanet Survey Satellite (TESS) by NASA was launched in April 2018 as an all-sky survey for transiting exoplanets. The \SI{24}{\deg}$ \times $\SI{96}{\deg} field of view observes each sector for about 27 days and sweeps across the sky. This creates some observations lasting longer than 27 days due to the overlapping sectors. The four telescopes (each with a field of view of \SI{24}{\deg^2}) measure \SI{105}{mm} in diameter and the cameras are sensitive for a wavelength range of \SIrange{600}{1000}{\nano\meter}. This is slightly redder than the \textit{Kepler} bandpass due to the mission's differing priorities: detecting transiting planets orbiting Sun-like stars for \textit{Kepler} and small, cool stars for TESS. The full frame cadence measures 30 minutes while the nominal cadence for transit detections two minutes. The TESS Asteroseismic Science Consortium (TASC) coordinates 20 seconds \enquote{short cadence} observations of bright asteroseismic target stars.

However, observations with such a short time span of 27 days (and up to one year but only around the ecliptic pole, e.g., \SI{1.7}{\percent} of the sky) do not allow an extensive search of sub-stellar companions using the pulsation timing method on their own. But they can be used as additional observations to existing data sets, especially the EXOTIME project data. Without additional ground-based observations, a joint $ O-C $ analysis of \textdelta~Sct data from \textit{Kepler} and TESS is likely to fail due to the large observational gap of about seven years.

\subsection{sdB stars}

TESS observations of V1636~Ori and V541~Hya are already discussed in chapter~\ref{sec:exotime}. The data are taken at two minutes cadence and thus do not allow the investigation of p-modes. 

\begin{figure}[ht!]
	\centering
	\begin{subfigure}[b]{\textwidth}
		\centering
		\input{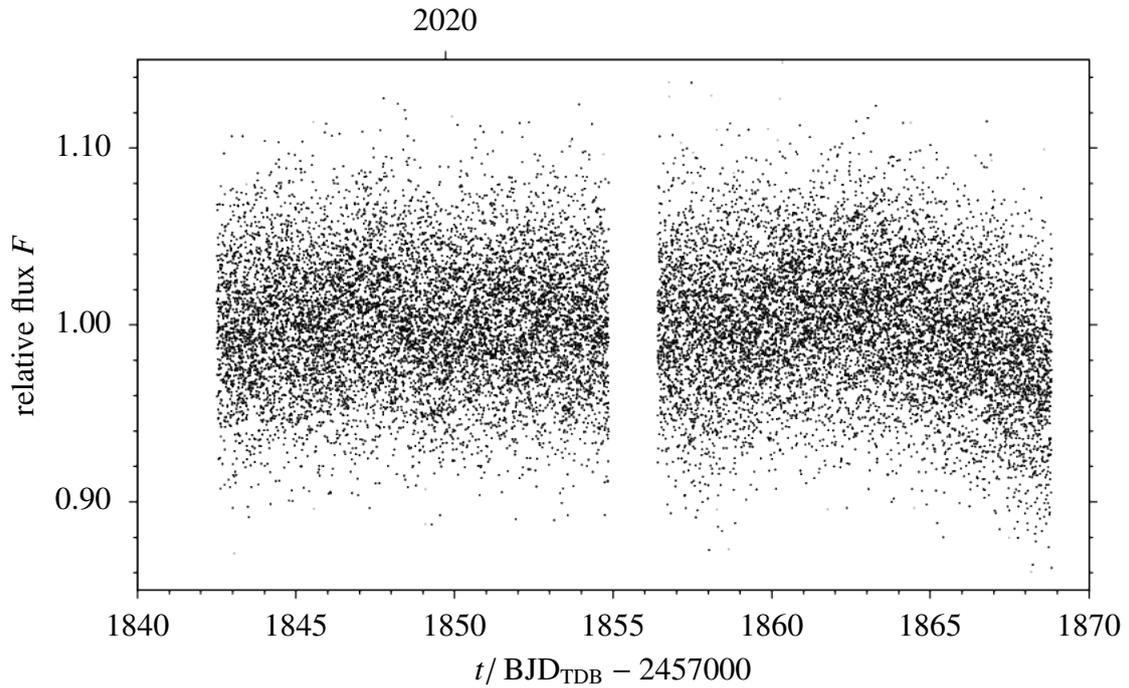}
		\caption{DW~Lyn light curve.}
		\label{fig:dwlyn-lc-tess}
	\end{subfigure}
	\vspace{10pt}
	\begin{subfigure}[b]{\textwidth}
		\centering
		\input{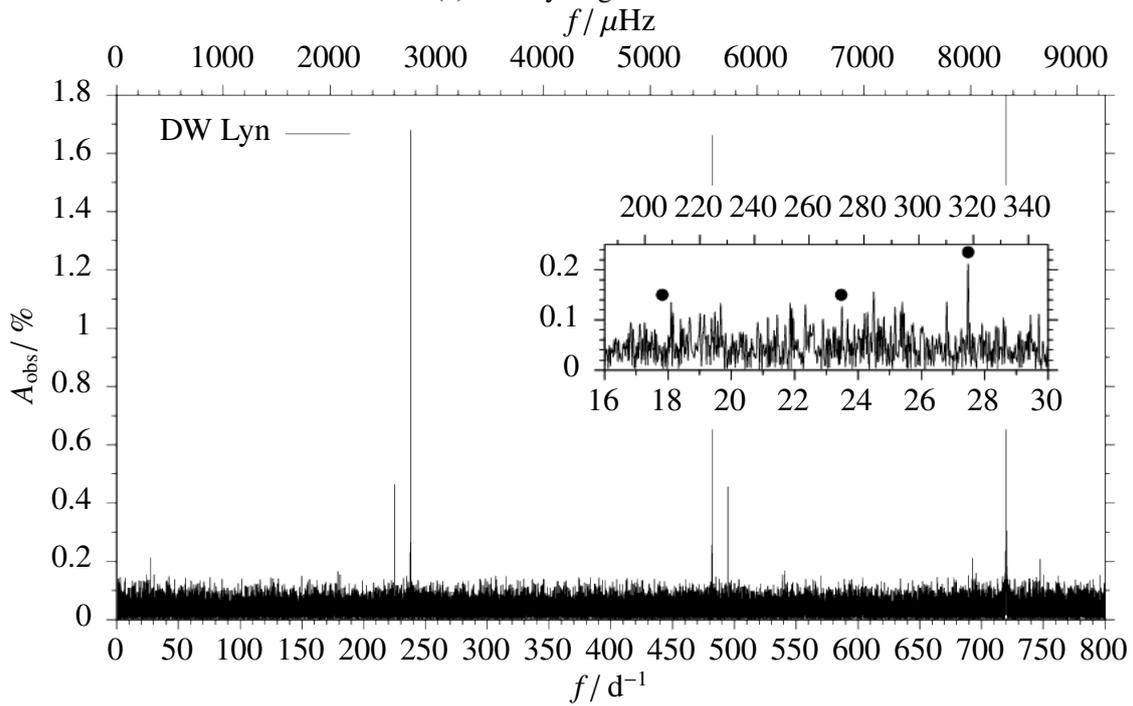}
		\caption{DW~Lyn amplitude spectrum.}
		\label{fig:dwlyn-power-tess}
	\end{subfigure}
	\caption[TESS DW~Lyn data.]{TESS DW~Lyn data. \textit{Top:} Light curve. Grey points are considered outliers and partially exceeding the plotting range. \textit{Bottom:} Amplitude spectrum. The inset shows a magnification of the g-mode regime. Modes reported by \citet{lutz_light_2008} are indicated as dots.}
	\label{fig:dwlyn-tess}
\end{figure}

DW~Lyn was observed by TESS with a cadence of $ \unit[120]{s} $ between December 25, 2019 and January 20, 2020. Fig.~\ref{fig:dwlyn-tess} shows the light curve and amplitude spectrum. Besides the \SI{120}{\second} alias at \SI{720}{\per\day}, the spectrum shows $ f_1 $ and $ f_2 $ at about \SI{238}{\per\day} and \SI{225}{\per\day} respectively, as well as their Nyquist alias above \SI{360}{\per\day} \citep[c.f.][]{murphy_super-nyquist_2013}. For the g-mode regime, \citet{lutz_light_2008} reported pulsations at \SIlist{271.7;318.1;206.3}{\micro\hertz} (\SIlist{23.47;27.48;17.83}{\per\day}). The TESS data show only the pulsation at \SI{318.1}{\micro\hertz} as a detection of about two times above the noise level. While the \SI{271.7}{\micro\hertz} pulsation is visible but not above the noise level, the \SI{206.3}{\micro\hertz} pulsation is not present in the TESS data set. Since we have no data between the last EXOTIME observations in 2013 and the TESS observations in 2020, the time span for a joint $ O-C $ analysis is too large. 

V391~Peg will be observed by TESS and the data can be used to extend the set of observation from \citet{silvotti_sdb_2018}. In order to span the observational gaps between 2013 and 2020, we performed additional ground-bases observations, as described in chapter~\ref{sec:targets}.
\cleardoublepage
	
\part{Summary}\label{sec:summary}
\chapter{Discussion} \label{sec:discussion}
Studying post main-sequence stars offers a window into the fate of planetary systems, including our own solar system. Can planets survive the final stages of stellar evolution and how do they in turn affect their host star's evolution? One discussed formation scenario of single sdB stars involves the influence of giant planets during the red giant phase of their progenitor. Observations can put constrains on the presence of sub-stellar companions orbiting sdB stars. 

Chapter~\ref{sec:exotime} investigated a unique set of long term ground-based observations of the four rapidly pulsating sdB stars DW Lyn, V1636 Ori, QQ Vir and V541 Hya. The main goals of the EXOTIME project are the determination of the evolutionary state of these stars and the constraints on sub-stellar companions to draw conclusions on possible formation scenarios.

The individual secular period drifts for different modes per target, derived from the $ O-C $ measurements, do not always agree with each other. These results therefore need to be interpreted with caution with regard to the evolutionary state of the star and hold only for the assumption of a linear change in period. For DW~Lyn, $ \dot{P}_{f1} $ is comparable to values derived from stellar models, and place the star in the evolutionary phase of core \chem{He} burning. $ \dot{P}_{f2} $ on the other hand, can be explained by the apparent mode splitting, whose origin cannot be confirmed with certainty. A possible explanation is the slow decay of one pulsation mode and an excitation of a mode with close frequency but not resolvable by the observations. Additionally, amplitude and phase variations indicate a mode beating effect. As shown in section~\ref{sec:testing}, two close - but not necessarily resolvable, pulsation modes can cause such effects. Measurements for the period changes of V1636~Ori need to be cautioned by the same arguments. For QQ~Vir, the identified radial order and degree of modes do not allow a direct comparison with stellar models but the sign of the $ \dot{P} $ measurements places QQ~Vir in the state of \chem{He} burning. The $ O-C $ measurements of V541~Hya show no significant change of period. If these results are not affected by observational limitations but originate from stellar evolution, the star might be in the transition phase between core \chem{He} burning and contraction. 

\begin{figure}[ht!]
	\centering
	\input{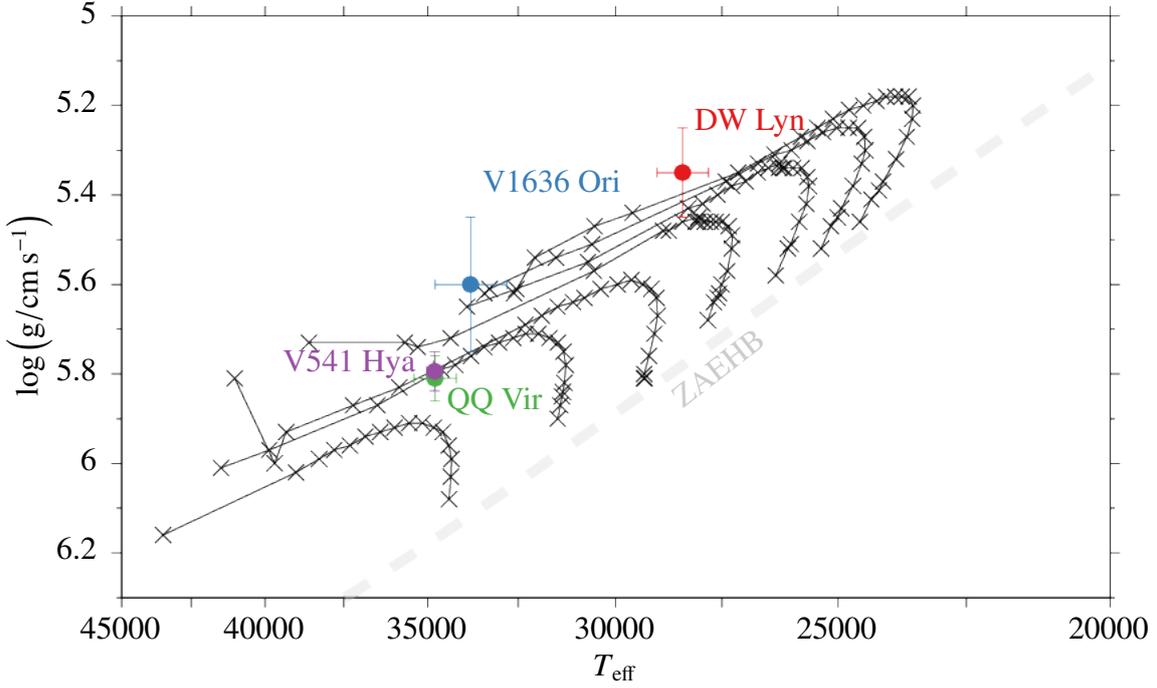}
	\caption[EHB evolutionary tracks of sdB stars.]{EHB evolutionary tracks covering the $ \log g - T_{\text{eff}}$ region where sdB stars are observed. Data points taken from Appendix B of \citet{charpinet_adiabatic_2002} (crosses) and Table~\ref{tab:stellar} (circles). The seven models correspond, respectively, from left (high $ T_{\text{eff}} $) to right (low $ T_{\text{eff}} $), to objects with a ZAEHB envelope mass $ M_{\text{env}} =$ 0.0001, 0.0002, 0.0007, 0.0012, 0.0022, 0.0032, and 0.0042 $ M_{\odot} $. The core mass is $M_{\text{c}} = 0.4758 M_{\odot} $ for all sequences except the first and third (from the left), which have $M_c = 0.4690 M_{\odot} $. For a positive (negative) slope in the models, the change in pulsation period is positive (negative). See \citet{charpinet_adiabatic_2002} for a detailed explanation.}
	\label{fig:disc:logg-teff}
\end{figure}

Atmospheric measurements of all four stars can place the stars in the $ \log g - T_{\text{eff}} $ plane relatively close to stellar evolutionary models of \citet{charpinet_adiabatic_2002}, as illustrated in Fig.~\ref{fig:disc:logg-teff}. This independently confirms the tentative interpretations above. DW~Lyn, V1636~Ori and QQ~Vir overlap each with several models of different envelope mass in the phase of core \chem{He} burning within their measured uncertainties. Although $ T_{\text{eff}} $ and $ \log g $ measurements of V541~Hya agree with two models in that same evolutionary phase, the measurements agree within $ \SI{3}{\sigma} $ with an evolutionary model in the transition between core \chem{He} burning and contraction. 

In principle, proper motions of stars can give an alternative explanation for the apparent change in pulsation period, as described by \citet{pajdosz_non-evolutionary_1995,pajdosz_effect_1995}. They derive the effect of proper motion to be 
\begin{align*}
\frac{\dot{P}_{\text{pm}}}{\si{\second\per\second}} = 2.430 \times 10^{-18} \frac{P}{\si{\second}} \frac{\mu}{ \si{''\,yr}^{-1}} \frac{d}{\si{pc}},
\end{align*}
where $ \mu $ is the proper motion in arcseconds per year and $ d $ the distance in parsec. With measurements available from the Gaia \citep{gaia_collaboration_gaia_2018}, the derived values are about three orders of magnitude smaller than the measured $ \dot{P} $ and thus neglectable. 

After subtraction of the long term trends in the $ O-C $ measurements, the residuals show no clear evidence for signatures of sub-stellar companions. The comparison with synthetic $ O-C $ curves, indicates a companion-induced signal in the main pulsation mode of DW~Lyn. But this is not confirmed by the independent measurement of the second pulsation mode, as would be expected. Furthermore, previously described periodic signals in the $ O-C $ measurements by \citet{lutz_exotime_2011} are ruled out to be caused by sub-stellar companions. Nevertheless, the data set is limited in detection sensitivity. Depending on the observational gaps, the data cannot set limitations for possible companion orbits shorter than approximately 30 days (DW~Lyn and V1636~Ori), 4 days (QQ~Vir) or 120 days (V541~Hya), or longer than the total length of observations, roughly 1500 days. Photometric precision additionally limits the detection threshold to companions signals larger than about \SI{1}{\second}. This corresponds to a companion with a mass of $ \SI{1}{M_{\jupiter}} $ ($ \SI{0.5}{M_{\jupiter}} $) and an orbital period of about \SI{0.5}{years} (\SI{1.5}{years}). 

Chapter~\ref{sec:other-targets} enlarges the target selection and applies the $ O-C $ analysis to one example. Pulsations of \textdelta~Sct  stars are driven by the \textkappa~mechanism, allowing stable pulsations over long periods of time. Planet detection methods, like the transit or radial velocity method, cannot be easily applied to those stars, which favours the pulsation timing analysis and thus opens a new parameter space in the exoplanet detection. The \textdelta~Sct pulsator KIC~7917485 was found to be orbited by a sub-stellar companion, whose signal can be recovered in the pulsation arrival times.

\chapter{Outlook} \label{sec:outlook}

As chapter~\ref{sec:other-targets} applied the $ O-C $ analysis to one example observation beyond the EXOTIME project, there are more extensive long time series of high cadence data already available, or with upcoming space missions. The next sections highlight data sets of interest. 

\section{\textit{K2} mission}

After the failure of two reaction wheels on the \textit{Kepler} space craft, its mission could be extended from 2014 onward as \textit{K2}. The space craft was balanced against the radiation pressure from the sun, which limits each campaign to a duration of about 80 days. Thus, \textit{K2} was not staring at only one field in the sky but sweeping along the ecliptic and could observe many more targets for signs of transiting exoplanets. \citet{armstrong_k2_2015} created a catalogue of variable stars from the \textit{K2} observations and categorized the targets into different classes of pulsators. From this catalogue, about 500 variables are classified as \textdelta~Sct stars. Additionally, 33 sdB stars were observed by \textit{K2} \citep{reed_review_2018}. Although the observations span only 80 days, they can set constraints on close in sub-stellar companions. 

\section{TESS}

The TESS mission aims to detect transiting exoplanets around bright stars. Asteroseismic targets are observed with a cadence of 20 seconds if they are bright enough. Most of the EXOTIME targets are already observed or are going to be observed. Due to the cadence of two minutes for these stars and the long duration in between the EXOTIME observations, the data cannot be used for a common $ O-C $ analysis. Nevertheless, they provide constraints on g-modes. For the upcoming V391~Peg observations, the gaps to ground based observations are not too big. With the planetary candidate claimed by \citet{silvotti_giant_2007} but cautioned later by \citet{silvotti_sdb_2018}, the new observations by ground based telescopes and TESS are expected to clarify the interpretations. 

For other asteroseismic targets, ether sdB or \textdelta~Sct stars, an $ O-C $ analysis is reasonable only for targets with overlapping sectors, i.e., more than 27 days of observing time. 

\section{PLATO}

PLAnetary Transits and Oscillations of stars (PLATO) is an ESA mission scheduled for launch in 2026. Its primary goal is to discover and characterize terrestrial exoplanets orbiting bright solar-type stars, especially  planets in the habitable zone. Bright host stars allow the determination of the planets' masses using the radial velocity method from ground-based observations and thus a mean density. This can be used to constrain planetary interior models. Additionally, bright targets allow transmission spectroscopy observations in order to measure chemical abundances of the planetary atmospheres. A secondary goal of the PLATO mission is asteroseismic observations to determine stellar masses, radii, and ages. These parameters are crucial in order to correctly determine the planetary masses, radii and their evolution.

\begin{figure}[tp]
	\centering
	\includegraphics[width=0.9\textwidth]{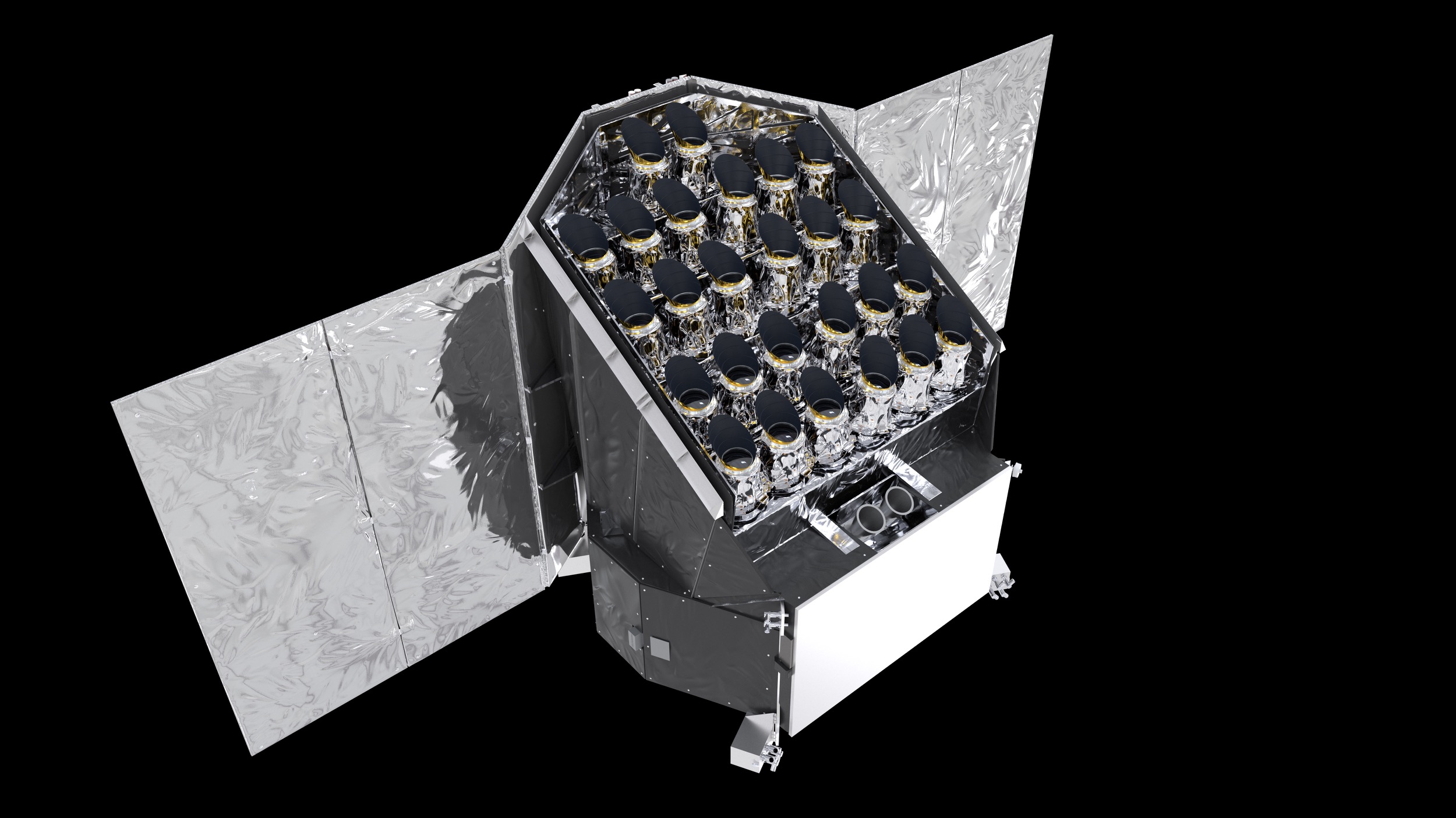}
	\caption[Artist's impression of ESA's PLATO spacecraft.]{Artist's impression of ESA's PLATO spacecraft. Credit: ESA/ATG medialab, CC BY-SA IGO 3.0}
	\label{fig:sum:plato}
\end{figure}

PLATO's payload module consists of 24 cameras, each with an aperture of \SI{120}{\milli\meter} and a \SI{1100}{\deg^2} field of view (see Fig.\ref{fig:sum:plato}). The cameras are arranged in four groups of six, where each group has the same field of view but their angle is slightly offset - this allows a total field view of about \SI{2250}{\deg^2} per pointing. The camera arrays overlap in some parts, resulting in different sensitivities over the field. They will observe with a cadence of \SI{25}{\second}. Additionally, two \enquote{fast} cameras will observe with a cadence of \SI{2.5}{\second}. PLATO's nominal life time sums up to four years but could be in orbit for double the time. For most of the time it will \enquote{stare} to one region in the sky. Additionally, an \enquote{step and stare} phase, comparable to \textit{K2} is possible.

With the explicit goal of asteroseismic observations, PLATO is expected to collect many data on evolved pulsating stars allowing for an investigation with the pulsation timing method.
The pipeline developed for this work can be easily adopted and used in an automated way in order to process many targets and search for sub-stellar companions. 

\appendix
\cleardoublepage
\part{Appendix}

\bibliographystyle{thesis} 
\bibliography{biblio}


\chapter*{Acknowledgements\markboth{Acknowledgements}{Acknowledgements}}
\addcontentsline{toc}{chapter}{Acknowledgements}
This thesis would not have been possible without the help of many people who guided and supported me during my time as a PhD student.

First I would like to thank my supervisor Sonja Schuh who advised and guided me trough this project and had always an open ear for me, despite her work as the IMPRS coordinator. 
A thank you goes also to the other members of my TAC Laurent Gizon and Stefan Dreizler for the productive discussions and ways to see some things differently.

I would also like to thank Roberto Silvotti, who helped to improve the EXOTIME paper.

The students of the IMPRS formed a great community at the institute. Without the friendships formed out of this group, my time at the institute would have been a very different experience. 
I tried gave my best to contribute to the great dynamics of the group and hope this spirit will be kept alive by many following generations of students!

Special thanks goes to my fellow musicians for the great time over the last years. Starting with the (in)famous successful \enquote{Quiet Sun} project, together with Theodosios Chatzistergos (literally a god on his guitar), Ankit Barik, Earl Bellinger and Alessandro Cilla (who left us way too early!), growing bigger for some gigs with Sudharshan Saranathan, Nils Gottschling, David Marshall, Robin Thor, Keaton Bell and James Kuszlewicz, and finally merging with the great MegaGauss band, including additionally Helge Missbach, Katja Karmrodt, Abbey Ingram, Christian Baumgartner, Daria Mokrytska, Tanayveer Bhatia, Marius Pfeifer, Ann-Kathrin Lohse (Atti) and Paula Wulff. No matter which problems would trouble me, during the band practise I could throw them all against the drum heads and cymbals. Keep on rockin'! 

I would like to thank Michelle for her support and proofreading.

And special thanks goes to my family which always supported me on my way, listened and offered good advice.

\vspace{12pt}
This work is founded by the International Max Planck Research School for Solar System Science at the University of G\"ottingen (IMPRS, Solar System School) and the Volkswagen Foundation (project grant number VWZN3020).
This work makes use of observations from the LCOGT network.
This work includes data collected by the Kepler mission. Funding for the Kepler mission is provided by the NASA Science Mission directorate.

\chapter*{Scientific contributions\markboth{Scientific contributions}{Scientific contributions}}
\addcontentsline{toc}{chapter}{Scientific contributions}
\large{\textbf{Refereed publications}}
\normalsize
\begin{itemize}
	
	\item Mackebrandt, F., M. Mallonn, J. M. Ohlert, T. Granzer, S. Lalitha, A. Garc\'{i}a Mu\~{n}oz, N. P. Gibson, et al. 2017. "Transmission Spectroscopy of the Hot Jupiter TrES-3 b: Disproof of an Overly Large Rayleigh-like Feature". A\&A 608 (December): A26. \url{https://doi.org/10.1051/0004-6361/201730512}.
	\item Mackebrandt, F., S. Schuh, R. Silvotti, S.-L. Kim, D. Kilkenny, E. M. Green, R. Lutz, et al. "The EXOTIME Project: Signals in the $ O-C $ Diagrams of the Rapidly Pulsating Subdwarfs DW Lyn, V1636 Ori, QQ Vir, and V541 Hya." A\&A 638 (June 1, 2020): A108. \url{https://doi.org/10.1051/0004-6361/201937172}.
	\item Mallonn, M., I. Bernt, E. Herrero, S. Hoyer, J. Kirk, P. J. Wheatley, M. Seeliger, et al. 2016. "Broad-Band Spectrophotometry of HAT-P-32 b: Search for a Scattering Signature in the Planetary Spectrum". Monthly Notices of the Royal Astronomical Society 463 (November): 604-14. \url{https://doi.org/10.1093/mnras/stw1999}.
	\item Schwope, A. D., F. Mackebrandt, B. D. Thinius, C. Littlefield, P. Garnavich, A. Oksanen, and T. Granzer. 2015. "Multi-Epoch Time-Resolved Photometry of the Eclipsing Polar CSS081231:071126+440405". Astronomische Nachrichten 336 (2): 115-24. \url{https://doi.org/10.1002/asna.201412151}.
	\item Strassmeier, K. G., I. Ilyin, E. Keles, M. Mallonn, A. J\"arvinen, M. Weber, F. Mackebrandt, and J. M. Hill. "High-Resolution Spectroscopy and Spectropolarimetry of the Total Lunar Eclipse January 2019." A\&A 635 (March 1, 2020): A156. \url{https://doi.org/10.1051/0004-6361/201936091}.

\end{itemize}

\noindent\large{\textbf{Conference contributions}}
\normalsize
\begin{itemize}
	\item "The stellar pulsation timing method to detect substellar companions": XXX IAU General Assembly 2018, Vienna (Poster)
	\item "The stellar pulsation timing method to detect substellar companions": Annual Meeting of the Astronomische Gesellschaft 2017, G\"ottingen (Talk)
	\item "The stellar pulsation timing detection method for substellar companions": Rocks \& Stars II Conference 2017, G\"ottingen (Talk) 
	\item "The stellar pulsation timing detection method for substellar companions": The PLATO Mission Conference 2017: Exoplanetary systems in the PLATO era, Warwick (Poster) 
	\item "The stellar pulsation timing method to detect substellar companions": 2nd Advanced School on Exoplanetary Science 2017, Vietri sul Mare (Poster) 
	\item "The Stellar Pulsation Timing Detection Method for Substellar Companions": Planetary Systems Beyond The Main Sequence II 2017, Haifa (Poster) 
\end{itemize}


\end{document}